\documentclass[reprint,
preprintnumbers,
superscriptaddress,
prd,
nofootinbib,
amsmath,amssymb, aps,
floatfix,
onecolumn
]{revtex4-2}
\usepackage[T1]{fontenc} % if needed
\makeatletter
\let\H@refstepcounter\refstepcounter
\makeatother
\usepackage{graphicx}
\usepackage{xcolor}
\usepackage{subcaption}
\usepackage{hyperref}
\hypersetup{
    colorlinks=true,       % True: colors the text; False: boxes the text
    citecolor=blue,        % Color for citation links (e.g., [1])
    linkcolor=black,       % Color for internal links (ToC, equations, etc.)
    urlcolor=teal          % Color for external web URLs
}
\usepackage{cleveref}

\newcommand{\quijote}{\texttt{Quijote} }

\newcommand{\bq}{\boldsymbol q}
\newcommand{\bx}{\boldsymbol x}
\newcommand{\bp}{\boldsymbol p}
\newcommand{\bk}{\textbf{k}}
\newcommand{\bn}{\textbf{n}}
\newcommand{\br}{\textbf{r}}
\newcommand{\bPsi}{\boldsymbol{\Psi}}

\newcommand{\ihmpc}{\,h{\rm Mpc}^{-1}}
\newcommand{\mpch}{\,h^{-1}\rm Mpc}
\newcommand{\be}{\boldsymbol e}
\newcommand{\flamingo}{\texttt{FLAMINGO} }
\newcommand{\kmax}{k_{\rm max}}

\newcommand{\appropto}{\mathrel{\vcenter{
  \offinterlineskip\halign{\hfil$##$\cr
    \propto\cr\noalign{\kern2pt}\sim\cr\noalign{\kern-2pt}}}}}

\begin{document}
\preprint{MIT-CTP/6131}

\title{Field-level modeling of halo and galaxy alignments}
\author{Nickolas Kokron}
\email{kokron@ias.edu}
\affiliation{School of Natural Sciences, Institute for Advanced Study, 1 Einstein Drive, Princeton, NJ 08540, USA}

\author{Mikhail M. Ivanov}
\email{ivanov99@mit.edu}
\affiliation{Center for Theoretical Physics -- a Leinweber Institute, Massachusetts Institute of Technology, Cambridge, MA 02139, USA} 
\affiliation{The NSF AI Institute for Artificial Intelligence and Fundamental Interactions, Cambridge, MA 02139, USA}

\author{James M. Sullivan}
\thanks{Brinson Prize Fellow}
\affiliation{Center for Theoretical Physics -- a Leinweber Institute, Massachusetts Institute of Technology, Cambridge, MA 02139, USA} 
\affiliation{The NSF AI Institute for Artificial Intelligence and Fundamental Interactions, Cambridge, MA 02139, USA}

\begin{abstract}
We develop an effective field theory (EFT) of intrinsic alignments at the level of spin-2 field fluctuations of large-scale structure tracers, such as the shapes of galaxies and dark matter halos. Our perturbative model
includes the non-linear treatment of the Zel'dovich displacement, and contains 
all necessary shape operators through the third order.
We apply our field-level model to the full 3D distribution of galaxy shapes, whose sky 
projections are a known
contaminant of cosmic shear measurements.
% of which contamination from intrinsic alignments is a line-of-sight projection. 
Our field-level EFT model achieves a better description of the helicity fields of dark matter halo shapes than the standard non-linear alignment (NLA) and tidal torquing + tidal alignment (TATT) models in the literature, as evidenced by improved correlations with the simulations' initial conditions. 
We perform precision measurements
of field-level bias parameters
by means of a generalized least-squares problem, for which we derive an optimal estimator and analytic covariance matrices.
% We then show the problem of field-level bias parameter inference can be cast as a generalized least squares problem, 
% This estimator is applied to 
We measure the full set of linear, quadratic, and cubic shape bias coefficients as a function of halo mass in the \quijote N-body simulations. We then study the intrinsic alignments of galaxies in the \flamingo hydrodynamical simulation,
focusing on observationally relevant samples: spirals, ellipticals, 
luminous red galaxies (LRG) and emission-line galaxies (ELG).
We detect linear and nonlinear alignment coefficients at high significance for LRG, spiral and elliptical galaxies.
We also find evidence for non-zero 
density-weighted tidal bias coefficient of the simulated ELG.  
% certain quenched and star-forming galaxies. 
This work supports the notion that EFT-based models of alignments will be paramount in optimally extracting constraints from future shape correlations, and lays the foundation for the development of simulation-based priors for future galaxy shape samples.
\end{abstract}
\maketitle
\makeatletter
\makeatother
\tableofcontents
\section{Introduction}
\label{sec:intro}
The shapes of dark matter halos and galaxies in the Universe are sensitive to the distribution of large-scale dark matter density fluctuations. The spatial correlations of shapes -- their `intrinsic alignments' -- have been the subject of interest to cosmologists for decades~\cite{Catelan_2001}. They represent both a significant contaminant to weak lensing surveys~\cite{Hirata:2004gc}, as these correlations are inevitably present in any measurement of cosmic shear or `galaxy--galaxy lensing', but they are also a cosmological signal of interest. As galaxy redshift and imaging surveys have become more powerful, large catalogs of galaxies with spectroscopic redshifts and high-quality shape measurements have been produced. 3D catalogs of galaxy shapes are promising probes of the BAO feature~\cite{Xu_2023}, can help pin down the nature of dark energy~\cite{Shim_2025}, and are potential probes of interesting new physics such as anisotropic non-gaussianities or parity violation~\cite{Akitsu_2021, Kurita_2023, kurita2025parityviolationgalaxyshapes}. \par 
Beyond spectroscopic samples, Rubin Legacy Survey of Space and Time~\cite{Mandelbaum:2018ouv}, the Roman Space Telescope~\cite{Eifler_2021,committee2025romanobservationstimeallocation, cao2026fisherforecastscosmologicalyields} and Euclid~\cite{laureijs2011euclid, euclid2025} space telescope will produce extremely dense samples of high-precision galaxy shapes. While these surveys will deliver shape catalogs of unprecedented quality, the degree to which intrinsic alignments can potentially contaminate weak lensing surveys is still in question. The depth to which modern lensing surveys will image galaxies (maximizing number density) far exceeds our spectroscopic follow-up capabilities, which are an important ingredient in data-driven alignment mitigation strategies. There have been many proposals to mitigate this effect, including selecting on subsets of galaxies which have smaller known alignment amplitudes (although only consistent with zero given measurement uncertainties)~\cite{McCullough_2026}, calibration and extrapolation of scaling relations from spectroscopic samples~\cite{siegel2025intrinsicalignmentdemographicsnextgeneration}, or alternative samples which can have their alignment amplitude directly calibrated from dedicated observing campaigns~\cite{derose2026steelingweaklensingsource}. \par 
So far, these efforts have predominantly focused on isolating and calibrating the \emph{linear} alignment amplitude of shapes with the dark matter distribution -- governed by its correlation with the large-scale tidal field. However, the symmetries of the Universe allow a much greater phenomenology of dependence of shapes, motivating `non-linear' alignment models which include additional degrees of freedom such as contributions from tidal torquing, density-weighting, and so on. As imaging surveys become more powerful, our sensitivity to these additional alignment contributions becomes larger, and current direct calibration methods are not necessarily optimal to infer their signatures. This is because direct calibration relies on two-point functions, which are relatively insensitive to these contributions (they are `1-loop' in the parlance of perturbation theory). From the perspective of analytic theory, many of the proposed models in the literature are not complete. EFT methods~\cite{Vlah_2020, Vlah_2021, bakx2023effectivefieldtheoryintrinsic, chenkokron2024, Maion_2024} are rigorous completions of the original crop of phenomenological models~\cite{Catelan_2001, Blazek_2019}. They have been tested rigorously for two point functions and, partially, for three point functions~\cite{Akitsu_2023, vedder2026threepointintrinsicalignmentsgalaxies, Bakx_2025, Bakx_2026}. Notably, the EFT formulations of intrinsic alignments suffer from a large amount of `nuisance' parameters -- both due to the large number of degrees of freedom and the completely uncertain time evolution of the signal for galaxies. Employing maximally conservative priors on the amplitudes of these effects often significantly degrades constraining power on physics of interest. As we see it, the current status of intrinsic alignments presents some challenges: 1) direct calibration cannot extend to samples of galaxies representative of Roman, LSST, especially for their higher-order alignments and 2) it is unclear whether the full basis of operators in the EFT is needed to model galaxy alignments, and 3) to what degree do alignment models capture the correlation of shapes with the underlying dark matter density v.s. simply over fit to correlation functions. \par 
\begin{figure}
    \centering
    \includegraphics[width=1\linewidth]{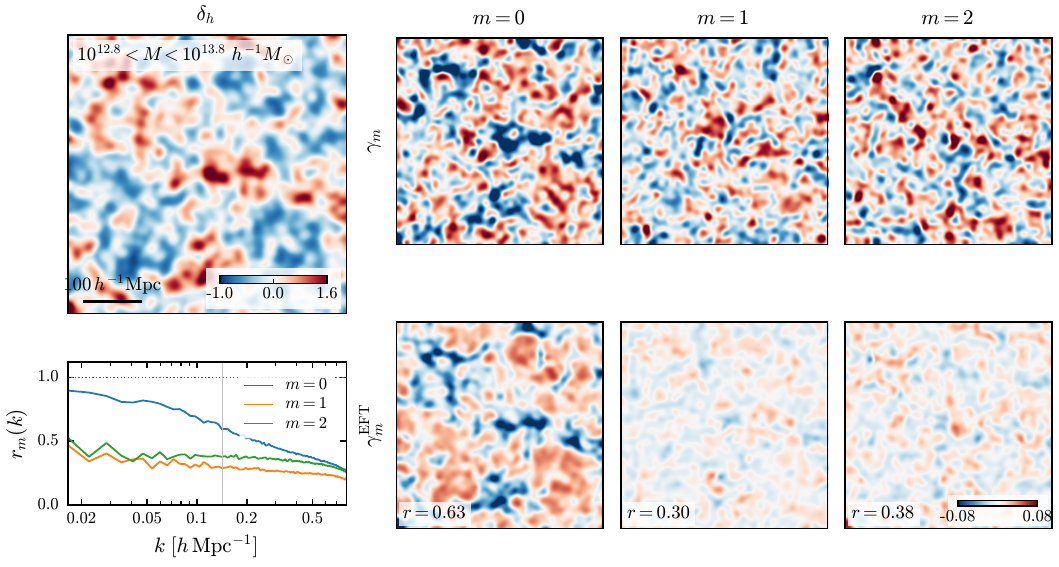}
    \caption{A visual summary of the field-level shape distribution. Visualization of the halo density (top left) and halo shape helicity fields (top right) for halos in $M_h \in [10^{12.8}, 10^{13.8}]M_\odot h^{-1}$, projected across $\approx 100 {\rm Mpc} h^{-1}$, smoothed with a Gaussian filter on $R=7 \,{\rm Mpc}h^{-1}$. The bottom-right panels show the best-fit field-level EFT model, as well as the cross-correlation coefficients between the simulation panels at this smoothing scale. For $m=\{1, 2\}$, stochastic noise dominates the field, leading to a lower cross-correlation coefficient. The bottom left panel shows the $k$-dependent cross-correlation coefficients between the EFT reconstructions and actual halo shape fields, and the vertical line corresponds to $k=1/R$.}
    \label{fig:halofield}
\end{figure}
The question of whether an EFT-based model of large-scale structure summary statistics adequately captures the full correlations or simply fits to said summaries is not unique to intrinsic alignments, it is a concern for all tracers of large-scale structure. To go beyond agreement at the level of summary statistics, there has been significant recent development in \emph{field-level} perturbative models. The key principle of field-level modeling is that, by describing the entirety of the distribution of the tracer in question, all summary statistics are naturally well-described. Pioneering work in perturbative field-level models for the case of galaxy clustering was carried out in \cite{Schmittfull_2019}, who developed a perturbative field-level forward model and applied it to study the distribution of halo number density fluctuations (see also \cite{Schmidt_2019, Cabass_2020, eft_like_2020, schmidt2021} for perturbative field-level models that use a different construction). This construction has since been extended to redshift space~\cite{schmittfull2020modeling}, galaxy densities in hydrodynamical simulations~\cite{ivanov2024millenniumastridgalaxieseffective, Shiferaw_2025}, the HI distribution at low redshifts and during reionization~\cite{Qin_2022, Obuljen_2023}, and
the Lyman-$\alpha$ forest~\cite{de_Belsunce_2026}. Beyond testing the validity of perturbative models in describing a tracer, by finding the set of EFT coefficients which minimize the field-level error, this approach is also a powerful estimator of said EFT coefficients. Measuring how EFT coefficients vary across different physical simulations of said tracers allows us to construct a mapping between model-dependent parameters in simulations and the general EFT space, understanding how different physics impacts EFT coefficients~\cite{barreira2021galaxy, ivanov2025fullshapeanalysissimulationbasedpriors, zhang2025hodinformedprioreftbasedfullshape,sarkar2026simulationbasedpriorshibias}. The measured values of coefficients can be used to calibrate priors for subsequent analyses in real data, enhancing the constraining power of EFT-based analyses~\cite{chudaykin2026reanalyzingdesidr15}. \par 
The focus of this paper is to develop the field-level formalism for the effective theory of intrinsic alignments, building on the recent development of a Lagrangian theory for IA ~\cite{chenkokron2024}, whose power spectrum was calculated at one-loop order. We aim to test the ability of the EFT in describing the full distribution of correlations of dark matter halo and galaxy shapes, beyond linear theory, applying it to the large \quijote suite~\cite{Villaescusa_Navarro_2020} of dark matter simulations as well as the recently publicly-released \flamingo suite of cosmological hydrodynamical simulations~\cite{Schaye_2023, Helly_2026}. A visual summary of the field-level distribution of halo shapes, and best fits from the EFT model, is shown in \Cref{fig:halofield}. The use of this field-level model for alignments will also allow us to extract the values of the coefficients in the theory at high statistical significance even when limited by the volumes available, in complementary fashion to bispectrum-based measurements~\cite{Akitsu_2023}, as well as `probabilistic estimators'~\cite{Maion_2025}, which have been previously used in large-volume dark matter only simulations. The techniques, simulations and samples we utilize in this work are distinct from past investigations of intrinsic alignments in \flamingo, such as \citep{herle2026assemblybiasredshiftevolution, herle2026intrinsicalignmentsflamingosimulations, vedder2026threepointintrinsicalignmentsgalaxies, van_Heukelum_2026}. The past works have mostly focused on massive halos which host luminous red galaxies, as well as emphasizing the largest volume \flamingo box, at $V=(2.8 {\rm Gpc})^3$. The statistical power of field level estimators will allow us to study extended samples of galaxies which are only resolved by the lower volume \flamingo simulation. \par 
This paper is structured as follows: in \S~\ref{sec:methods} we review the theory of spin-2 biased tracers of the density field, constructing the set of 11 operators at cubic order, introducing the \emph{shifted} basis, as well as the helicity decomposition we will use throughout this work to simplify measurements. We then discuss the field-level model of spin-2 tracers, and the associated \emph{transfer functions} in \S~\ref{sec:shapetransfers}. In \S~\ref{sec:galileons} we discuss degeneracies in the spin-2 expansion that are different from those of scalar biases. Two key results from this section are the re-definition of the basis into the spin-2 analog of the Galileon basis often used in the scalar bias expansion, as well as uncovering a previously unknown redundancy that eliminates one operator, $s^2 s_{ab}$, from the basis, leading to 10 total operators as opposed to 11. In \S~\ref{sec:tffit} we define a generalized least-squares problem which leads to an estimator for the shape bias coefficients, as well as their uncertainties. \S~\ref{sec:simulations} introduces the data we will analyse in this paper -- the large-volume \quijote ensemble dark matter simulations and the \flamingo hydrodynamic simulations. We discuss how shapes are measured in halos and galaxies for both simulations, as well as the different definitions of tracer shapes we will study in this paper. \S~\ref{sec:haloalign} applies the field-level estimator to study the shape statistics of dark matter halos in both simulations, while \S~\ref{sec:galaxyalign} studies the shape statistics of various galaxy samples in \flamingo. \S~\ref{sec:conclusions} summarizes our findings. \par 

Technical details are relegated to a series of appendices: \Cref{app:tfortho} discusses the construction of the orthogonalized transfer functions used in this work from first principles, \Cref{app:lagbasis} validates the spin-2 basis and helicity decomposition numerically against predictions from Gaussian field correlators, \Cref{app:tensorrecon} discusses how we independently recreate corrupted data from the \flamingo simulation, \Cref{app:tferror} validates the analytic expressions for transfer function and parameter covariance matrices derived in this work, \Cref{app:transfers} shows an extended set of transfer functions for various halo masses that are not shown in the main work, and finally \Cref{app:conversion} derives the conversion between the shape bias coefficients used in this work and the alignment parameters customarily adopted in the weak lensing community. 
\section{Spin-2 theory of biased tracers -- Recap and conventions}
\label{sec:methods}

We begin by reviewing how we will define the distribution of shapes in this work, the correlations we will study, as well as the construction of the EFT model used in this work. Halo and galaxy shapes are spin-2 tracers of large-scale structure. In \S~\ref{sec:simulations} we specify our exact scheme for how we construct a \emph{field} of halo shapes from a catalog. For now, suppose a spin-2 field of shapes, $I_{ab} (\bx)$, exists. The shape field can be decomposed into trace and trace-free parts
\begin{equation}
    I_{ab} (\bx) = \frac{\delta_{ab}^K}{3} \delta_I (\bx) + \gamma_{ab} (\bx).
\end{equation}
The trace of an object's quadrupole tensor is related to its total size, and so $\delta_I (\bx)$ is dubbed the \emph{size fluctuation field}. Correspondingly, $\gamma_{ab} (\bx)$ is the \emph{shape fluctuation field}. It is $\gamma_{ab}$ which we will concern ourselves with in this work. $\gamma_{ab}$ is a symmetric, trace-free, spin-2 field. Guided by the equivalence principle, we can write a basis for the large-scale cosmological dependence of $\gamma_{ab}$ in terms of all tensor products which can be built from the Hessian of the large-scale gravitational potential $\partial_a \partial_b \Phi$, which decomposes into trace and trace-free fields 
\begin{equation}
\label{eqn:phideriv}
    \partial_a \partial_b \Phi = \frac{\delta_{ab}^K}{3} \delta_{\rm lin} + s_{ab}, 
\end{equation}
where $\delta_{\rm lin}$ is the linear matter density contrast and $s_{ab}$ is the traceless tidal tensor
\begin{equation*}
    s_{ab} = \underbrace{\left (\frac{\partial_a \partial_b}{\partial^2} - \frac{\delta_{ab}^K}{3} \right )}_{\equiv \mathcal{D}_{ab}} \delta_{\rm lin}.
\end{equation*}
Note that we have not specified the reference frame in which the expansion is constructed -- either the late-time Eulerian frame with coordinates $\bx$, or at some earlier time in the Universe (which we call the Lagrangian coordinates $\bq$). We will drop the `lin' subscript from $\delta$ for the remainder of this paper. The two most commonly-adopted intrinsic alignment models, `nonlinear linear alignment' (NLA)~\cite{Catelan_2001, 2007NJPh....9..444B} and `tidal alignment and tidal torquing' (TATT)~\cite{Blazek_2019}, can be derived from counting powers of $\partial_a \partial_b \Phi$. The shape of an object, $\gamma_{ab}$, is often defined as the quadrupole tensor of its observed constituents (e.g. luminosity for galaxies, dark matter particle content for dark matter halos) -- which is by definition traceless and symmetric in the $(a,b)$ indices. Thus, to linear order in the Hessian of the potential $\partial_a \partial_b \Phi$ the only dependence that can exist is 
\begin{equation}
    \label{eqn:LA}
    \gamma_{ab} (\bx) = c_s s_{ab} (\bx), 
\end{equation}
where $c_s$ is the \emph{linear alignment coefficient}. It is often defined as `$A_1$' in the weak lensing literature and the conversion between $c_s$ and $A_1$ is given in \Cref{app:conversion}. Choosing $s_{ab}$ to correspond to the \emph{linear} tidal field gives the `linear alignment model', but the traceless shear operator $\mathcal{D}_{ab}$ can also be applied to the \emph{nonlinear} density field. This is equivalent to stating that the shapes of galaxies are immediately aligned and correlated with the present-day tidal field. The next tensor that one can construct from $\partial_a \partial_b$ is its square 
\begin{equation*} 
\partial_{a} \partial_c \Phi \partial_c \partial_b \Phi = \frac{\delta_{ab} \delta^2}{9} + \frac{2}{3} s_{ab} \delta+ (s^2)_{ab}. 
\end{equation*}
which generates two new traceless tensor dependences -- $(\delta s)_{ab}$ and $(s^2)_{ab}$. This density-weighted contribution is naturally expected to be present, for we only measure the shapes of galaxies where there are galaxies. The second term is naturally generated from tidal torquing theory~\cite{1984ApJ...286...38W}. The fields can also be generated from taking products of the trace and trace-free components of $\partial_a \partial_b \Phi$ to second order. There is no dynamical reason for their coefficients to be related, and so their inclusion generates the full `TATT' model 
\begin{equation}
    \label{eqn:tatt}
    \gamma_{ab} (\bx) = c_s s_{ab} (\bx) + c_{\delta s} (\delta s)_{ab} (\bx) + c_{s^2} (s^2)_{ab} (\bx).
\end{equation}
There is one additional trace-free tensor that can be generated from the current fields at second order. Note that the following zero-mean, second-order, scalar field can be written
\begin{equation}
    \mathcal{G}_2 (\bx) = {\rm Tr}[(s^2)_{ab}](\bx) - \frac{2}{3} \delta^2 (\bx).
\end{equation}
This field in the context of scalar biasing theory is called the second-order \emph{Galileon} field. Applying the tidal operator to this Galileon, $\mathcal{D}_{ab} \mathcal{G}_2$, generates a second-order traceless, symmetric spin-2 field which is distinct from the other two terms. This algorithmic construction of fields (products of current tensors + applying $\mathcal{D}_{ab}$ to new scalars at a given order) generates the full spin-2 basis of shape fields~\cite{Vlah_2020}. \par
Extending this construction to third order in $\partial_a \partial_b \Phi$, there are 11 fields that can be written down for the shape bias basis:
\begin{align*}
    1^{\rm st} \text{ order: } &s_{ij}\\
    2^{\rm nd} \text{ order: } &t_{ij},\ s^2_{ij}, \ \delta s_{ij}\\
    3^{\rm rd} \text{ order: }  &L^{(3)}_{ij},\ (st)_{ij},\ \delta t_{ij},\ s^3_{ij},\ s^2 s_{ij}, \delta^2 s_{ij}, \ \delta s^2_{ij}.
\end{align*}
The Galileon term is normalized as $t_{ij} = \frac27 \mathcal{D}_{ij} \mathcal{G}_2 = \frac27  \mathcal{D}_{ij} \left ( s^2 - \frac23 \delta^2\right ) $, and is commonly called the \emph{velocity shear} tensor in the literature~\cite{gomes2026tidalalignmenttidaltorquing}. It is equivalently defined by the trace-free part of the gradient of the second-order displacement computed in lagrangian perturbation theory (LPT) -- $t_{ab} \propto {\rm TF} [\partial_a \Psi_{2b}]$. The operator
$L^{(3)}_{ij}$ is equivalently built from the third order LPT displacement $L^{(3)}_{ab} \propto {\rm TF}[\partial_a \Psi_{3b}]$. We have neglected higher-derivative operators in the above expression, but at one-loop order only a higher-derivative tidal tensor $\nabla^2 s_{ab}$ field contributes. We can write this basis for the protohalo shape field at Lagrangian coordinates $\bq$, and then match it to the late-time shape fluctuation field by displacing the fields by the same Lagrangian displacements $\mathbf\Psi_{n}$ used to build the basis.\par 
Before continuing onward let us make some remarks about this basis. When studying the IA power spectrum at one-loop order, the cubic basis can be further simplified. Operators which are only built from powers of $s_{ij}$ or $\delta$ (so-called `Zeldovich operators'), are fully degenerate with the linear shear tensor\footnote{This can be seen by cross-correlating with these fields with the linear matter density contrast. Wick contractions in configuration-space show these correlators resolve exactly to products with structure $\propto \sigma^2 \langle s_{ab} \delta \rangle (\bk)$.} and can be neglected. 
From the three remaining cubic operators in the expansion, $L_{ij}^{(3)}, (st)_{ij}, \delta t_{ij}$, only two are non-degenerate at the one-loop power spectrum level~\cite{Vlah_2020, chenkokron2024} due to a constraint that appears for the tadpole momentum configuration of the one-loop diagram. However,
all three are non-trivial
at the field level. Another remark we make, but which we will expand on in \S~\ref{sec:galileons}, is that there are in fact only 10 independent operators in the cubic shape bias basis, instead of the usual 11 that are written down. A constraint that was written down in the one-loop context in Ref.~\cite{Vlah_2020} extends to the entire field level, which leads to finding the $s^2s_{ij}$ is redundant and can be dropped from the basis.
\par 
Having enumerated the standard bias shape basis, we write the complete perturbative expansion for the shape fluctuation field in terms of the allowed operators\footnote{In principle, there are ``mixed'' contributions e.g. $\gamma_{ab}\supset \delta\epsilon_{ab}$, but these no have observational consequences
at the power spectrum level. They may be important for the higher-order statistics, see e.g.~\cite{Ivanov:2021kcd,Bakx:2025pop} for the example of the galaxy density bispectrum.} 
\begin{equation}
\label{eqn:shapebias}
    \gamma_{ab}  = \sum_{\mathcal{O} \in {s,\delta s,s^2,\cdots}} c_\mathcal{O} \mathcal{O}_{ab} + \epsilon_{ab}
\end{equation}
where all 10 operators receive a coefficient $c_{\mathcal{O}}$ and $\epsilon_{ab}$ is the \emph{shape stochasticity field}, which by definition is uncorrelated with any of the deterministic contributions. 
The symmetries of the EFT constrain the correlations functions 
of $\epsilon_{ab}$. In particular,
in the limit $k\to 0$ the only
form allowed by symmetries is 
\begin{equation}\label{eq:Poiss}
    \langle \epsilon_{ab}(\bk)\epsilon_{cd}(\bk')\rangle =(2\pi)^3\delta_D(\bk+\bk')\left(\delta_{ac}^K\delta_{bd}^K+\delta_{ad}^K\delta_{bc}^K-\frac{2}{3}\delta_{ab}^K\delta_{cd}^K\right)P_{\epsilon}\,, 
\end{equation}
where $P_{\epsilon}$ is a constant.
It is parameterized as
$P_\epsilon =\sigma^2_\gamma/\bar n$, where $\sigma^2_\gamma$
is the shape noise 
dispersion and $\bar n$
is the galaxy number density.

\Cref{eqn:shapebias} can be defined either in late-time, Eulerian coordinates $\bx$ or in the Lagrangian coordinates $\bq$ in the initial conditions of the Universe. The Eulerian and Lagrangian shape fields of halos can be formally related to each other through the displacement $\bPsi (\bq)$ 
\begin{equation}
\label{eqn:advectionintegral}
    \mathcal{\gamma}_{ab} (\bx) = \int d^3q \,\delta^D (\bx - \bq - \bPsi(\bq))R_{ac} (\bq) R_{bd} (\bq) \mathcal{\gamma}^L_{cd} (\bq),
\end{equation}
where $R_{ac} \equiv \delta_{ac} + \partial_a \Psi_c$ is an active alignment contribution which distorts halo shapes from their initial proto-halo distribution, and $\gamma^L$ is the shape fluctuation field of \emph{proto-halos}~\cite{chenkokron2024}. A common assumption that is made in alignment theories is to assume ``passive evolution'' models of halo shapes. In the passive alignment picture, the distortion matrix $R_{ac}$ is neglected -- halo shapes are frozen in at early times and their centers are displaced~\cite{Schmitz_2018}. When passive alignment is assumed, predictions can be made between the coefficients of the Lagrangian bias expansion and the Eulerian bias expansion, similar to the `co-evolution' relations that have been derived for the bias of galaxy density fluctuations~\cite{Abidi:2018eyd}.\par 
The ``active" v.s. ``passive" distinction does not apply to the basis fields of the bias expansion themselves, which are defined in the Eulerian or Lagrangian frames and connected by a displacement vector. In this paper, we adopt the ``shifted operator" basis used to study field-level perturbative EFT, first defined in \cite{Schmittfull_2019}. Shifted operators are defined by treating the Zeldovich displacement non-perturbatively, but treating higher-order contributions in perturbation theory. In Fourier space a shifted shape operator is defined as
\begin{align}
    \tilde{\mathcal{O}}_{ab} (\bk) = \int d^3q e^{-i \bk \cdot \bq} e^{-i \bk \cdot \bPsi_{1}(\bq)}\mathcal{O}_{ab}(\bq),
\end{align}
where a tilde applied to an operator denotes it is in the shifted basis. \par 
We implement the tensor basis at the field level as follows: given the linear initial conditions used to seed a simulation, $\delta_{\rm IC} (\bq)$, we construct the traceless tidal tensor $s_{ab} (\bq)$ through a Fourier-space application of the $\mathcal{D}_{ab}$ operator. Given $\delta$ and $s_{ab}$ we generate the rest of the basis by taking appropriate configuration space tensor products, and applying $\mathcal{D}_{ab}$ in Fourier space as needed. All tensor fields of the cubic basis are generated in this form. Cubic fields are `renormalized' with respect to their overlap with the linear tidal field $s_{ab}$, such that the implemented cubic operators are
\begin{equation}\label{eq:b1ren}
\begin{split}
    & s^3_{ab}\to  s^3_{ab} - \frac{7}{15}\sigma^2 s_{ab},\quad \delta^2 s_{ab}\to \delta^2 s_{ab} - \sigma^2 s_{ab}, \quad \delta s^2_{ab}\to \delta s^2_{ab}  \\
    &st_{ab}\to st_{ab} - \frac{8}{315} \sigma^2 s_{ab}, \quad  \delta t_{ab}\to \delta t_{ab} - \frac{8}{105} \sigma^2 s_{ab},  
\end{split}
\end{equation}
and $L^{(3)}_{ab}$ is also not renormalized. For cubic operators, the underlying component fields are filtered in order to prevent aliasing from affecting the fields on large scales. For memory purposes, given we will focus on perturbative scales, we construct the basis on downsampled initial conditions. Our fiducial simulation has $L = 1000 \mpch $ and $N=1024$ grid cells, but the basis fields are constructed from $N=512$ linear grids downsampled in Fourier space (in order to preserve the phases of the field) to $N=256$ where transfer functions are measured. As such, the Nyquist frequency of our fiducial analysis is $k_{\rm Nyq} \approx 0.8 \ihmpc$. To build shifted operators, we sample displacements from the full resolution initial conditions and downsample them to the resolution of the Lagrangian operators. The displacement field is used to shift the Lagrangian shape tensors to their Eulerian coordinates. Each of the 6 components of the tensor are shifted simultaneously. The numerical implementation is the same as \cite{Kokron_2021}, except with perturbative displacements in lieu of the full $N$-body ones. We refer to that paper for further details. 
\par 
The two-point correlations of shape operators are in principle rank 4 tensors. In Ref.~\cite{Vlah_2020} it was shown that tensor correlators simplify significantly if the fields are initially decomposed into the helicity basis of tensor spherical harmonics 
\begin{equation}
    \mathcal{O}_{ab} (\bk)= \sum_{m=-2}^2 Y_{\ell=2,\,ab}^m (\hat{\bk}) \mathcal{O}^m (\bk). 
\end{equation}
Each $m$-mode is orthogonal, and so this basis diagonalizes the power spectra of tensor fields
\begin{equation}
    \langle \mathcal{O}_m (\bk) \mathcal{O}'_{m'}(\bk') \rangle = (2\pi)^3 \delta^K_{mm'} \delta^D(\bk + \bk') P^{m}_{\mathcal{O}\mathcal{O}'} (k).
\end{equation}
In the helicity basis, shape power spectra are simplified to five scalar spectra. In the absence of parity violation, the positive and negative $m$ are the same~\cite{kurita2025parityviolationgalaxyshapes}, and only three spectra are required to describe the two-point statistics of the shape field. The spherical tensors are defined in Fourier space in the following way: For every $\bk$, given a generic unit vector $\hat{\boldsymbol{n}}$, we construct an orthonormal basis  $\{\hat{\bk}, \hat{\be}_1, \hat{\be}_2\}$ where
\begin{equation}
\label{eqn:orthoe}
    \hat{\be}_1 = \frac{\bk \times \hat{\bn}}{|\bk \times \hat{\bn}|}, \quad \hat{\be}_2 = \hat{\bk} \times \hat{\be}_1.
\end{equation}
From this orthonormal basis we construct the helicity basis, composed of raising and lowering vectors $\{\hat{\bk}, \hat{\be}_+, \hat{\be}_- \}$ with $\hat{\be}_{\pm} = \frac{1}{\sqrt{2}}(\hat{\be}_1 \pm i \hat{\be}_2)$. We now define the spherical tensors as symmetric, trace-free outer products of $\hat{k}, \hat{e}_\pm$:
\begin{equation}
\label{eqn:sphericaltensors}
    Y_{\ell=2, ab}^{m=0} (\bk) = \sqrt{\frac{3}{2}} \left ( \hat{k}_a \hat{k}_b - \frac{1}{3} \delta_{ab} \right),\quad Y_{\ell=2, ab}^{m=\pm1}(\bk)  = \sqrt{\frac{1}{2}} \left ( \hat{k}_a \hat{e}^{\pm}_b + \hat{k}_b \hat{e}^{\pm}_a\right), \quad  Y_{\ell=2, ab}^{m=\pm2}(\bk) =\hat{e}^{\pm}_a\hat{e}^{\pm}_b.
\end{equation}
To construct these basis tensors on the lattice we pick a random vector $\bn$ which is not co-linear with any of the wavenumbers $\bk$ in the Fourier grid of the simulation\footnote{Poorly picking a vector $\bn$ which lies on the Fourier grid leads to a singular definition of the basis due to the cross products in \Cref{eqn:orthoe}.} and implement \Cref{eqn:orthoe,eqn:sphericaltensors} using the resulting array of basis vectors. Projecting the shifted operators with the $Y^m_{\ell = 2, ab}$ lattice generates the helicity components of different shifted operators.\par 
We show validation of our Lagrangian basis implementation, and the field-level helicity decomposition, against analytic theory predictions in \Cref{app:lagbasis}. 
\textbf{Note:} Henceforth, every index pair $ab$ or $cd$ in this work is implicitly assumed to be symmetrized and trace-free in the pair.

\section{The transfer functions of shapes}
\label{sec:shapetransfers}
A catalog of tracer shapes and their positions can be used to generate the shape fluctuation field $\gamma_{ab} (\bk)$. We decompose the field-level expression for intrinsic alignments (\Cref{eqn:shapebias}) into the helicity basis by projecting the tensor expression with $Y^{m;*}_{ab} (\bk)$
\begin{equation}
    \gamma_m(\bk) = \sum_{\mathcal{O}} c_{\mathcal{O}} \tilde{\mathcal{O}}_{m}(\bk) + \epsilon_m (\bk).
\end{equation}
For any set of coefficients $c_{\mathcal{O}}$ the statistics of $\epsilon_m(\bk)$ quantify how good of a description the deterministic component of the model is to the full shape fluctuation field. 
Coefficients $c_{\mathcal{O}}$
are estimated by minimizing 
the difference $|\gamma_m(\bk) - \sum_{\mathcal{O}} c_{\mathcal{O}} \tilde{\mathcal{O}}_{m}(\bk) |$,
which is equivalent to the minimization of the \emph{error power spectrum}
\begin{equation}
    \langle \epsilon_{m}(\bk)\epsilon_{m'}(\bk') \rangle = (2\pi)^3 \delta^{D}(\bk + \bk') \delta_{mm'}^K P_{\rm err}^{m}.(k),
\end{equation}
There are, in total, three error power spectra $P_{\rm err}^{m}(k)$ due to the diagonality in $m$ and the equivalence of $\pm m$ spectra. Minimizing $P_{\rm err}(k)$ as a function of scale is a central objective of the perturbative field-level program, so it is worth understanding how it works for the case of spin-2 fields in a simplified case. \par 
For intuition, let's assume we only have a linear shape bias field, so 
\begin{equation}
\label{eqn:linbiasfield}
    \gamma_m (\bk) = c_s \tilde{s}_m (\bk) + \epsilon_m (\bk).
\end{equation}
In order to achieve the best possible field-level fit to the measured shape field, we can promote the coefficient $c_s$ to a \emph{transfer function} $\mathcal{C}_{s;m} (k)$. 
This can be thought of as a 
EFT-improved version of the 
model, which includes all necessary 
higher-derivative counterterms
in the model. 
The resulting error power spectrum for this model is given by 
\begin{equation}
    P_{\rm err}^{m}(k) = \langle|\gamma_m(\bk) - \mathcal{C}_{s;m}(k)\tilde{s}_m(\bk)|^2 \rangle.
\end{equation}
Minimizing this error power spectrum for each $k$-bin leads to an estimator for the transfer function 
\begin{equation}
\label{eqn:lineartransfer}
    \hat{\mathcal{C}}_{s;m}(k) = \frac{\langle \tilde{s}_m(\bk) \gamma_m^*(\bk) \rangle}{\langle \tilde{s}_m(\bk)\tilde{s}_m^*(\bk)\rangle}.
\end{equation}
Several comments are in order. 
First, 
the transfer functions of spin-2 tracers are $m$-dependent. In fact, we will show in the next section that each helicity is sensitive to a different combination of shape bias coefficients. Note that the existence of higher-$m$ transfer functions for the linear shape bias is a consequence of the shifted basis -- $\langle \tilde{s}_m(\bk)\tilde{s}_m(\bk)\rangle$ has no $m=1$ or $m=2$ contribution without advection~\cite{chenkokron2024}. 

The second important observation
is that the error field defined as 
$\epsilon_m=\gamma_m -\gamma^{\rm model}_m  $
by construction includes higher-order deterministic contributions not included
in $\gamma^{\rm model}_m$. If the model is 
sufficiently 
accurate, we expect the 
error power spectrum to match the Poisson-like expectation~\eqref{eq:Poiss}, which predicts a uniform $m$-independent 
$P_{\rm err}=\sigma^2_\gamma/\bar n$ on large scales. The amplitude of $P_{\rm err}$
in the $k\to 0$ limit is affected by both 
higher-order EFT operators, and physical 
effects related to halo exclusion~\cite{Baldauf_2013}. The former contribution can be removed by explicitly including the 
higher-order bias operators in the field-level
shape model, which we discuss now. 

\par 
This extension to higher-order operators is implemented by performing a helicity-dependent orthogonalization of the operators, 
\begin{equation}
    \gamma_m (\bk) = \sum_{\mathcal{O}} \mathcal{C}_{\mathcal{O};m} (k) \widetilde{\mathcal{O}}^{\perp}_m + \epsilon_m (\bk),  
\end{equation}
such that every new operator that is included is by construction uncorrelated with previous members of the basis. An advantage of performing this orthogonalization in the helicity basis is that the process becomes fully analogous to the standard approach of scalar fields from Ref.~\cite{Schmittfull_2019}, but with five copies for each operator. The field-level decomposition of galaxy shapes is thus described by three transfer functions per operator, 
\begin{equation}
\label{eqn:transferdef}
    \hat{\mathcal{C}}_{\mathcal{O};m} (k) = \frac{\langle \widetilde{\mathcal{O}}^{\perp}_m \gamma_m \rangle (k)}{\langle |\widetilde{\mathcal{O}}^{\perp}_m|^2 \rangle (k)}, 
\end{equation}
after averaging over $\pm m$ modes. Some details of the numerical orthogonalization are discussed in \Cref{app:tfortho}. \par 
\subsection{Transfer functions of shifted shape fields}
We can glean some more intuition for the shape transfer functions, and their m-dependence, from perturbation theory. There are some differences compared to the usual theory for scalar fields -- namely, the parameter dependence of higher-helicity transfer functions. \par 
We begin by recalling how perturbation theory can be used to understand corrections to the field-level galaxy bias expansion. Ref.~\cite{Schmittfull_2019} expresses the halo density field in shifted operator form as 
\begin{equation}
\label{eqn:densshift}
    \delta_h (\bk) = \int_{\bq} e^{-i \bk \cdot \bPsi_1} \left [ 1 + \delta_h^L + i k_a (1 + \delta_h^L) \partial_b \Psi_1^a \bar{\Psi}^b - (1 + \delta_h^L) \nabla \cdot \bar{\bPsi} - \bar{\bPsi} \cdot \nabla \delta_h^L \right ] e^{-i \bk \cdot \bq},
\end{equation}
where $\bar{\bPsi} = \bPsi_2 + \bPsi_3 + \cdots$. This relation arises from writing the Eulerian halo distribution with a fully nonlinear shift, and expanding the perturbative displacements out of the exponent. We can do the same expansion for shapes, but there is a key difference: there is no conservation principle to convert between Eulerian and Lagrangian shapes, unlike the relation $(1+\delta_h(\bx))d^3x = (1+\delta_h^L(\bq))d^3q$ for number densities. We \emph{define} the Eulerian shape $\gamma (\bx)$ to be the result of an advection process 
\begin{equation}
    \gamma_{ab} (\bx) = \int d^3q \delta_D(\bx - \bq - \bPsi(\bq)) \gamma^L_{ab} (\bq), 
\end{equation}
where $\gamma^L_{ab} (\bq)$ was the proto-halo field expanded in the shape bias basis. For a given sample of halo shapes, we can find values for the Lagrangian coefficients $c_\mathcal{O}$ such that this equivalence is approximately satisfied, but there is no guarantee they are related to the coefficients of an Eulerian expansion. This is the ``passive advection'' picture, often assumed in the literature~\cite{Schmitz_2018}, and which is equivalently formulated as a conservation ansatz 
\begin{equation}
\label{eqn:passiveadvection}
    \gamma_{ab}(\bx)d^3x = \gamma^L_{ab}(\bq) d^3q.
\end{equation}
From the above equation, `co-evolution relations' for the coefficients can be derived~\cite{Schmitz_2018, Taruya_2021}, but it is not clear to which degree they should hold in practice, as they neglect the contribution from active $R_{ab}(\bq)$-type perturbations. This passive advection principle assumes that the Lagrangian shapes of halos merely undergo parallel transport. We can derive the helicity-dependence of the $\tilde{s}$ transfer function from this passive advection picture, assuming only Zeldovich displacements. Let us revisit the linearly biased shape field from \Cref{eqn:linbiasfield}, neglecting the stochasticity term for now. After shifting, we have
\begin{align}
    \tilde{\gamma}_{ab} (\bk) &= c_s \tilde{s}_{ab} (\bk) = c_s \int d^3q e^{-i\bk \cdot \bq}e^{-i \bk \cdot \bPsi_1(\bq)} s_{ab} (\bq). 
\end{align}
Expanding the $\bPsi_1(\bq)$ exponent term and integrating by parts we arrive at 
\begin{equation}
    \tilde{\gamma}_{ab} (\bk) = c_s \left [ s_{ab}  - \bPsi_{1} \cdot \nabla s_{ab} + \delta s_{ab} \right ](\bk).
\end{equation}
Thus, the linearly biased shifted tracer picks up a density-weighted contribution and therefore a $m \neq 0 $ helicity, as argued at the beginning of this section\footnote{Note that no $m=1$ helicity is generated from the shift at this order.}.\par
Instead of expanding in the $\bPsi_1$ shift, let us instead expand higher-order contributions, and keep the Zeldovich displacements resummed. Imposing \Cref{eqn:passiveadvection} allows for an analogous expression to \Cref{eqn:densshift} to be derived for galaxy shapes. It is given by (now $i,j$ indices denote displacements and $a,b$ denote shape indices as in Ref.~\cite{chenkokron2024})
\begin{equation}
    \label{eqn:shapeshift}
    \gamma_{ab} (\bk) = \int_{\bq} e^{-i \bk \cdot \bPsi_1} \left [ \gamma_{ab}^L (\bq) + i  k_i \gamma_{ab}^L \partial_j \Psi_1^i \bar{\Psi}^j - \gamma_{ab}^L \nabla \cdot \bar{\bPsi} - \bar{\bPsi} \cdot \nabla \gamma_{ab}^L \right ] e^{-i \bk \cdot \bq}.
\end{equation}
The lack of a `$1+$' implies that the second term in square brackets will always produce an operator larger than one-loop. The third and fourth terms will only get contributions from $\bPsi_2$, which is in contrast to the scalar density scenario where terms such as $\nabla \cdot \bPsi_3$ also show up. At one-loop order, the only contributions come from the third and fourth terms
\begin{align*}
&\gamma_{ab}^L \nabla \cdot \bPsi_2 \to \frac{3}{14}c_s^L \mathcal{G}_2 s_{ab},\\
&\bar{\bPsi} \cdot \nabla \gamma_{ab}^L \to c_s^L \bPsi_2 \cdot \nabla s_{ab}.
\end{align*}
Both of these terms contribute at one-loop order with the same-amplitude for its coefficient, but note that $\mathcal{G}_2 s_{ab}$ is an existing cubic operator in the basis (it can be written in terms of $\delta^2 s_{ab}$ and $s^3_{ab}$, whereas the shift $\Psi_2 \cdot \nabla s_{ab}$ is a new term in the expansion (with a coefficient fixed by $c_s^L$). \par 
From the perturbative decomposition of the shape field in the shifted basis, we compute transfer functions for shape operators. We first project the shape tensor into the helicity basis 
\begin{equation}
    \gamma_m (\bk) = Y^{ab,*}_m (\hat{\bk}) \gamma_{ab}(k), 
\end{equation}
from which we get transfer functions for every $m$. At second order in shape bias, and including only the 2LPT correction, the linear transfer function follows from \Cref{eqn:lineartransfer}
\begin{align}
    \mathcal{C}_{s;m} (k) &= \frac{\langle \tilde{s}_m \sum_O c_O \tilde{O}_m \rangle }{\langle \tilde{s}_m \tilde{s}_m \rangle}\nonumber \\
    &= c_s^L + \alpha_m k^2 + c_{s^2}^L \frac{\langle \tilde{s}_m^2 | \tilde{s}_m \rangle}{\langle \tilde{s}_m \tilde{s}_m \rangle} + c_{\delta s}^L \frac{\langle [\widetilde{\delta s}]_m | \tilde{s}_m \rangle}{\langle \tilde{s}_m \tilde{s}_m \rangle}  + c_{t}^L \frac{\langle \tilde{t}_m | \tilde{s}_m \rangle}{\langle \tilde{s}_m \tilde{s}_m \rangle} + c_s^L  \frac{\langle \tilde{S_3}_m | \tilde{s}_m \rangle}{\langle \tilde{s}_m \tilde{s}_m \rangle}, \label{eqn:linstransfer}
\end{align}
where $\tilde{S}_{3,ab} \equiv \tilde{s}_{ab}^{\rm 2LPT} - \tilde{s}_{ab}^{\rm ZA} \approx  \widetilde{\Psi_2 \cdot \nabla s_{ab}} + \frac{3}{14}\widetilde{\mathcal{G}_2 s_{ab}} $, and we have included a broad-band term $\alpha_m k^2$, which acts as a counter-term at high-$k$. Note that by defining $\tilde{S}_3$ as a difference of displacements this contribution is automatically renormalized at one-loop order -- $\langle \tilde{S}_3 s \rangle = 0$. This expression elucidates the argument of the previous section that transfer functions for a given operator will be different depending on their helicity. Since every observable shape field is generically density-weighted\footnote{Since shapes can only be observed in locations where there are objects, the observed shape field is really $(1+\delta_h (\bx))I_{ab}(\bx)$.}, even the linear transfer function will carry a significant helicity-dependence. 
\subsection{Example of the helicity-dependence of transfer functions}
\label{subsec:m2hel}
As discussed, the shifted Lagrangian tidal tensor generates a density-weighted contribution to the final shape anisotropy field. 
This produces a helicity-dependence of the transfer functions. 
In addition, the $m=2$
transfer functions have strong 
scale-dependence even in the 
$k\to 0$ limit. 
Let us discuss this effect in detail. 
The density-weighted contribution to the linear transfer function is given by 
\begin{equation}
\label{eqn:lintransferterm}
    \mathcal{C}_{s;m}(k) \supset \frac{\langle [\widetilde{\delta s}]_m | \tilde{s}_m \rangle (k)}{\langle \tilde{s}_m \tilde{s}_m \rangle (k)}.
\end{equation}
On general scaling grounds we expect this transfer function will not asymptote to a constant at low-$k$. Note that at one loop, the un-shifted $(\delta s)_m$ operator will be the main contribution. For the numerator contribution, we can define the angle between the raising vector $\epsilon^\pm_i$ and an internal loop momentum $\bp$ as $\alpha \equiv \epsilon_a^+ \hat{p}_a$. A direct projection finds
\begin{equation*}
    (\delta s)_{m=2}(\bk) \ni \int_{\bp} \alpha^2 \delta_{\bp} \delta_{\bk - \bp}.
\end{equation*}
For the linearized shift, the higher-helicity contributions necessarily come from 
\begin{align*}
    \tilde{s}_{ab} (\bk) &\ni i \bk \cdot [\bPsi s_{ab}](\bk) \\ 
    &\ni \bk \cdot \int_{\bp} \left [ \frac{\bp}{p^2} S_{ab} (\widehat{\bk - \bp}) + \frac{(\bk - \bp)}{(\bk - \bp)^2} S_{ab}(\hat{\bp}) \right ] \delta_{\bk - \bp} \delta_{\bp}.
\end{align*}
Applying $Y^{m=2}_{ab}(\bk)$ to this field, the remaining term is
\begin{equation}
    \tilde{s}_{m=2} (\bk) \propto \int_{\bp} \frac{\alpha^2 k^2}{(\bk - \bp)^2} \delta_{\bk - \bp} \delta_{\bp},
\end{equation}
In the low-$k$ limit, we thus see $\tilde{s}_{m=2} \propto k^2$. The numerator, therefore, has to scale as $\langle (\delta s) \tilde{s}\rangle_{m=2} \propto k^2$ in the $k\to0$ limit. For the auto-spectrum, in this limit, we have $\langle |\tilde{s}_{m=2}|^2\rangle \propto k^4$. Thus this transfer function has a contribution which \emph{diverges} as $k^{-2}$.\par
\begin{figure}
    \centering
    \includegraphics{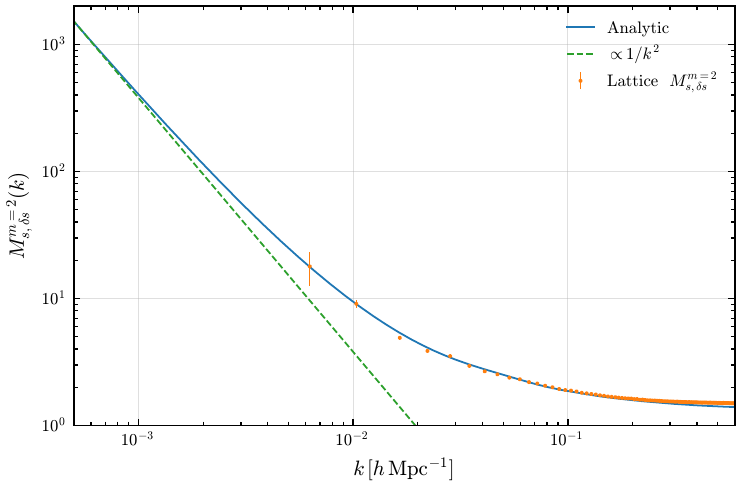}
    \caption{Calculation of the $\delta s$ contribution to the $m=2$ linear transfer function. The blue line shows the analytic calculation of \Cref{eqn:lintransferterm}, while the orange shows the measurement of this ratio from field-level realizations of shifted operators. }
    \label{fig:analyticratio}
\end{figure}

 
Beyond asymptotic scalings, the $k$-dependence of this transfer function contribution can be calculated in perturbation theory. To first order in the shift $\mathbf \Psi_1$, the numerator of \Cref{eqn:lintransferterm} is the Fourier transform of configuration-space correlation functions:
\begin{equation*}
    \langle \widetilde{(\delta s)}_{ab} | \widetilde{s}_{cd} \rangle = \int_{\bq} e^{- i \bk \cdot \bq } i k_i \left ( U_i (q) C_{abcd}(q) + E_{ab}(q)B_{icd}(q) \right ).
\end{equation*}
When projected onto the $m=2$ helicity, the numerator is\footnote{The functions $U_i(q), C_{abcd}(q), E_{ab}(q), B_{icd}(q), \mathcal{J}_n(q)$ and so on are standard quantities in LPT. Their definitions can be found in \citep{Carlson_2012, chenkokron2024} for example. We omit their definition in this work for brevity.} 
\begin{align*}
    \langle \widetilde{(\delta s)}_{m=2} | \widetilde{s}_{m=2} \rangle = \int_{\bq }  e^{- i \bk \cdot \bq } i k & \left [ -\frac{4}{35} (\mathcal{J}_1 \mathcal{J}_4 - (35 \mathcal{J}_6 + 14 \mathcal{J}_8 + \mathcal{J}_9) U(q)) P_1(\mu) \right. \\ 
    &+ \frac{8}{45} (\mathcal{J}_1 \mathcal{J}_4 - (9 \mathcal{J}_8 + \mathcal{J}_9) U(q)) P_3(\mu)\\ 
    &\left. -\frac{4}{63}(\mathcal{J}_1 \mathcal{J}_4 - \mathcal{J}_9 U(q))P_5(\mu) \right], 
\end{align*}
where $P_\ell (\mu)$ are Legendre polynomials of order $\ell$. The denominator of \Cref{eqn:lintransferterm} was computed in resummed form in \cite{chenkokron2024}, however the implementation in \texttt{spinosaurus} is numerically unstable at low-$k$. Instead, we derive the first set of corrections to the $m=2$ helicity, giving 
\begin{align*}
    \langle \widetilde{s}_{ab} | \widetilde{s}_{cd} \rangle \supset -\frac{1}{2}\int_{\bq} e^{-i \bk \cdot \bq} \left [ \frac{1}{2} k_i k_j A_{ij}(q) C_{abcd}(q) +  k_i k_j B_{iab} (q) B_{jcd} (q) \right ]. 
\end{align*}
This expression admits a similar decomposition into standard LPT configuration-space correlators when the $m=2$ projection is taken. In \Cref{fig:analyticratio}, we show the contribution of \Cref{eqn:lintransferterm} computed analytically, as well as measured empirically from the correlations of shifted tensor operators on the lattice. The measured and calculated ratios are in good agreement, and validate that at low-$k$ the linear transfer function will be highly sensitive to the coefficient $c_{\delta s}$. This figure also shows that the helicity transfer functions are qualitatively different from the scalar transfer functions considered in past works. The inevitable density-weighting already leads to a low-$k$ scale dependence in the transfer function for $m=2$, unseen in other field level models of tracers. \par 
In general, for a purely quadrupolar operator ($s_{ab}, t_{ab}$) the higher-$m$ helicity transfer functions are sensitive to operators at one higher order. The $L^{(3)}_{ab}$ operator, having only $m=\{0, \pm1\}$\footnote{Since $\mathbf \Psi_3$ has a curl mode, even the Lagrangian $L^{(3)}_{ab}$ tensor contains an $m=1$ helicity, but it can only contribute to an observed signal at two loops in the power spectrum.} modes will similarly be sensitive to quartic operators through its $m=2$ transfer function. 
 \par 
\subsection{Introducing Tidal Galileons}
\label{sec:galileons}
In the EFT for spin-2 tracers, as in the scalar case, there exist certain degeneracies between the operator basis that hold at the field level. The first one is the degeneracy between cubic operators and the linear tidal field at the level of the one-loop power spectrum, which 
are understood as a renormalization
of the linear bias~\cite{McDonald_2009}.
We have already fully taken this into account for our cubic operators by subtracting from the Lagrangian operators their proper overlap with $s_{ab}$, see Eq.~\eqref{eq:b1ren}.\par 

 However, there exist other degeneracies at the field level that show up already for the quadratic operators. This was first discussed in \cite{Akitsu_2023}, where it was shown that for the $m=1$ helicity 
\begin{equation*}
\label{eqn:m1degen}
    (s^2)_{m=1}(\bk) = \frac{1}{3}(\delta s)_{m=1} (\bk) = \frac{1}{3}(\Psi \cdot \nabla s)_{m=1},
\end{equation*}
and so this channel cannot disentangle contributions from these operators. 
A related question that one can ask, then, is whether the angular content of tensor operators in the low-$k$ limit is degenerate or not, independent of helicity. The degeneracy highlighted in this section is a field-level one, not restricted to any specific observable. The purpose of this section is to discuss the $k\to0$ structure of other operators in the shape expansion. \par 
Since we are interested in the $k \to 0$ analysis we will neglect the contributions from the $\bPsi_1$ shifts. We generically write a spin-2 field as 
\begin{equation}
    \mathcal{O}_{ab} (\bk) = \int_{\bp_1,\cdots,\bp_n}(2\pi)^3 \delta^D(\bk - \bp_{1..n})\,F_{ab}(\bp_1, \cdots,\bp_n) \delta_{\bp_1}\cdots\delta_{\bp_n} . 
\end{equation}
For quadratic operators, the $k \to 0$ limit considers the angular mode-coupling kernel $F_{ab}$, evaluated for a $\bp, -\bp$ configuration, while for cubic operators we study these kernels when their wavenumbers are forced onto a triangle. Let us first consider the low-$k$ structure of quadratic operators before looking at the general structure of a cubic operator in this limit.   
We define $S_{ab}(\bk) = \hat{k}_a \hat{k}_b - \delta_{ab}/3$ as the Fourier-space analog of $\mathcal{D}_{ab}$. Then note that for the $[ s^2]_{ab}$ operator
\begin{align}
    s_{ac} s_{cb} (k \to 0)&= \int_{\bp} S_{ac}(\bp)S_{cb} (\bp) \delta_{\bp} \delta_{\bp} =\frac{1}{3} \int_{\bp} S_{ab}(\bp) \delta_{\bp} \delta_{\bp} = \frac{1}{3} (\delta s)_{ab }(k \to 0) 
\end{align}
So we see that in the $k \to 0$ limit there is an exact degeneracy between the $s^2_{ab}, \delta s_{ab}$ operators, and this holds for \emph{all helicities} as opposed to the prior $m=1$ analysis. 
This degeneracy is analogous 
to a degeneracy between the $\delta^2$
and $s^2$ operators present for the galaxy bias expansion. This degeneracy is removed
if one uses a Galileon tidal basis, 
which has a nice property that $\langle
\delta^2 \mathcal{G}_2
\rangle \to 0$ as $k\to 0$. 
Let us now use the same approach for the shape bias: instead of working with the degenerate $s^2, \delta s$ combination we can instead define the \emph{tidal Galileon} \begin{equation*} 
\mathcal{G}^s_{ab} \equiv s^2_{ab} - (\delta s)_{ab}/3,
\end{equation*} 
which is the contribution from the tidal torquing operator explicitly deprojected from the density-weighting contribution. The parameter for the tidal Galileon is the same as the usual tidal torquing operator, $c_{\mathcal{G}_s} = c_{s^2}$, but rotating to this basis leads to $c_{\delta s} \to c_{\delta s} - c_{\mathcal{G}_s}/3$, and so comparing amplitudes of density-weighting coefficients with the literature must be done with care. The tidal Galileon $\mathcal{G}^s$ also contains the following degeneracy for $m=0$: 
\begin{equation}
    \mathcal{G}^s_{m=0} (\bk) = \frac{7}{8} t_{m=0} (\bk)
\end{equation}
so there is a degeneracy with the velocity shear $t_{ab}$ operator for $m=0$ (up to shifts), but not for other helicities (since $t$ is strictly quadrupolar). It can also be shown that to first order in the shift 
\begin{equation}
    \mathcal{G}^s_{m=2} (\bk) = - \tilde{s}_{m=2} (\bk),  
\end{equation}
so rotating the quadratic terms into the tidal Galileon basis introduces to the $m=2$ channel of the $\tilde{s}$ transfer function a perfectly degenerate contribution\footnote{A priori this is not a terrible situation to be in. One can argue $m=0$ $\tilde{s}$ channel cleanly measures $c_s$, the $k^{-2}$ dependence of the $m=2$ measures $c_{\delta s}$ and so any remaining constant contribution in principle pins down $c_{\mathcal{G}^s}$.} to $c_s$. 
All in all, using the Galileon basis, the only field with a constant power spectrum in the $k\to 0$ limit is $(\delta s)_{ab}$. Since we include this operator in our model, the auto spectrum of this operator does not 
contribute to $P_{\rm err}^m$
even though its large-scale power spectrum 
is flat. 

\par 
Do similar degeneracies exist for the cubic fields? For the cubic scalar bias operators, the mode coupling kernels necessarily decompose cubic operators in this limit into a universal form: 
\begin{equation*}
    F^{(3)}_i(\bk - \bp_1 - \bp_2, \bp_1, \bp_2)^{p_1, p_2 \to \infty} = A_i + B_i f(\bp_1, \bp_2), 
\end{equation*}
where $f(\bp_1, \bp_2)$ is the same for all cubic operators~\cite{bakx2026twolooprenormalizationrunninggalaxy}. There is a richer degeneracy structure for the spin-2 case in this same kinematic limit. For the cubic shape operators, there are four distinct angular structures in their mode-coupling kernels: 
\begin{equation}
    \mathcal{S}_{ab} = \sum_i S_{ab}(\bp_i);\, \mathcal{S}^{(\mu)}_{ab} = \sum_i \mu_{jk}^2 S_{ab}(\bp_i);\, \mathcal{\chi}_{ab} = \sum_{i<j} \mu_{ij} [\hat{\bp}_{i_a} \hat{\bp}_{j_b}];\, \mathcal{\chi}^{(\mu)}_{ab} = \sum_{\rm cyc} \mu_{ik}\mu_{kj} [\hat{\bp}_{i_a} \hat{\bp}_{j_b}],
\end{equation}
where $i,j,k$ indices correspond to combinations of the three momenta contained in the cubic kernel, and `cyc' denotes a cyclic sum over $i,j,k$. The six non-$L^{(3)}$ cubic operators, in the low-$k$ limit, decompose into combinations of these contributions. The simplest case is $\delta^2 s_{ab}:$
\begin{align}
    \delta^2 s_{ab} (k \to 0) &= \int_{\bp_1, \bp_2, \bp_3} (2\pi)^3\delta^D(\bp_{123}) \left [ \frac{1}{3}\sum_{i=1}^3 S_{ab}(\bp_i) \right ] \delta_{\bp_1}\delta_{\bp_2}\delta_{\bp_3}\\
    \delta^2 s_{ab} (k \to 0) &\ni \frac{1}{3} \mathcal{S}_{ab}.
\end{align}
and for the additional three Zel'dovich fields we have 
\begin{align*}
    &s^2 s_{ab} \to \, \, \mathcal{S}^{(\mu)}_{ab}/3 - \mathcal{S}_{ab}/9\\ 
    &\delta s^2 \to \, \chi_{ab}/3 \,- 2\mathcal{S}_{ab}/9\\ 
    &s^3_{ab} \to \chi^{(\mu)}_{ab}/3 - \chi_{ab}/3 + \, \mathcal{S}_{ab} /9.
\end{align*}
This decomposition hides an additional degeneracy that exists at the field level: the operators $s^2 s_{ab}$ and $s^3_{ab}$ are equivalent. This is a consequence of the Cayley-Hamilton theorem, which for all $3\times3$ square matrices $A$ states 
\begin{equation}
    A^3 - ({\rm tr} A) A^2 + \frac{1}{2} \left ( ( {\rm tr} A)^2 - {\rm tr} (A^2)\right) A - \det (A) I_3 = 0,
\end{equation}
where the 0 on the right-hand side is understood as a matrix of zeros. Applying this to $A = s_{ab}$, noting that ${\rm tr}(s_{ab}) = 0$, we find 
\begin{equation}
    s^3_{ab} - \frac{1}{2} s^2s_{ab} - \det (s)\delta_{ab}^K= 0.
\end{equation}
The trace of this equation implies $s^3 /3 = \det(s)$. From this the trace-free part of this equation reads 
\begin{equation}
    s^3_{ab} - \frac{s^3}{3} \delta_{ab}^K = \frac{1}{2} s^2 s_{ab},
\end{equation}
and so the $s^3$ and $s^2 s_{ab}$ operators are equivalent to each other when trace-free components are taken into account. This implies the original set of four Zeldovich operators is in fact three, and the shape bias basis to cubic order contains 10 independent operators\footnote{This also implies the four angular structures written down should be linked. Indeed, it can be shown that $\chi^{(\mu)} = \chi + \mathcal{S}^{(\mu)}/2 - \mathcal{S}/2$ so this also implies there are only three unique angular structures.}. \par 
The non-Zeldovich operators do not generate new angular dependencies, and in fact suffer from an analogous `full degeneracy' that $\{s^2,\delta s\}$ do: 
\begin{equation*} 
[st]_{ab} (k \to 0) = \delta t_{ab} (k \to 0)/3,
\end{equation*} 
and its low-$k$ angular structure is 
\begin{equation*}
\delta t_{ab} (k \to 0) \propto \mathcal{S}_{ab}^{(\mu)} - \mathcal{S}_{ab},
\end{equation*}
which is strictly contained within the $s^2 s_{ab}$ and $\delta^2 s_{ab}$ fields\footnote{If one rotates $s^2 s_{ab} \to \mathcal{G}_2 s_{ab}$ then it shares the same low-$k$ structure that $\delta t_{ab}$ does.}. The three Zel'dovich operators carry distinct weightings of these angular structures -- they are not necessarily degenerate in this limit. The $\delta t$ and $st$ operators, however, are strictly degenerate in this sense. In fact, if one re-defines $s^2 s_{ab} \to \mathcal{G}_2 s_{ab}$, then $(\mathcal{G}_2 s_{ab}, \delta t_{ab}, st_{ab})$ are completely indistinguishable in this limit, and only higher-$k$ corrections can break this degeneracy. We can thus rotate the $\delta t, st$ operators into their cubic tidal Galileons 
\begin{align*}
    &\mathcal{G}^{(3a)}_{ab} \equiv st_{ab} - \frac{1}{3} \delta t_{ab}, \\
    &\mathcal{G}^{(3b)}_{ab} \equiv \delta t_{ab} - \frac{2}{7} \mathcal{G}_2 s_{ab},
\end{align*}
such that they have no overlap with the four angular structures. These tidal Galileons are also, by construction, normal-ordered with the linear density field, i.e. their $P_{13}$ contributions
do not renormalize $c_s$.
Given the tidal galileon $\mathcal{G}^s_{ab}$, we can also rotate the Zel'dovich cubic fields into operators composed of Galileons
\begin{align*}
    &s^2 s_{ab} \to \mathcal{G}_2 s_{ab} \\
    &\delta s^2 _{ab} \to \delta \mathcal{G}^s_{ab} \\
    &s^3_{ab} \to \mathcal{G}^s_{ac} s_{cb},
\end{align*}
however eliminating $s^2 s_{ab}$ from the basis also implies we do not need $\mathcal{G}_2 s_{ab}$. In this basis there is one `pure Zel'dovich' operator, two Galileon-based operators, and three non-Zeldovich operators which are composed of $t_{ab}$ or $L^{(3)}_{ab}$. The $L^{(3)}_{ab}$ operator has a unique angular decomposition (with an external $m=0$ and $m=1$ mode). 
The operators with constant $k\to0$ auto-spectra are $\delta^2 s_{ab}, \delta \mathcal{G}^s_{ab}, (\mathcal{G}^s s)_{ab}$. The corresponding two-loop
$P_{33}$ terms
would contribute to $P_{\rm err}^m$ if we did not include them
in our model. This is one reason
to model the shape bias 
through the cubic order,
even though the state-of-the-art 
one-loop shape power 
spectrum computation
can be reproduced with
only quadratic operators at the field level, whilst all the contributions from the cubic operators are captured by the transfer functions. 
In light of the above degeneracies, we will measure transfer functions and constrain parameters of the 10-parameter tidal Galileon basis, instead of the usual Lagrangian basis used in \cite{chenkokron2024}:
\begin{align}\label{eq:gal_basis}
    1^{\rm st} \text{ order: } &s_{ab}\\
    \nonumber 2^{\rm nd} \text{ order: } &\delta s_{ab}, \mathcal{G}^{s}_{ab}, t_{ab}\\
    \nonumber 3^{\rm rd} \text{ order: }  &\delta^2 s_{ab}, \delta \mathcal{G}^s_{ab}, (\mathcal{G}^s s)_{ab}, \mathcal{G}^{(3a)}_{ab}, \mathcal{G}^{(3b)}_{ab}, L^{(3)}_{ab}
\end{align}
This basis isolates the contributions to low-$k$ spectra, organizes operators in terms of their angular contributions, and significantly simplifies the structure of the mixing matrix $\mathcal{M}$. In practice, the implementation of this Galileon basis is realized by applying a rotation to the original local Lagrangian basis described in \S~\ref{sec:methods}. This rotation makes the `non-Zeldovich` cubic operators ($\delta t, st, L^{(3)})$ have zero renormalization, while the remaining cubic operators are still appropriately renormalized. \par
\begin{figure}[ht!]
    \centering
    \includegraphics[width=\linewidth]{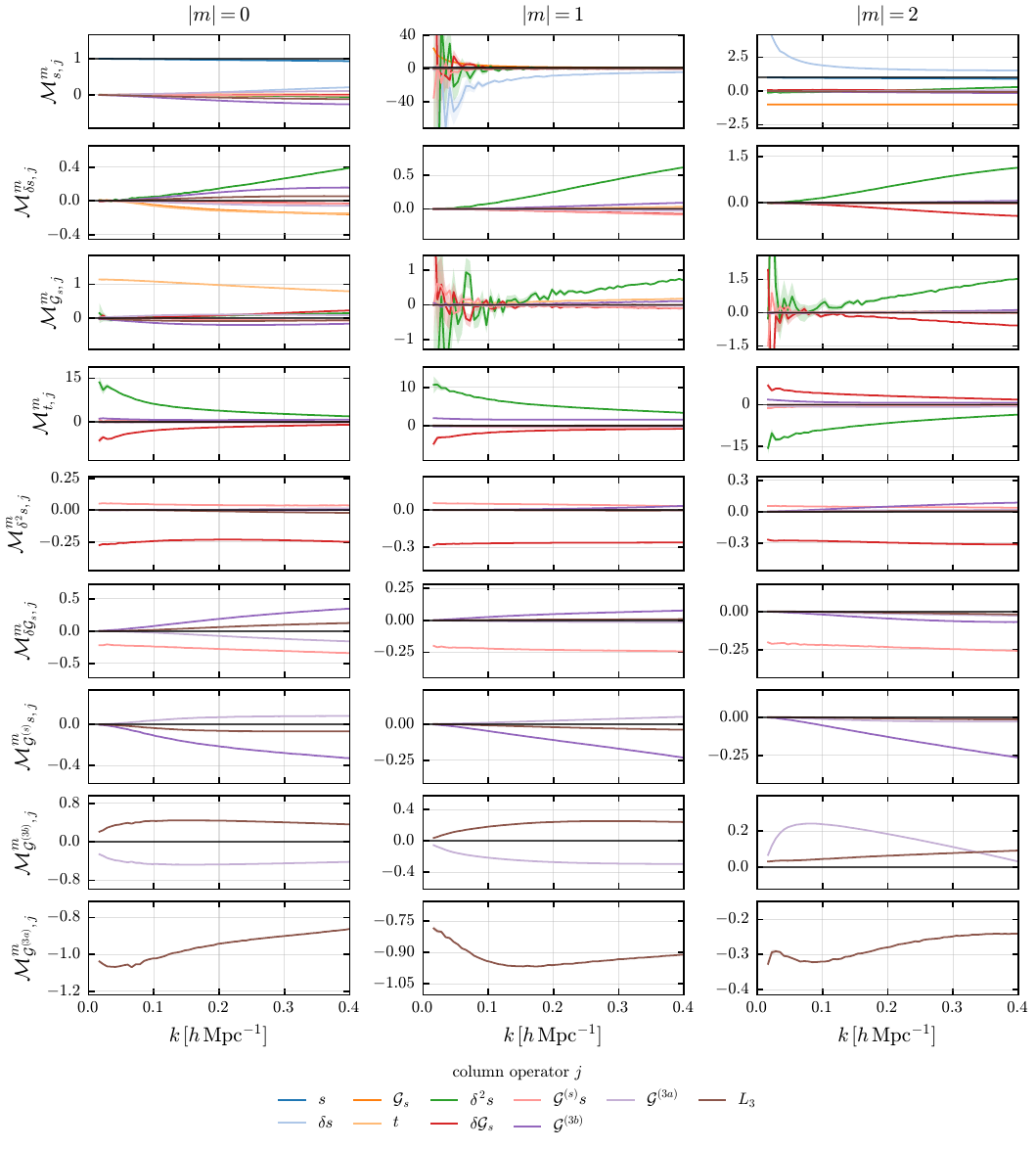}
    \caption{Entries of the mixing matrix \Cref{eqn:designmatrix}, for each helicity. The $\propto k^{-2}$ enhancement of the linear $\mathcal{M}_{s, \delta s}^{(m=2)}$ matrix derived in the text is clearly shown. The curves correspond to the templates which can compose each measured transfer function to cubic order. The shaded bands denote uncertainty from a $V=50\,({\rm Gpc}/h)^3$ volume.}
    \label{fig:designmat}
\end{figure}

 
Extending the relation between the transfer function measured for one orthogonalized field and the shape bias coefficients which contribute to it from the linear case (\Cref{eqn:linstransfer}) to higher orders leads to the construction of a \emph{mixing matrix} 
\begin{equation}
\label{eqn:designmatrix}
    \mathcal{M}_{m;ij}(k) = \frac{\langle \widetilde{\mathcal{O}}_i^\perp \widetilde{\mathcal{O}}_j\rangle(k)}{\langle |\widetilde{\mathcal{O}}_i^\perp|^2\rangle(k)},
\end{equation}
which describes correlations between orthogonalized and non-orthogonalized fields. The mixing matrix for transfer functions is shown in \Cref{fig:designmat}. Several of the analytic degeneracies we have discussed in this section are visible in this figure: the low-$k$ enhancement for the $m=2$ $\tilde{s}$ transfer function is visible, the low-$k$ equivalence between $t_{m=0}$ and $\mathcal{G}^s_{m=0}$, as well as the degeneracies between cubic operators (which extend beyond low-$k$). We also note a strong, two-loop degeneracy is visible in the mixing matrix for the $t_{ab}$ operator: the cubic $\delta^2 s_{ab}$ and $\delta \mathcal{G}^s$ operators exhibit a strong low-$k$ scaling which changes sign depending on helicity. This can be seen in the measured transfer functions for halos -- in \Cref{fig:transfer_mbin3} the curves corresponding to the $\mathcal{C}_{t;m=0}(k)$ and $\mathcal{C}_{t;m=2}(k)$ have opposite signs, irreconcilable with there being no influence of higher-order operators in the low-$k$ regime of the measured transfer function, but plausibly captured by the contributions from the aforementioned cubic operators. We can thus see that, unlike in the case of scalar operators where analytic intuition can be built for all elements of the mixing matrix~\cite{Schmittfull_2019,Ivanov:2024hgq}, the influence of higher-loop contributions for higher-order/helicity transfer functions makes such an interpretation difficult. This complex structure motivates us to construct the full mixing matrix from lattice correlators of shifted fields.

Finally, let us note that, the last three operators 
in the last line of Eq.~\eqref{eq:gal_basis}, one linear combination of coefficients is redundant, leaving two independent combinations at the level of the one-loop shape power spectrum.
However, they are not
redundant at the level of the shape trispectrum, which we effectively probe
in our field-level analysis~\cite{Nguyen_2024, akitsu2025cosmologyinferenceperturbativeforward, spezzati2025equivalencefieldlevelinferenceconventional}.

\subsection{Estimating coefficients from transfer functions}
\label{sec:tffit}
Having settled on a suitable basis which is minimally degenerate, we turn to measuring the coefficients of this basis. The transfer functions are related to the shape bias coefficients through a generalized least-squares (GLS) linear regression problem 
\begin{equation}
\label{eqn:matrixsystem}
   \vec{\mathcal{C}}_m(k) = \mathcal{M}_m(k) \cdot \vec{c}_L, 
\end{equation}
where note that the mixing matrix $\mathcal{M}_{m}(k)$ defined in the previous section plays the role of the design matrix in the GLS problem. The projector property of the orthogonalized transfer functions implies $\mathcal{M}$ is 1 in the diagonal. The perturbative expansion of higher-order displacements (\Cref{eqn:shapeshift}) modifies the matrix. The lowest-order inclusion comes from the $S_3$ tensor induced by the difference between the shifted tidal tensor in the Zeldovich approximation and the same operator shifted with higher-order dynamics. In this case, the matrix is shifted by a column vector
\begin{equation}
\label{eqn:2lptshift}
    \mathcal{M}_{ij} \to \mathcal{M}_{ij} + \frac{\langle \tilde{S}_3 | \mathcal{O}^\perp_i\rangle}{\langle |\mathcal{O}_i^\perp|^2\rangle} \delta^K_{0 j},
\end{equation}
and in the case of higher-loop shifts (with coefficients that are not $c_s$) the extension is straightforward. In this work, we opt to only include the 2LPT shift in the $s_{ab}$ transfer function, which is the only contribution which enters at one loop in the mixing matrix. We also introduce broad band `counterterm' contributions such that the most generic model for a given transfer function is given by 
\begin{equation}
    \mathcal{C}_{i;m}(k) = \sum_j c_j\mathcal{M}_{m;ij}(k) + \alpha_{m;i} k^2 + \beta_{m;i}k^4, 
\end{equation}
and the $\alpha_{m;i}, \beta_{m;i}$ terms are included in the linear regression by extending the $\mathcal{M}$ matrix appropriately. \par 
By construction, the transfer functions have no cross-covariance across different helicities and operators. We also neglect $k$ mode-by-mode covariance, as we do not find empirical evidence for substantial cross-$k$ covariance over the range of scales considered in this work. The bias coefficients can thus be estimated by solving the weighted least-squares problem of \Cref{eqn:matrixsystem} with weights given by a diagonal precision matrix (\Cref{eqn:lstsq}). While for large ensembles of dark matter-only simulations, such as the \quijote suite, we can empirically measure the scatter of transfer functions, this is not possible for the \flamingo suite. We thus implement an analytic approximation to the transfer function variance. \par 
For a given field-level realization, the shifted operator fields are fixed. It is the statistics of $\epsilon_m$ that sets the scatter in the field-level bias relation. If we assume $\epsilon_m$ is Gaussian, then the uncertainty in estimating the transfer function is given by\footnote{A similar expression for the linear transfer function of galaxy clustering was first noted in \S VI.D of Ref.~\cite{Schmittfull_2019}.}
\begin{equation}
\label{eqn:tferr}
    \sigma^2 (\mathcal{C}_{i;m})(k) = \frac{a_m}{N_{\rm modes} (k)} \frac{P_{\rm err,m}(k)}{\langle |\tilde{O}_{i;m}^\perp|^2\rangle (k)}, 
\end{equation}
where $a_m = \{1, 1/2, 1/2\}$ for $m=\{0,1,2\}$ respectively and $N_{\rm modes}(k) = 4\pi (k/k_{\rm F})^2$ for $\Delta k = k_F = 2\pi/L_{\rm box}$. We derive and validate \Cref{eqn:tferr} against the scatter in the transfer functions of halo shapes in \Cref{app:tferror}. Given weights from \Cref{eqn:tferr}, the linear regression formulation of the perturbative field-level analysis immediately yields an analytic expression for shape bias coefficients
\begin{equation}
\label{eqn:lstsq}
    \hat{c}_L = (\mathcal{M}^T \cdot W \cdot \mathcal{M})^{-1}  \cdot \mathcal{M}^T \cdot W \cdot \mathcal{C},
\end{equation}
with $W(k) \equiv [\sigma^2 (\mathcal{C}_{i;m})(k)]^{-1}$. This estimator can be thought of as a rigorous, Fourier space-based formalization of the `direct alignment field fit' estimator presented in Ref.~\cite{Samuroff_2021}. Both $\mathcal{C}$ and $\mathcal{M}$ are measured directly from a realization, and the regression problem is solved through NumPy's \texttt{np.linalg.lstsq} routine\footnote{To prevent some low-$k$ noise from affecting coefficient inference when including cubic coefficients, we in practice perform an iterative GLS fit. A value of $c_s$ inferred from the low-$k$, $m=0$ linear transfer function. This value is used as a prior for the subsequent full-transfer function fit. For the largest box sizes this result is equivalent to the full fit, but for certain \flamingo cross-correlations a spurious contamination of $L_3$ at low-k could affect the inference.}. This formulation also leads to an expression for the Fisher matrix of the parameters
\begin{equation}
\label{eqn:paramfisher}
    F = \mathcal{M}^T \cdot W \cdot \mathcal{M},
\end{equation}
 Uncertainties for the field-level shape bias coefficients follow directly from inverting this matrix, while correctly marginalizing over the impact of including the large number of broadband terms in this fit. This analytic parameter covariance is compared to the ensemble \quijote covariance in \Cref{app:tferror} and shown to agree qualitatively and quantitatively.\par
In Ref.~\cite{Schmittfull_2019} the components of the operator mixing matrix \Cref{eqn:designmatrix} were calculated in perturbation theory assuming low-$k$ arguments. However, for their box size certain assumptions underlying their perturbative calculation did not hold\footnote{For example, when deriving the $\delta^2$ transfer function the contribution from the $\mathcal{G}_2$ operator was neglected under the argument $\langle \widetilde{\delta^2}^{\perp} | \widetilde{\mathcal{G}}_2\rangle / \langle \widetilde{\delta^2}^{\perp}|\widetilde{\delta^2}^{\perp}\rangle  \to 0$ at low-$k$. In practice, at $k=0.1\ihmpc$ this ratio is $\approx 0.3$, a non-negligible contribution.}. For this reason, we build the components of $\mathcal{M}_{ij;m}(k)$ directly from correlators of shifted operators measured directly from the initial conditions that correspond to the simulation being analyzed in question. This ensures issues such as discreteness effects and IR-cutoffs from small box sizes (in the case of hydrodynamical simulations) are naturally taken into account, and adds a degree of cosmic variance cancellation since the correlators which enter the transfer functions are imbued with the same noise as the shape power spectra.\par 
The fiducial analysis choice we make in this work is to fit the linear and quadratic transfer functions to $\kmax = 0.4 \ihmpc$, while only fitting the cubic transfer functions to $\kmax=0.07 \ihmpc$,  not including broad band parameters for cubic transfer functions. Finally, we find strong evidence that the $m\neq 0$ transfer function for $L^{(3)}$ is contaminated by even higher order contributions (due to the density-weighting argument discussed in the previous section) and so we do not include them in the linear regression. This fiducial array of analysis choices allows for stable inference of the $c_i$ coefficients out to $\kmax = 0.4 \ihmpc$, while maintaining $\chi^2 / {\rm dof} \approx1$ for all halo masses in question.  At higher $\kmax$ the coefficients begin to shift significantly, as does the $\chi^2$, indicating a potential breakdown of the cubic model at the field level. \par 
\section{Simulations and shape definitions}
\label{sec:simulations}
In this work we study the shapes of particles bound to dark matter halos across two sets of simulations: the dark matter-only \quijote suite of simulations and the \flamingo suite of hydrodynamical simulations. We restrict ourselves to $z=0.5$, and defer the study of redshift trends in the field-level model to future work. \par 
For \quijote, we use halo catalogs derived from \texttt{MPI-Rockstar}~\cite{Tokuue_2026}, which uses a phase space-based approach to halo finding and calculates the inertia tensors of output halos. We run \texttt{MPI-Rockstar} on the ``High Resolution'' suite, which has $N=(1024)^3$ particles for a cubic box with size $L=1000 \mpch$. The parameter files used to run \texttt{MPI-Rockstar} are identical to the publicly available \quijote-\texttt{Rockstar} catalogs to ensure consistency with this sample. We have modified \texttt{MPI-Rockstar} to output several definitions of halo shape, in order to understand how the definition of halo shape influences the values of the shape biases. The default inertia tensor computed by \texttt{MPI-Rockstar} is a \emph{reduced} and \emph{iterative} inertia tensor~\cite{Allgood:2005eu}. That is, for every bound particle within a sphere of radius $R_{\rm vir}$ from a halo center the inertia tensor is initially defined as 
\begin{equation}
\label{eqn:redtensor}
        I_{ab}^{\rm red} = \frac{1}{N} \sum_n \frac{(\bx_n - \bar{\bx})_a(\bx_n - \bar{\bx})_b}{|\bx_n - \bar{\bx}|^2},
\end{equation}
where $\bar{\bx}$ is the halo center and $\bx_n$ is the position of a constituent particle. In an iterative tensor calculation, the boundaries associated with the ellipsoid defined by \Cref{eqn:redtensor} are used to define a subset of particles, and the reduced tensor is re-computed. \texttt{MPI-Rockstar} by default will compute this procedure until convergence, or up to 10 iterations. We also compute non-iterative shape tensors (using only the first spherical pass), for both the reduced inertia tensor and the `standard' inertia tensor 
\begin{equation}
    I_{ab} = \frac{1}{N} \sum_n (\bx_n - \bar{\bx})_a(\bx_n - \bar{\bx})_b,
\end{equation}
as well as iterative `standard' tensors. That is, we will look at four definitions of halo shapes in \quijote, spanned by the 2x2 matrix of `Reduced / Standard' and `Iterative / non-Iterative'. We use $50$ \quijote-HR boxes of the 100 available in the suite. The resulting volume of $V=50 ({\rm Gpc}h^{-1})^3$ is among the largest volumes used for a study of halo alignments. When a halo shape definition is not explicitly reported, we will have adopted the fiducial `standard'+`iterative' shape definition for that halo finder. \par 

\begin{figure}
    \centering
    \includegraphics[width=1\linewidth]{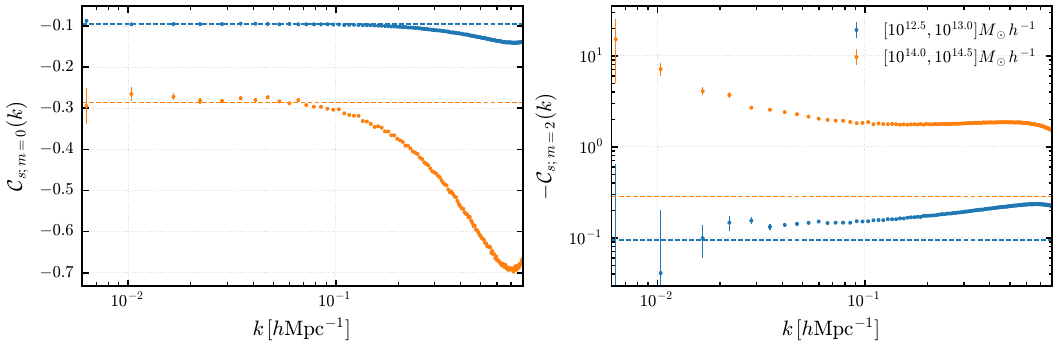}
    \caption{Linear transfer functions, $\hat{\mathcal{C}}_{s;m}(k)$, for the $m=0$ and $m=2$ helicity of \quijote-HR halos in two mass bins. The dashed lines indicate the estimate of $c_s$ for the sample. The error bars are uncertainty on the mean from 50 \quijote boxes. The second panel is multiplied by -1 so values are shown on a log axis.}
    \label{fig:linshape}
\end{figure}

\begin{figure}[ht!]
    \centering
    \includegraphics[width=\linewidth]{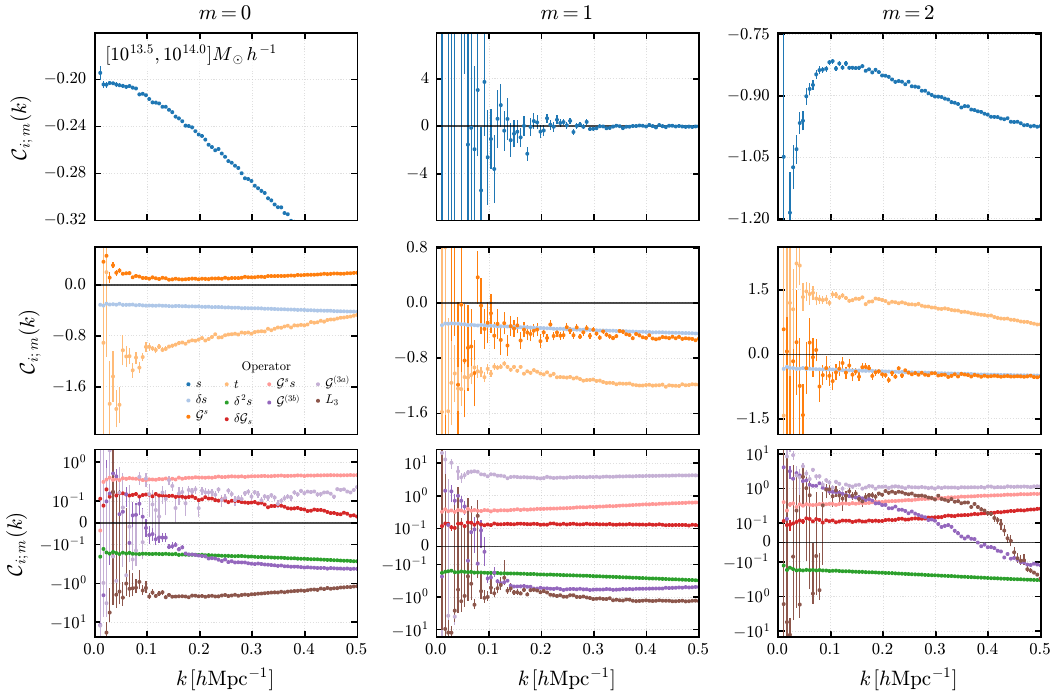}
    \caption{Full set of transfer functions, $\hat{\mathcal{C}}_{i;m}(k)$ for halos with mass $M \in [10^{13.5}, 10^{14}]M_\odot h^{-1}$. Each column shows a different helicity, while rows show the linear, quadratic, and cubic operator transfer functions, respectively. The points are averaged over 50 \quijote realizations, and the uncertainties reported are uncertainties on the mean. }
    \label{fig:transfer_mbin3}
\end{figure}

In \Cref{fig:linshape} we show the $m=0$ and $m=2$ linear transfer functions for two different mass bins, estimated in the \quijote suite, for the standard iterative shape definition. The $\propto k^{-2}$ excess for the $m=2$ helicity, derived in \Cref{subsec:m2hel} is clearly visible in the data for the more massive halo sample.\par 
Turning to the \flamingo suite, in this work we restrict ourselves to a subset of the entire suite of simulations. The use of \flamingo allows us to establish a baseline between the shapes of CDM halos (and compare to \quijote), but also allows us to study the shapes of galaxies in a hydrodynamical simulation (defined as the inertia tensors of stars bound to a halo). Since the \flamingo simulation adopts a different cosmology, a different halo finder, and slightly different algorithms for determining iterative halo shapes we do not necessarily expect 1:1 agreement with the \quijote results. Specifically, we will measure the shapes of halos and galaxies from the \texttt{L1\_M9\_DMO, \texttt{L1\_M9}} and \texttt{L1\_M8} simulations. \texttt{L1} denotes these simulations correspond to a box length of $L=1000{\rm Mpc}$, or $L=681 \mpch$. The \texttt{M9} suite corresponds to a gas particle mass of $m_{\rm gas} \approx 10^9M_\odot$ and CDM particle mass of $m_{\rm CDM} = 5.65 \times 10^9 M_\odot$. \texttt{M8} simulations are higher resolution, corresponding to $\{m_{\rm gas}, m_{\rm CDM}\} = \{ 1.34, 7.06\}\times 10^8 M_\odot$. There are many publicly available definitions of inertia tensors in the \flamingo data, but due to a bug in selecting which halos would have their tensors computed, the vast majority of shapes of stellar components have not been recorded\footnote{\url{https://dataweb.cosma.dur.ac.uk:8443/flamingo/support/support_issues.html\#missing-inertia-tensors} . Empirically, we find that sub-selecting on \texttt{M9} halos with more than 20 star particles results in $>95\%$ of the publicly available halos having no shape measurement.}. We also found evidence for systematic differences between how the shapes of dark matter halos were calculated\footnote{The definition of aperture used to compute halo shapes appears to have changed from $r_{1/2}$ to $10 \times r_{1/2}$.} -- for a fixed halo mass bin, the resulting linear transfer functions are irreconcilably different between \texttt{L1\_M9\_DMO} and \texttt{L1\_M8\_DMO} (but comparable for the non-DMO boxes). To this end, we have independently implemented the shape-measuring pipeline used in \flamingo, making sure we reproduce the stellar and dark matter shapes which were recorded. In \Cref{app:tensorrecon} we detail how we reconstruct a subset of the missing inertia tensors, specifically non-iterative tensors from the `bound subhalo' halo definition used by \texttt{HBT-HERONS}~\cite{moreno2025assessingsubhalofinderscosmological} + \texttt{SOAP}~\cite{McGibbon_2025} (\flamingo's default halo finder and code to compute halo properties), finding strong numerical agreement and validating our reconstruction.  \par 
We generate \flamingo initial conditions by running \texttt{monofonIC}~\cite{Michaux:2020yis, Hahn_2020} using the publicly available config files used to run the simulations. The cosmology is the same as the fiducial DES Y3 cosmology, and the phases are set by \texttt{panphasia}~\cite{jenkins2013panphasiauserguide} using the identical descriptor. Two key differences are that we set \texttt{zstart = 0.5}, matching the redshift of the fiducial snapshot in this work, and we generate the `CDM+baryon' $\delta_{\rm cb}$ at $N_{\rm cell} = 900$ instead of the fiducial $1800$ value, from which the Lagrangian basis fields are computed. 

Given a catalog of halo positions and inertia tensors, we produce the trace-free shape fluctuation field, $\gamma_{ab}(\bx)$, from the tensor component fields. Each of the 6 components of the inertia tensor are used as cloud-in-cell weights when depositing the halo catalogs, compensating for the CIC window. The fields are then mean-subtracted and normalized by the average of the trace field
\begin{equation}
    \gamma_{ab} (\bx) = \frac{I_{ab}(\bx) - \langle I_{ab} \rangle}{{\rm Tr}\langle I_{ab}\rangle},
\end{equation}
which is consistent with normalizations of halo shape fields from prior works~\cite{Akitsu_2023}. In \Cref{fig:halofield} we show both the halo number density for a given mass bin, as well as the inverse Fourier-transformed helicity fields for $m=\{0,1,2\}$. A strong visual correlation between the $m=0$ shape field and the density field is visible, and by construction the $m=\{1,2\}$ components are uncorrelated with the scalar density -- probing truly non-linear dependence on the dark matter distribution. \par

\begin{figure}
    \centering
    \includegraphics[width=1\linewidth]{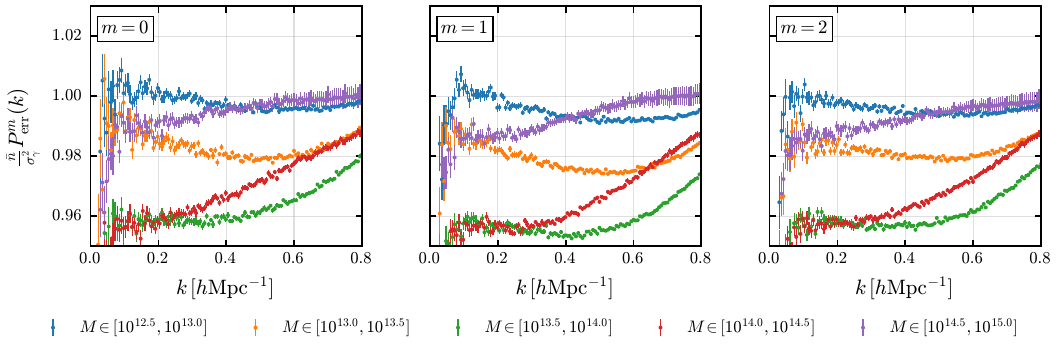}
    \caption{Error power spectra of \quijote halo shapes, re-scaled by the expected shape noise amplitude $\sigma^2_\gamma / \bar{n}$. The downturn in the lowest $k$-bins comes from the small number of available $k$-modes relative to degrees of freedom for those bins, so the transfer function overfits to the measured shape field.}
    \label{fig:haloperr}
\end{figure}
\section{Alignments of dark matter halos}
\label{sec:haloalign}
We begin by studying the field-level distribution of 3D shapes of dark matter halos. Halos are selected by mass, in bins of width 0.5 dex, with lower bin edges $\log_{10}( M_{\rm min} /M_\odot h^{-1}) = \{12.5, 13.0, 13.5, 14.0, 14.5\}$. The transfer functions for the $M\in [10^{13.5}, 10^{14}]M_\odot h^{-1}$ mass bin are shown in Fig.~\ref{fig:transfer_mbin3} as an illustrative example. Transfer functions for all other mass bins are shown in \Cref{app:transfers}. We first present the error power spectra of the best-fit transfer functions for each simulation, and compare the performance of the field-level EFT against popular IA models (the `nonlinear linear alignment' model and `tidal alignment and tidal torquing' models). We will then discuss the best-fit coefficients extracted from these measurements. Finally, we will use the large ensemble of \quijote simulations to compare the parameters inferred from transfer functions to the parameters inferred from fitting the four power spectra of halo shapes (the $\langle \gamma_{m=0} \delta_h \rangle$ spectrum and the three helicity `II' spectra). 

\subsection{Error power spectra}
Having fit transfer functions to the shape fluctuation field which minimize the error power spectrum, we construct the helicity-dependent residual field $\epsilon_m (\bk)$. If the field-level model is fully correlated with the shape fluctuations, then $\epsilon_m$ is fully characterized by its shape noise contribution. Shape noise is the spin-2 generalization of the standard Poisson shot noise which sets the amplitude of the error power spectrum for scalars. For a given sample of shapes, the dispersion of shape tensors $\sigma^2_\gamma$ is computed as 
\begin{equation}
    \sigma^2 _\gamma = \frac{1}{5N_h} \sum_{n=0}^{N_h} \gamma_{ij,n}\gamma^{ij}_n,
\end{equation}
where $\gamma_n$ is the trace-removed, mean-trace-normalized moment of inertia tensor for that object. The symmetries of the spin-2 shape expansion constrain the error power spectrum to carry a specific helicity dependence~\cite{chenkokron2024}: 
\begin{align}
\label{eqn:mperr}
    &P_{\rm err}^{m=0}(k) = A_0 \frac{\sigma^2_\gamma}{\bar{n}}+ k^2A_{21} + \cdots\\
    &P_{\rm err}^{m=1}(k) = A_0 \frac{\sigma^2_\gamma}{\bar{n}}+ k^2 (A_{21} + A_{22}) + \cdots \\
    &P_{\rm err}^{m=2}(k) = A_0 \frac{\sigma^2_\gamma}{\bar{n}}+ k^2 (A_{21} + 4 A_{22}) + \cdots
\end{align}
Any significant low-$k$ scale-dependence is a sign of a breakdown or insufficiency in the field-level model used to describe this tracer. \par 

\begin{figure}
    \centering
    \includegraphics[width=\linewidth]{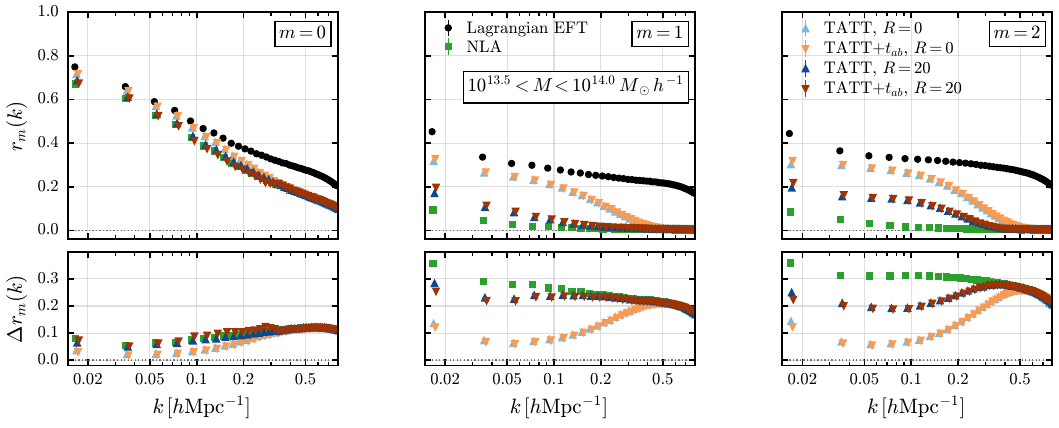}
    \caption{Comparison of cross-correlation coefficients achieved for Eulerian IA models and the 10-parameter field-level EFT. Each model is slightly horizontally offset so the points do not overlap. The $m=1$ and $m=2$ NLA curves at low-$k$ show the transfer function procedure fitting to the small number of phases available at low-$k$. No Eulerian IA model achieves as good of a field-level fit as the Lagrangian EFT. Uncertainties shown are on the mean $r_m$, averaged over 50 \quijote realizations.}
    \label{fig:eulerperr}
\end{figure}
In \Cref{fig:haloperr} we show the error power spectra, normalized by the shape noise amplitude, for the five halo mass bins in question. The reported uncertainties are the uncertainties on the mean from averaging over $N=50$ simulations. We see that deviations from `Poisson' stochasticity are very small -- on the order of 5\%. This is in contrast to scalar clustering where deviations from $\bar{n}P_{\rm err} \approx 1$ are on the order of 30\%~\cite{Kokron:2021faa}. The presence of sub-Poisson stochasticity hints at the imprint of the `halo exclusion' mechanism manifesting itself in the alignments of halo shapes. The amplitude is non-monotonic with halo mass, which is the same behavior observed for halo clustering -- there exists an interplay between the amplitude of the exclusion contribution and the amplitude of the shot noise. For the two lowest mass bins (densest samples), the error power spectra appear to increase at low-$k$, potentially inconsistent with the EFT prediction that they should be flat. All halo helicity $P_{\rm err}$ trend back towards the shape noise prediction at smaller scales.

\subsection{Comparison with Popular IA Models }

% \par 
It is also interesting to ask the question of whether this full field-level EFT is required in order to describe the halo shape field. To this end, we compare our field-level pipeline to two standard models commonly adopted in the intrinsic alignment literature -- the nonlinear alignment model (NLA) and TATT (tidal alignment and tidal torquing)\cite{Catelan_2001, 2007NJPh....9..444B, Blazek_2019}. The NLA model is defined by defining the halo shape field as being 100\% correlated (up to stochasticity) with the nonlinear tidal field of matter 
\begin{equation}
    \gamma_{ab} (\bx) = c_s \left (\frac{\partial_a \partial_b}{\partial^2} - \frac{\delta_{ab}^K}{3} \right ) \delta_m^{\rm NL} (\bx) + \epsilon_{ab}(\bx).
\end{equation}
It is immediately apparent that this model cannot work at the field level -- it predicts zero power for $m=1$ and $m=2$ helicities. We promote the NLA model to one with a free transfer function 
\begin{equation*}
    c_s \mathcal{D}_{ab} \delta_m (\bk) \to \mathcal{C}_{s;m}(k) \mathcal{D}_{ab} \delta_m (\bk)  
\end{equation*}
and find the Eulerian transfer function which minimizes the $m=0$ error power spectrum. We will similarly consider the Eulerian extensions of NLA by adding the Eulerian `TATT' operators 
\begin{equation*}
    (\delta s)_{ab} (\bx), (s^2)_{ab}(\bx),
\end{equation*}
and the `velocity shear' contribution $t_{ab} (\bx)$. Instead of being constructed from the shifted fields, we compute these operators by applying a Gaussian smoothing to the linear density field of the simulation on a scale $R$, and building the operators from products of the smoothed linear field. We pick $R=0 \mpch$ (smoothed only on the IC grid scale) and $R=20 \mpch $ (following \cite{Abidi:2018eyd, Akitsu_2023}). This implementation is consistent with how galaxy surveys such as DES and HSC currently implement their intrinsic alignment models~\cite{sanchezcid2026darkenergysurveyyear}, where the linear contribution is sourced from the nonlinear matter density field but the higher order terms are computed perturbatively. Each of these additional Eulerian operators also have their shape biases promoted to transfer functions.\par 
\Cref{fig:eulerperr} shows the comparison between NLA, TATT and the field-level EFT for the $M_h \in [10^{13.5}, 10^{14.0}]M_\odot h^{-1}$ mass bin. The specific metric for comparison being shown is the helicity-dependent cross-correlation coefficient, 
\begin{equation}
    r_m (k) = \frac{P^m_{\gamma, {\rm fit}}(k)}{\sqrt{P_{\gamma}^m(k)P_{\rm fit}^m(k)}}.
\end{equation}
For $m=0$, all Eulerian models decorrelate from the true halo shape fluctuation field at a lower wavenumber than the Lagrangian model. This disagreement is worse for the higher helicities, where the NLA model does not correlate with $m=1,2$ beyond numerical artifacts of the transfer function minimization. The extended Eulerian models present some cross-correlation with the true shape field, but they rapidly decorrelate past $k \gtrsim 0.1 \ihmpc$. 

The bottom panel shows the per-wavenumber improvement in cross-correlation coefficient $\Delta r_m$
of different helicities from using the field-level EFT compared to that specific perturbative model. 
The improvement of EFT
is the largest
w.r.t. the NLA model 
for $m=1,2$ helicities
because this model does not 
correctly capture them. 
The second worst model 
is TATT with the $R=20~h^{-1} {\rm Mpc}$
cutoff. The reason for the poor 
performance in this case is that
this aggressive cutoff 
removes too many quasi-linear 
modes which otherwise are 
well modeled by EFT. 
This is particularly 
important for the 
displacements that control the 
non-linear evolution of the BAO wiggles. Finally, the third worst model is un-filtered TATT, whose 
$\Delta r_{m=1,2}$
is small on large scales, but 
becomes sizeable for $k>0.2~\ihmpc$.

Since the addition of $t_{ab}$
does not affect $\Delta r_m$
significantly, the extra error of
this model 
does not seem to be related to missing shape field operators. 
Turning now to 
the $m=0$ sector, we see that 
filtered TATT and NLA models have a roughly similar $\Delta r_{m=0}$
deficit at low $k$, while the un-filtered 
TATT has no appreciable 
deficit there. The situation changes for $k>0.2~\ihmpc$
where all the models have a roughly similar $\Delta r_{m=0}$.
Since the models differ in their
operator content, it is suggestive 
that the degradation of the cross-correlation coefficient is driven 
by the lack of IR resummation
in these models, which is naturally captured in the field-level EFT. 

The upshot of our comparison is that full EFT outperforms all popular models
at the field level, and this result is not 
driven by additional parameters.
Indeed, having 
a correct implementation
of the displacement 
is crucial for
an optimal cross-correlation coefficient. 
The non-linear displacement of the EFT fields 
is captured by the Zel'dovich 
field, which does not involve any free parameters. 
At  the same time, 
the filtered and unfiltered TATT 
models have a very different performance as well, even though they are characterized by the same set of parameters. Our results thus strongly motivate the use of the full EFT calculations to model the intrinsic 
alignments in the 
cosmic shear data.

\subsection{Extracting coefficients from transfer functions}
\begin{figure}
    \centering
    \includegraphics[width=\linewidth]{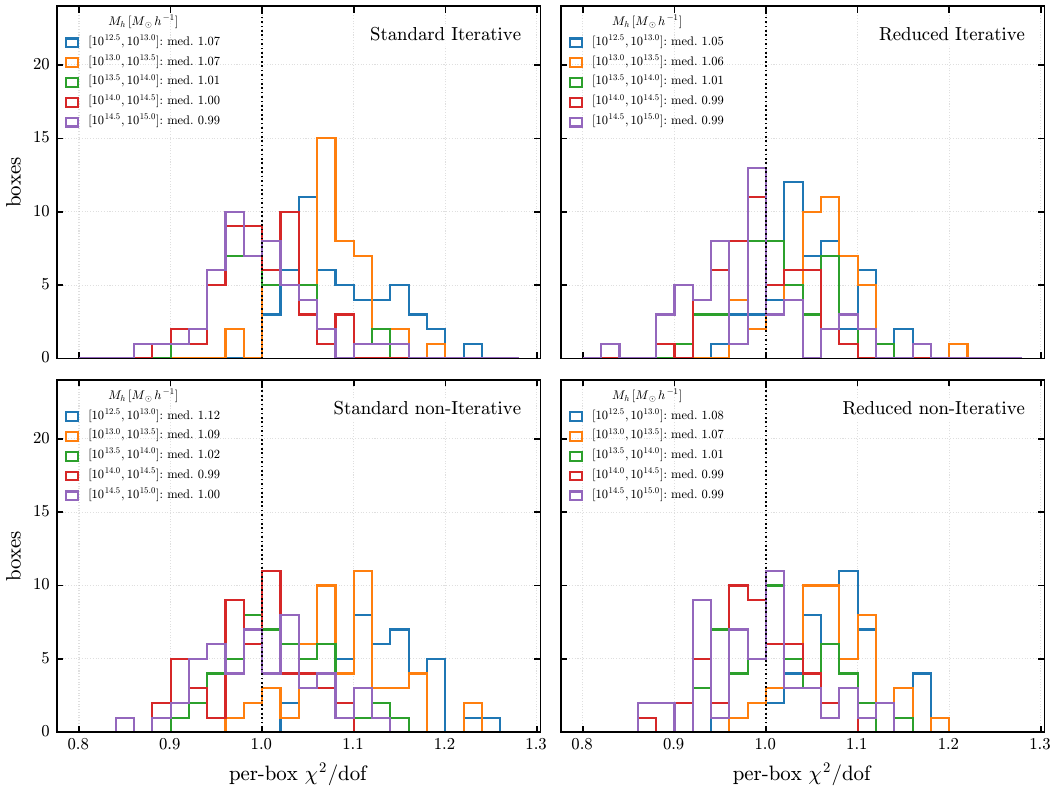}
    \caption{Distribution of $\chi^2$, normalized by degrees-of-freedom, for the \quijote mass bins. Each panel shows the distribution for a different definition of halo shape. For each mass bin we also quote the median reduced $\chi^2$ from the ensemble.}
    \label{fig:chi2distrib}
\end{figure}
The error power spectra obtained from the field-level EFT show the transfer functions are adequately capturing the structure of halo shape correlations. We now turn to extract the shape bias coefficients from the transfer functions of dark matter halo shapes. We focus on the coefficients $\{c_i\}$, and their results for both \flamingo and \quijote. Shape biases are obtained by solving the least-squares problem of \Cref{eqn:matrixsystem}. We restrict the design matrix to include the fiducial choice of scales discussed in \S~\ref{sec:tffit}: $\kmax = 0.4 \ihmpc$ for the linear and quadratic channels -- with two helicity-dependent broad-band terms, and $\kmax=0.07 \ihmpc$ for the cubic channels, while only fitting a constant amplitude. When applying these fits over the full suite of \quijote-HR boxes, we find the distributions of $\chi^2/{\rm dof}$ to have median reduced $\chi^2$ near 1, with a scatter of $\pm 0.05$, for all shape definitions and mass bins, with ${\rm dof} = 920$. This is shown in Fig.~\ref{fig:chi2distrib}. Note that the lowest mass bin has the most heavy-tailed distribution -- this is the bin with the lowest shape noise, and so transfer functions are measured at higher significance than for the more massive bins. The fit to the \emph{mean} transfer function has a very high $\chi^2$, but in this case the uncertainties are so small the transfer functions will be highly sensitive to two-loop operators not included in our model. We note in passing that the mean transfer functions can be fit to $\chi^2/{\rm dof} \sim 1.5$ if we restrict cubic transfer functions to $\kmax = 0.05 \ihmpc$ and linear+quadratic transfer functions to $\kmax = 0.2 \ihmpc$, but we will not comprehensively scan over fits to such a large volume in this work.  \par
\begin{figure}
    \centering
    \includegraphics[width=1\linewidth]{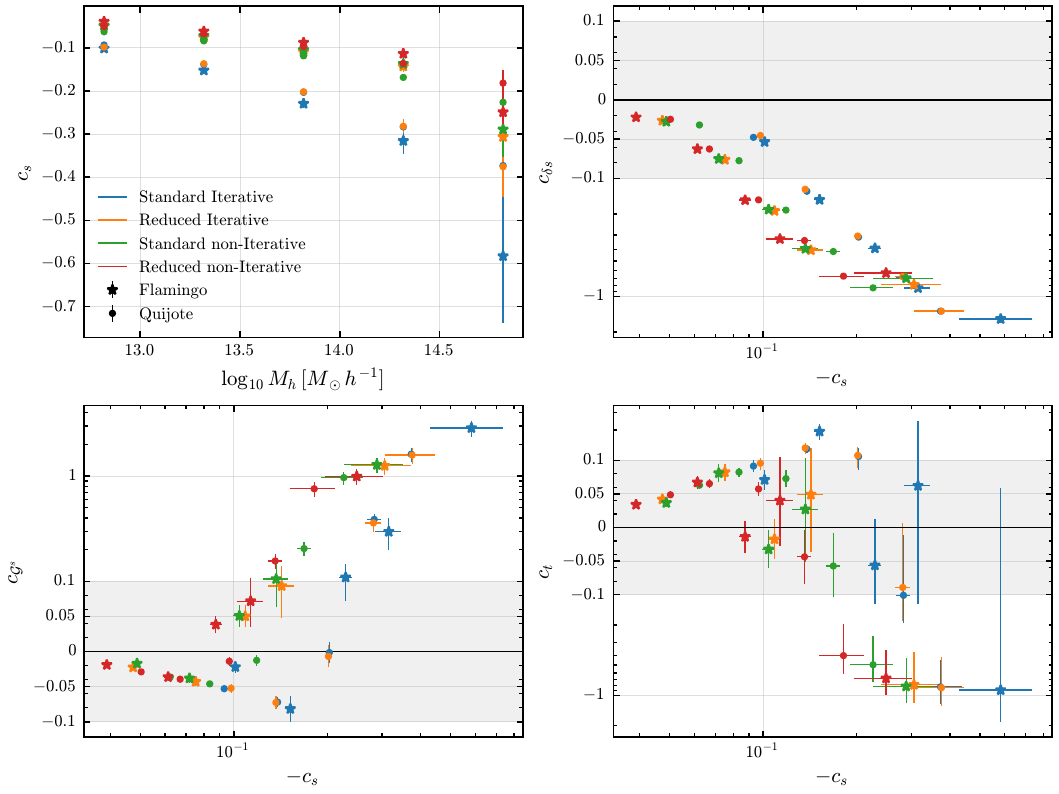}
    \caption{Measurement of linear and quadratic IA coefficients from field-level transfer functions, as a function of halo mass and halo shape definition. The uncertainties are estimated from the ensemble of $N=50$ simulations for \quijote, and analytically using the Fisher matrix of the least-squares problem for the DMO variant of \flamingo. The grey band denotes the region where the y-axis scale is linear, before switching to a logarithmic scale.}
    \label{fig:cparams}
\end{figure}

 
In Fig.~\ref{fig:cparams} we plot the linear and quadratic shape bias coefficients which result from the transfer function fit, across \quijote and \flamingo-DMO, for the four shape definitions considered in this work. The first panel shows the scaling of the linear alignment amplitude, $c_s$ with halo mass, while the other panels show how higher-order alignment values depend on $c_s$. The iterative halo shape definitions consistently lead to a higher alignment amplitude than the non-iterative definitions, consistent with past works on halo alignments~\cite{herle2026intrinsicalignmentsflamingosimulations}. We also observe broad consistency between \flamingo and \quijote alignment amplitudes, despite vastly different halo finding algorithms and different definitions for how bound particles are used to compute halo shape. The shape bias of dark matter halos is small compared to their clustering bias, $|c_s| \lesssim 0.4$ for most mass bins and shape definitions. Note that the parameter values we report are \emph{Lagrangian} shape biases, and so comparison with past measurements of higher-order shape biases (such as ~\cite{Akitsu_2023}) should be done with an appropriate conversion. 
This does not affect the linear coefficient but does affect higher-order coefficients. \par 
The density-weighted alignment coefficient $c_{\delta s}$ always has a negative sign, which grows in amplitude with $c_s$. The other two quadratic coefficients change sign as halo mass increases, at around $c_s \approx -0.1$, with only a minor dependence on specific mass bin or definition of halo shape. Even with the smaller volume of \flamingo the quadratic coefficients are detected at high statistical significance. The Lagrangian non-linear alignment amplitudes, when scaled by $c_s$, follow the same two `branches' which we find for the $c_s (M_h)$ dependence, where the type of halo shape (or the halo finder used) follows two broad relations. We also note $c_{s^2}$ and $c_{t}$ change sign as a function of halo mass / value of linear alignment coefficient. The broader trends (increasingly positive $c_s^2$ and negative $c_t$ for increasing halo mass) are consistent with the observations of \cite{Akitsu_2023}, who infer the trends in Lagrangian shape biases from deviations of their measured Eulerian biases from the co-evolution prediction. A careful comparison between their (Eulerian) bispectrum method and our Lagrangian field-level technique is of interest, but is not straightforward in practice. One effect that makes such comparison difficult is the use of an explicit cutoff in~\cite{Akitsu_2023},
which  significantly affects shape bias parameter measurements. The cutoff effect can be mitigated by an explicit computation 
of the renormalization-group evolution
of bias parameters, which 
we defer to future work.

\par 

\subsection{Cubic alignment coefficients}
\label{sec:cubicparams}
\begin{figure}
    \centering
    \includegraphics[width=1\linewidth]{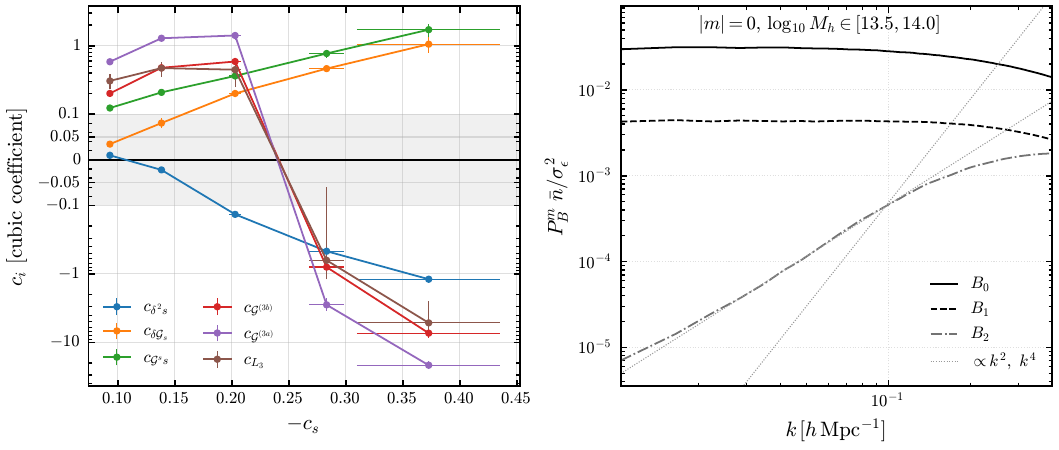}
    \caption{\emph{Constraints on cubic shape biases and impact on $P_{\rm err}$}. \emph{Left:} The values of cubic shape bias parameters, as a function of $c_s$, for \quijote dark matter halos. \emph{Right:} The auto contributions of the three invariant sub-spaces of cubic parameters, $\{C_i\}$ that are well constrained when fitting to the error power spectrum, normalized by the shape noise for that halo mass bin. Scalings as $k^2, k^4$ are shown in dotted lines as expected from analytic arguments.}
    \label{fig:cubicparams}
\end{figure}
We now turn to the cubic alignment coefficients inferred from our GLS fits. To our knowledge this is the first time cubic coefficients are measured in the literature -- in Ref.~\cite{chenkokron2024} their posteriors were consistent with zero when fit to the helicity auto-spectra of halos with a volume of $V=8\,(h^{-1}{\rm Gpc})^3$. The cubic parameters, for the \quijote halos with iterative unreduced inertia tensors, are shown in the left panel of \Cref{fig:cubicparams}. Their behavior is qualitatively similar to what is observed in the linear and quadratic sector -- increasing $|c_s|$ (or, equivalently, halo mass) increases their absolute value. For a subset of the cubic parameters (those which could potentially contribute at one loop to the power spectrum), their amplitudes coherently change sign from the bin at $M_h \in [10^{13.5}, 10^{14.0}]M_\odot h^{-1}$ to the $M_h \in[10^{14.0}, 10^{14.5}]M_\odot h^{-1}$. The parameters which contribute to $P_{\rm err}$ at low-k ($c_{\delta^2 s}, c_{\delta \mathcal{G}_s}, c_{\mathcal{G}_s s}$) are all measured at high significance and do not undergo sign changes. Given the coherent evolution of parameter values with halo mass, a possible hypothesis is that only a subset of coefficients are well-constrained, and the covariance between different cubic parameters leads to these shifts.\par 
To investigate this, we study the spectrum of the cubic parameter covariance matrix. Specifically, given the 10x10 parameter covariance matrix, $C$, inferred from the ensemble of bias estimates across the 50 Quijote boxes, we decompose the 6x6 sub-block of cubic parameters into its eigenvectors and eigenvalues
\begin{equation}
    C = U \Lambda U^T, \, C\cdot  u_k = \lambda_k u_k, 
\end{equation}
where $U \equiv u_{ik}$ is the matrix whose columns are the eigenvectors of $C$ and $\lambda$ the entries of the diagonal eigenvalue matrix $\Lambda$. For this decomposition, we can compute how well are the principal components of the cubic coefficients measured. The SNR for each principal component $k$ is defined as 
\begin{equation}
    {\rm SNR}_k = \frac{|u_{ik} c_i |}{\sqrt{\lambda_k}}. 
\end{equation}
In this decomposition, we find that for all mass bins, all eigenmodes are measured with ${\rm SNR} > 1$ for the volume of a single \quijote box. The SNR per component broadly grow as a function of mass, peaking for the $M_h \in[10^{14.0}, 10^{14.5}]M_\odot h^{-1}$ mass bin -- the next bin has too much noise compared to the amplitude of its cubic parameter contributions. While the individual eigenmodes change from one mass bin to the next, we find empirically that they change by a block-diagonal rotation with three fixed blocks. The sub-spaces which remain the same, roughly, are spanned by the modes $B_0 = {\rm span}\{u_0\}$, $B_1  = {\rm span}\{u_1, u_2, u_3\}$\footnote{That is, for mass bins $M, M'$, the eigenvectors $\{u_1, u_2, u_3\}$ are very close to orthogonal linear combinations of $\{u_1', u_2', u_3'\}$. From the covariance matrices estimated by 50 boxes, the cosine between different blocks does not exceed $\cos \theta_{BB'} \sim 0.1$, within the uncertainties from estimating this procedure from a sample of 50 simulations.} and $B_2  = {\rm span} \{u_4, u_5\}$. For example, we find empirically $u_0 \cdot c_i \approx c_{\delta^2 s} - c_{\delta \mathcal{G}_s}/3$ across all mass bins, with the dependence on other $c_i$ being consistent with numerical noise. This combination can be analytically understood from the fact $\delta \mathcal{G}_s$ contains $-\delta^2 s_{ab}/3$ within itself, and so this is the primary coefficient that is in fact constrained in the field level fit. Additionally, the modes which span the $B_2$ space is a 2D space nearly entirely composed of $c_{\mathcal{G}^{(3a)}}, c_{\mathcal{G}^{(3b)}}, c_{L_3}$. 

In general, we can define projector matrices $\Pi^B_{ab} = \sum_{k \in B} U_{ak} U_{bk}$ where the $k$ sum runs over modes in a given block subspace.
In the right panel of Fig.~\ref{fig:cubicparams} we show the contributions of these sub-spaces to the auto-spectrum of halo shapes for the $m=0$ helicity, for the mass bin of $M_h \in [10^{13.5}, 10^{14.0}]M_\odot h^{-1}$. Specifically, the quantities shown are the power in a block $B$\footnote{
To obtain mass-independent blocks we average each block projector, $\Pi_B$, over the five mass bins. We define the mass-averaged $B_0$ as the primary eigenvector, $u_0$, which is closely aligned to the $c_{\delta^2 s} - c_{\delta \mathcal{G}_s}/3$ direction and explicitly null out components to the eigenvector which are not these two, after noting the contamination is very small. A similar procedure is used to construct $B_2$ from ${\rm span}\{u_4, u_5\}$, and we again null out contributions unrelated to the cubic coefficients which contribute to $P_{13}$. The $B_1$ block is constructed as the complement of $B_0$ and $B_2$, to maintain the property that $\sum_B \Pi_B = \mathbb{I}$.}

\begin{equation*}
    P_{B}^{m} (k) = \Pi^B_{ab} \Pi^B_{dc} c_a c_d P_{bc}^m(k)
\end{equation*}

Note this is an illustration of how these combinations of coefficients contribute to the low-$k$ helicity spectrum but is not a comprehensive description -- we are not showing the cross-terms that contribute to the stochasticity e.g. $\langle \Pi^B_{ab}c_a O_b \gamma_h \rangle$, and the block fields themselves are not orthogonal, only the constrained parameter combinations. Additionally, the blocks correlate with the linear and quadratic coefficients to some degree (as expected for one-loop contributions). The biggest contribution to the block-auto contribution comes from $B_0$, and its amplitude is $\sim2.5\%$ of the total shape noise term. The other cubic sub-spaces contain differing combinations of the cubic bias coefficients, but we observe $B_1$ also contributes with a flat low-$k$ spectrum (albeit orthogonal to the $B_0$ contribution), and the $B_2$ mode corresponds to a term that scales as $k^2$. \par 
A particular correlation between operators that is large (and can be seen explicitly in Fig.~\ref{fig:paramcov}) is that of the cubic Galileons $c_{\mathcal{G}^{(3b)}}, c_{\mathcal{G}^{(3a)}}$. The cubic Galileons are defined to subtract specific $k^0$ overlaps from the $st_{ab}, \delta t_{ab}$ fields. However, their $k^2$ scaling is also the same. In principle the cubic basis can be further re-defined as
\begin{equation*} 
\mathcal{G}^{(3b)},\,  \mathcal{G}^{(3a)} \to \mathcal{G}^{(3a)} + \frac{1}{6} \mathcal{G}^{(3b)},
\end{equation*}
in which case one component scales as $\mathcal{O}(k^2)$ for its auto-spectrum while the other scales as $\mathcal{O}(k^4)$. This parameter combination is one of the directions contained in the `$B_2$' subspace -- along with the $c_{L_3}$ coefficient whose auto-power scales as $k^4$ for $m=0$. \par 
In conclusion, we are able to measure all six coefficients across mass bins in the field level analysis for halos, however a spectral decomposition of the marginalized parameter covariance matrix implies that three parameter sub-spaces dominate the overall field-level fit, two which contribute as constant low-$k$ amplitudes in the auto-spectrum, and one which is predominantly proportional to $k^2$ in the low-$k$ limit. 

\subsection{Comparison with fits to power spectra}

\begin{figure}
    \centering
    \includegraphics[width=\linewidth]{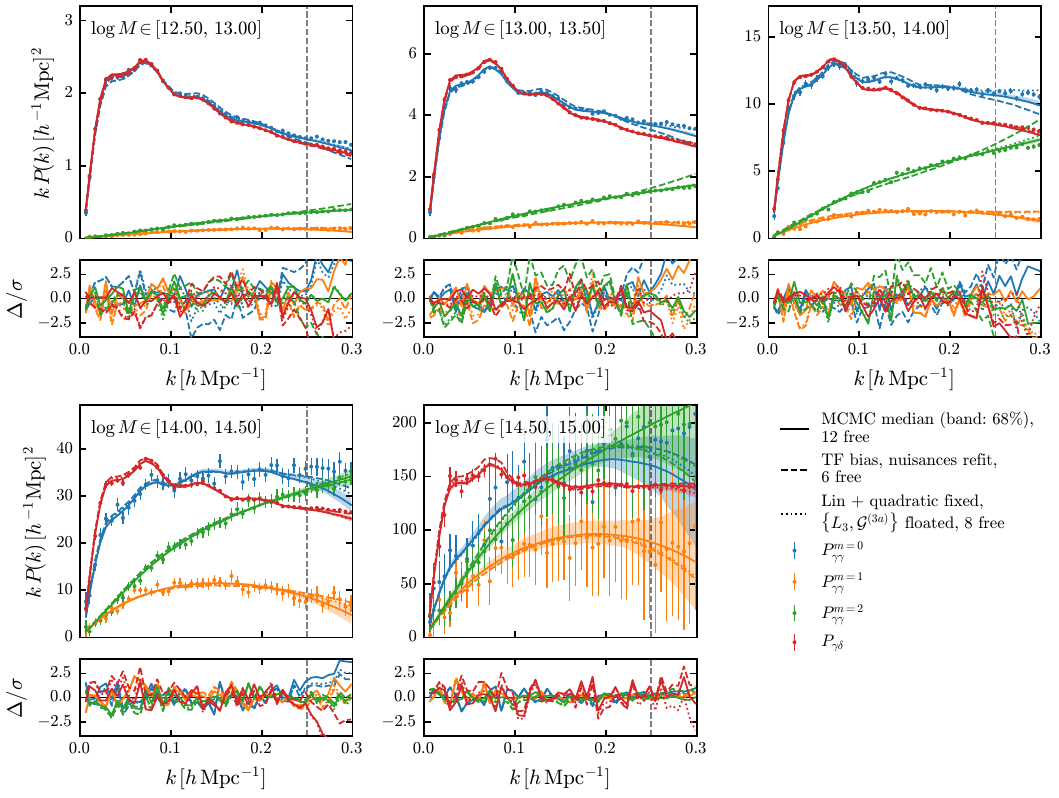}
    \caption{Fits to halo helicity power spectra and cross-spectra with model. Solid lines are the result from MCMC fits with a strict one-loop model. Dashed lines keep the biases fit to values from transfer functions, and refit the nuisance coefficients. The dotted lines are fits where a subset of the cubic coefficients and the nuisance parameters are free to vary. The shape--matter cross spectrum is normalized so its amplitude matches the m=0 helicity spectrum at $k=0.1 \ihmpc$. The bottom panel shows the residuals with respect to the uncertainties on the mean from $N=50$ simulations.}
    \label{fig:halopkfits}
\end{figure}
Are the best-fit values for $\{c_i\}$ consistent with fits to shape power spectra? In this section we compare with the perturbative 1-loop EFT of IA. For each sample of halo shapes, we measure four two-point correlators -- one shape--matter spectrum (GI) and the three helicity (II) components. For each mass bin, we fit to the data vector
\begin{equation}
    \boldsymbol{d} = \left (P_{\gamma \delta}(k_i), P_{\gamma\gamma}^{m}(k)\right ),
\end{equation}
where $P_{\gamma \delta} \equiv \langle \gamma^{m=0} \delta_{\rm NL} \rangle'(k)$ is the cross-spectrum between the $m=0$ component of the halo shape field and the non-linear matter density field. The covariance matrices are estimated from the ensemble of \quijote HR simulations. On the perturbative scales considered here we observe no discernible off-diagonal contribution to these covariances other than the cross-covariance between $P_{\gamma\delta}$ and $P_{\gamma \gamma}^{m=0}$ for a fixed $k$ mode. A Gaussian likelihood for the power spectra decomposes into an $m=0$ component and independent $m=\{1, 2\}$ components
\begin{equation}
    -\ln L \propto \sum_k \left (\br ^T C^{-1} \br \right )_{m=0} (k)+ \sum_{m=1}^2\sum_k \left ( \frac{\hat{P}_{\gamma \gamma}^m -  \bar{P}_{\gamma \gamma}^m}{\sigma_{\gamma \gamma, m}} \right )^2 (k), 
\end{equation} 
where the $m=0$ residual vector for a given k-mode is 
\begin{equation*}
    \br(k) = \left (\hat{P}_{\gamma \delta} - \bar{P}_{\gamma \delta},\hat{P}_{\gamma \gamma}^{m=0} -  \bar{P}_{\gamma \gamma}^{m=0} \right ) (k). 
\end{equation*}
The predictions for the shape-matter and shape-shape spectra follow from the bias expansion 
\begin{align*}
    \bar{P}_{\gamma \delta} (k) &= \sum_{i} c_i \langle \widetilde{\mathcal{O}}_i^{m=0} \delta_m \rangle (k) + \alpha k^2 P_{\tilde{s}\delta}(k) +  a_2 k^2, \\
    \bar{P}^{m}_{\gamma \gamma}(k) &= \sum_{i,j} c_i c_j \langle \widetilde{\mathcal{O}}_i^{m}\widetilde{\mathcal{O}}_j^{m} \rangle + \beta k^2 P_{\tilde{s}\tilde{s}}^m(k) + N_0 + N_m(k),
\end{align*}
where $\alpha, \beta$ are the counter-term contributions at one-loop (analogous to $\nabla^2 s_{ab}$) and $a_2$ is the leading stochastic contribution to the cross-spectrum. $N_0$ and $N_m$ are the helicity-dependent contributions detailed in \Cref{eqn:mperr}. 
 While the shifted operator cross-spectra $\langle \widetilde{\mathcal{O}}_i^{m}\widetilde{\mathcal{O}}_j^{m} \rangle$ can in principle be computed in perturbation theory, we opt to estimate them from the $N=50$ \quijote-HR simulations, using Zel'dovich displacements. 
 For the cross-spectra that contribute to $P_{\gamma \delta}$ we take the cross-spectrum of shifted operators with the non-linear matter density field, since this is the field the halo shapes are cross-correlated with. We only include correlations with cubic operators that are at most one loop. This model is formally equivalent to the one-loop spectrum used in \cite{chenkokron2024}, but the use of Zeldovich-only displacements for the basis implies the coefficient values should not be directly compared (as explicitly shown in \Cref{eqn:shapeshift}, higher-order displacements modify coefficient values). \par
In Fig.~\ref{fig:halopkfits} we show all four shape noise-subtracted spectra for \quijote, for five mass bins. The uncertainties shown are uncertainties on the mean, averaged over the full 50 $($Gpc/$h)^3$ volume. Three different fits to these shape spectra are shown per bin, and the bottom panel of each plot shows the residuals normalized by the uncertainty on the mean. Solid lines are a fit where all 12 parameters, including the shape biases, are allowed to vary. This fit is analogous to that of \cite{chenkokron2024}, and despite the large volume the residuals are closely scattered around zero with no large excursions. This full-freedom fit is done through markov chain monte carlo using \texttt{emcee}~\cite{2013PASP..125..306F}, in the Lagrangian shape bias basis. The shaded bands indicate the 68\% posterior predictive distributions from fitting to the full volume. We fit to slightly larger scales of $\kmax = 0.25 \ihmpc$, partially due to the larger volume and partially due to only using shifted operators that lack any power from small-scale displacements.\par 
We compare the full-freedom MCMC fit to maximum-likelihood fits where subsets of the shape bias parameters are constrained to parameters obtained from the transfer function fits. The dashed lines are fits where all 10 $\{c_i\}$ are kept fixed to the value from the transfer function fit, and only the six nuisance parameters related to noise and counter-terms are allowed to vary. Here, we identify some model mismatches for the lower mass bins, with residuals around the BAO at $k\sim0.15 \ihmpc$ reaching  $4\sigma$. These excess residuals are driven by the cubic sector, which implies that the inferred cubic parameters from the field-level fit are potentially biased. To check for this, we carry out a third fit where the linear and quadratic coefficients are kept fixed, and a subset of two cubic bias parameters have variable amplitudes (as expected for the one-loop power spectrum). These results are shown in the dotted lines. Recalibrating the cubic coefficients restores concordance around the BAO wiggles.\par 
In summary, we have carried out a comprehensive examination of the spin-2 field-level bias expansion when applied to dark matter halo shapes across two suites of simulations -- \flamingo-DMO and \quijote-HiRes -- and four different definitions of halo shape. We find that the field-level model produces error power spectra that are flat at low-k (and with up to 5\% lower amplitude than the Poisson shape noise prediction). No Eulerian field-level model, even with the full freedom of a transfer function, can achieve comparably low error power spectra. We then use the measured transfer functions to infer the Lagrangian shape bias coefficients $c_i$ as a function of halo mass, halo finder, and shape definition. We find that the quadratic non-linear alignment coefficients all grow in amplitude as the linear coefficient, $c_s$, increases. We also find that $c_{\mathcal{G}^s}, c_{t}$ change signs once, around the mass bin $M_h \in [10^{13.5}, 10^{14}] M_\odot h^{-1}$, for essentially all halo types and shape definitions considered. We then showed these same linear and quadratic coefficients inferred at the field-level can reproduce the set of shape auto-and-cross spectra as a function of mass, but the cubic coefficients must be recalibrated in order to achieve a good fit to the BAO of halo shapes.

\section{Galaxy shape alignments in hydrodynamic simulations}
\label{sec:galaxyalign}

\begin{figure}
    \centering
    \includegraphics[width=1\linewidth]{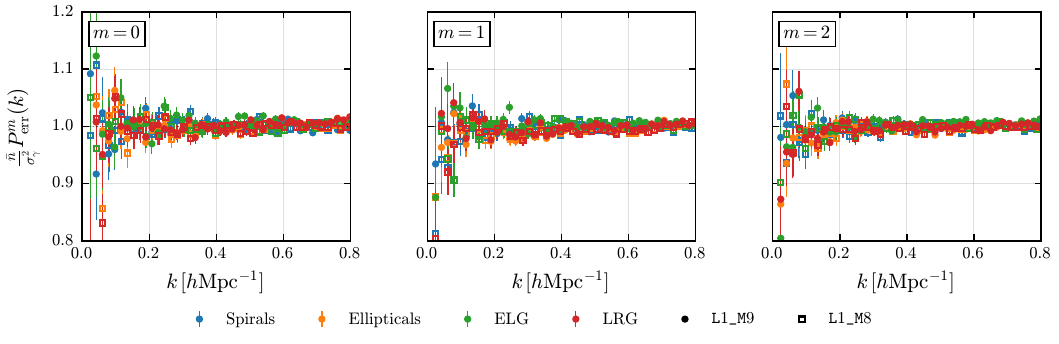}
    \caption{Error power spectra of galaxy shape samples from \flamingo considered in this work, normalized to their shape noise expectation. }
    \label{fig:galperr}
\end{figure}

Having studied and calibrated the field-level alignment machinery against dark matter halos, we turn to measurements of the transfer functions of galaxy shapes in hydrodynamical simulations.  We split galaxies in \flamingo into two separate categories: The first split is motivated by measurements of direct intrinsic alignments in spectroscopic galaxy surveys, such as BOSS and DESI. Alignments have been detected in spectroscopic catalogs of LRGs for nearly two decades~\cite{Hirata_2007}, primarily with galaxies at $z \lesssim 0.5$. Significantly larger number densities and volumes afforded by modern instruments have pushed the frontiers of direct alignment studies to higher redshifts, including for star-forming emission line galaxies (ELGs)~\cite{siegel2025intrinsicalignmentdemographicsnextgeneration}, and potentially new samples which can be used to calibrate photometric surveys~\cite{derose2026steelingweaklensingsource}. Reproducing LRG-like and ELG-like samples within a hydrodynamic simulation provides an interesting point of comparison to current data. In \flamingo, we define LRGs as galaxies which have a low specific star formation rate (sSFR $< 0.1 {\rm Gyr}^{-1}$) and large stellar mass ($M_* > 1.8\times 10^{11} M_\odot$). This selection gives a number density of $\bar{n}_{\rm LRG} = 6.8 (7.8) \times 10^{-4} \left (h {\rm Mpc}^{-1}  \right)^3$ for \texttt{L1\_M9} and \texttt{L1\_M8}, respectively, which closely matches reported DESI number densities for LRGs at $z\approx 0.5$. For ELGs, we cut on ${\rm sSFR} \geq 0.1 {\rm Gyr}^{-1}$, and cut on $M_* > 10^{11} M_\odot$. This selection gives $\bar{n}_{\rm ELG} = 4.0 (1.6) \times 10^{-4} \left (h {\rm Mpc}^{-1}  \right)^3 $, which are consistent with those from past studies of DESI-like galaxies in the \texttt{IllustrisTNG} simulation~\cite{Yuan_2022}. The decrease in the number density of ELGs from going to a higher mass resolution is driven by a lower star formation rate at fixed stellar mass in the higher-resolution \flamingo box. \par 
The second set of sample selections are based on a morphological split of the galaxies. We sub-select catalogs into `spiral' and `elliptical' galaxies (following~\cite{Delgado_2023}), by splitting galaxies along their fraction of baryonic kinetic energy in rotation, $\kappa_{\rm corot}$. Galaxies with $\kappa_{\rm corot} > 0.5$ are deemed rotationally supported, and an additional cut on ${\rm sSFR} > 0.1 {\rm Gyr}^{-1}$ selects for blue spiral galaxies. Setting $\kappa_{\rm corot} < 0.5$, a quenched selection of ${\rm sSFR} < 0.1 {\rm Gyr}$, and a stellar mass cut of $M_* > 2\times10^{11}$ yields a sample we define as the elliptical galaxy sample. This morphological selection of spiral galaxies is a proxy for selections of blue galaxies in cosmic shear samples, which optimize for number density in order to suppress shape noise. The number densities for the elliptical sample are similar to the LRGs -- $5.6 (6.4) \times 10^{-4} \left (h {\rm Mpc}^{-1}  \right)^3$ -- while the spiral galaxy selection possesses the property of having the highest number density of $6.0 (20) \times 10^{-4} \left (h {\rm Mpc}^{-1}  \right)^3$. The importance of the \texttt{L1\_M8} resolution is made clear here: the fiducial resolution does not properly resolve the host halos of spiral galaxies. \par 
We define galaxy shapes as the inertia tensor of the star particles bound to a given subhalo, once again reconstructing the SOAP inertia tensors for missing subhalos. Having defined the galaxy samples we proceed to make a second quality cut on their shape tensors. We find that certain galaxy shapes (when using iterative shape definitions) have nearly-singular inertia tensors. These discrepant galaxies contribute more than 40\% of the variance in the shape tensor for the sample, inflating the shape noise (through $\sigma_\gamma^2$) significantly. We thus filter the galaxy catalogs and selectively exclude the galaxies which contribute the most to the population's shape noise, until the value for $\sigma^2 _\gamma$ for the sample is converged to 5\%. In practice, this results in minor cuts to the number density of the sample -- at most 3\% of galaxies in the sample are removed (for the high resolution ELGs). While we have made measurements for all four shape definitions, we only report the results for iterative, standard stellar shapes (but find qualitatively similar SNR on coefficients for all shape definitions).\par 
\begin{figure}
    \centering
    \includegraphics[width=1\linewidth]{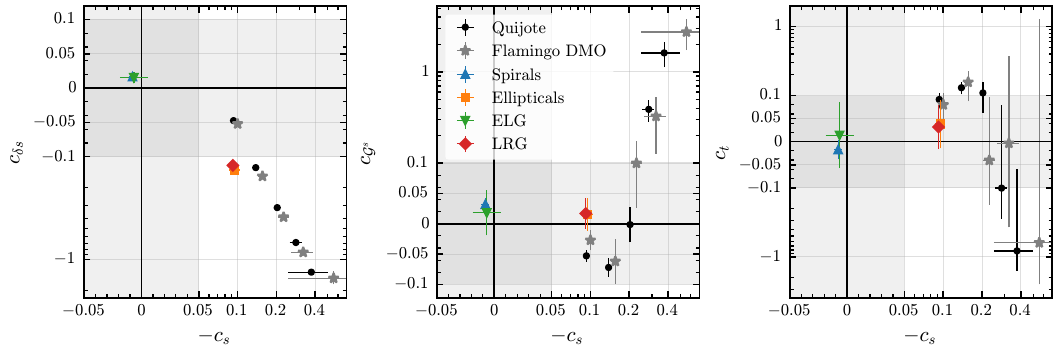}
    \caption{Shape bias coefficients for the four \texttt{L1\_m8} galaxy samples extracted from \flamingo in this work, compared to the dark matter halo samples. Uncertainties shown are 2$\sigma$, corresponding to the volume of a single realization. The shaded bands denote the region where the axes scale linearly, scaling logarithmically outside. }
    \label{fig:galparams}
\end{figure}
 
In \Cref{tab:flamingo_gal_bias} we report the inferred values of linear and quadratic shape biases for the \flamingo sample, from the field-level methodology developed in this work, using the same scale cuts and broad-band choices which were calibrated on the \quijote dark matter halo ensemble. Uncertainties on the measured transfer functions are given by \Cref{eqn:tferr}, and the error bars on parameters come from \Cref{eqn:paramfisher}. The error power spectra for all galaxy samples, shape definitions and helicities are consistent with the shape noise expectation, and this is explicitly shown for the standard iterative shapes in \Cref{fig:galperr}. In \Cref{fig:galparams}, we show the $c_i(c_s)$ relations for the subset of galaxies from \texttt{L1\_M8} and how they compare to these relations for dark matter halos in \flamingo and in \quijote. We have also tried to measure cubic alignment coefficients for galaxies, but only find a detection of a $c_{\delta^2 s}$ for the LRG/elliptical sample, which does not contribute at one loop in the power spectrum. We do not find any significant detections for any other coefficient or galaxy sample, and so we do not report them in \Cref{tab:flamingo_gal_bias}. 
\par 
For galaxies living in massive halos (LRGs / ellipticals), the inferred shape biases are similar to each other in terms of central values and uncertainties. Linear alignments are found to be within $c_s \in [-9.49, -8.34] \times 10^{-2}$ with uncertainties on the order of $\sigma_{c_s} \sim 4\times10^{-3}$ -- strong detections of $\sim 20\sigma$. The density-weighted alignment $c_{\delta s}$ is similarly detected at high statistical significance ($>40\sigma$). The inferred alignments are consistent across mass resolutions. For these galaxies (and fiducial shape definition) we find no strong evidence for tidal galileon or velocity shear contributions. \par 
The alignments of LRGs in \flamingo were previously studied in \cite{herle2026intrinsicalignmentsflamingosimulations, herle2026assemblybiasredshiftevolution} using the $L=2.8{\rm Gpc}$ \flamingo box (albeit not with fully equivalent definitions of what constitutes an LRG, and with projected iterative stellar shapes). From the posteriors shown in Figure 2 of \cite{herle2026intrinsicalignmentsflamingosimulations}, we take their fiducial value of $A_{IA} = 5 \pm 0.2$ for $z=0$ as a point of comparison. In \Cref{app:conversion} we show that this corresponds to an estimate of $10^2 c_s \approx -11.05 \pm 0.44$. Our inferred value of $-9.4 \times 10^{-2}$ is qualitatively consistent with this prior work (and the inferred redshift evolution for LRG alignment displayed in \cite{herle2026assemblybiasredshiftevolution}). What's notable is the size of the uncertainty: the 2.8 Gpc \flamingo box has $\sim20\times$ more volume than the fiducial box considered in this work, but the relative detections for LRGs alignments are comparable (${\rm SNR}^{(2.8 {\rm Gpc})} \sim 11 / 0.44 \sim 25$ and ${\rm SNR}^{(1.0 {\rm Gpc})} \sim 9.25/0.46 \sim 20$). This highlights the power of the field-level approach to probing intrinsic alignments. \par 
Turning to the blue samples, the statistical power of the field-level analysis allows us to make, for the first time, precise measurements of the alignments of blue galaxies in the \texttt{L1\_M8} \flamingo box. For ELGs, their low number density do not allow for a robust detection of linear alignment or tidal galileon contribution, but the central value is comparable to the spiral result. Direct measurements of intrinsic alignments for ELGs in DESI report non-detections (Table 3 of \cite{siegel2025intrinsicalignmentdemographicsnextgeneration}.) with $\sigma (A_{1}) \sim 1, \sigma(A_2)\sim 4, \sigma(b_{TA})\sim1$. A linear propagation of uncertainty using $A_1$ and $b_{TA}$ gives $\sigma(A_{1\delta}) \sim 1$ for this DESI ELG sample. Translating from \cref{tab:flamingo_gal_bias}, we find in \flamingo that Eulerian ELG alignment amplitudes are 
\begin{align*}
& A_1 =  -0.52 \pm 0.56, \\
&A_{1\delta} = -1.01 \pm 0.55, \\
&A_2 = 0.15 \pm 0.27.    
\end{align*}
We thus see that our inference is consistent with the current non-detections of alignment in ELG for these coefficients. Our measurements have smaller error bars than what the two-point function analysis in data is able to achieve, despite the larger effective volume of the ELG sample of $V_{\rm eff}^{\rm ELG} = 2 {\rm Gpc}^3$~\cite{DESI:2025yuk}. We find mild evidence ($3.5\sigma$) for a density-weighted alignment component in the ELG signal of $10^2 c_{\delta s} = 1.47 \pm 0.42$\footnote{Although one can see that the significance for the \emph{Eulerian} $A_{1\delta}$ is lower. This comes from  the formulae in \Cref{app:conversion}, with properly propagating uncertainties in $c_{\delta s}$ and $c_{\mathcal{G}^s}$.}, and the other quadratic contributions are consistent with zero given uncertainties. \par 
More interestingly, we now turn to the denser spiral galaxy sample. Here, we find a $>4\sigma$ detection of a \emph{linear} $c_s$ with opposite sign to the LRG alignment. This is consistent with past measurements of tangential alignment of disc galaxies in hydrodynamical simulations~\cite{Chisari_2015, zjupa2020intrinsicalignmentsillustristngimplications, van_Heukelum_2026}. However, in contrast with past work we note the sign of spiral alignment amplitude for \texttt{L1\_m8} spirals does not depend on the shape definition used. We also detect non-zero density-weighting contribution ($>13 \sigma$) \emph{and} non-zero tidal Galileon term ($> 6\sigma)$ -- which is equivalent to a non-zero `tidal torquing' contribution.  Using the formulae of \Cref{app:conversion}, the equivalent TATT amplitude constraints we find for the \texttt{L1\_M8} spirals are
\begin{align*}
& A_1 =  -0.76 \pm 0.16, \\
&A_{1\delta} = -0.77 \pm 0.16, \\
&A_2 = 0.34 \pm 0.08.
\end{align*}
It is appealing to ask the question of what amount of contamination relative to a cosmic shear data vector do these values of $A_i$ support. However, this requires a detailed set of measurements for such a sample across a broad range of redshifts, due to the width of the $n(z)$ of cosmic shear samples from photometric surveys. We will investigate the redshift evolution of galaxy alignments in \flamingo in upcoming work. \par 
The application of field-level perturbative IA modelling to hydrodynamical simulations, or simulations which use empirical prescriptions to imbue alignments~\cite{Van_Alfen_2024}, paves the way for the calibration of simulation-based priors (SBPs) for these probes in much the same way as has recently been proposed for galaxy densities and the Lyman-$\alpha$ forest~\cite{Ivanov:2024hgq,Ivanov:2024xgb,ivanov2025fundamentalphysicslymanalphaforest, sullivan2025highredshiftmillenniumastridgalaxies}. The utilities of SBPs for alignments are two-fold: they will enhance the constraining power from growth-of-structure constraints using alignments, and will also be important in bounding the allowed contamination of IA in cosmic shear measurements in the absence of direct calibration~\cite{chen2024lensinglowanalysisdesitimesdes}. 
\begin{table}[t]
  \centering
  \begin{tabular}{lccccc}
    \hline\hline
    Selection & $10^{2}\,c_{s}$ & $10^{2}\,c_{\delta s}$ & $10^{2}\,c_{\mathcal{G}^s}$ & $10^{2}\,c_{t}$ & $\chi^2/\mathrm{dof}$ \\
    \hline
    Spirals (\texttt{L1\_M8}) & $0.83 \pm 0.18$ & $1.65 \pm 0.12$ & $3.22 \pm 0.53$ & $-1.65 \pm 1.01$ & $1.09$ \\
    Ellipticals (\texttt{L1\_M8}) & $-9.25 \pm 0.46$ & $-13.46 \pm 0.31$ & $1.64 \pm 1.38$ & $4.32 \pm 2.64$ & $1.04$ \\
    ELG (\texttt{L1\_M8}) & $0.57 \pm 0.62$ & $1.47 \pm 0.42$ & $1.62 \pm 1.86$ & $1.31 \pm 3.57$ & $0.96$ \\
    LRG (\texttt{L1\_M8}) & $-8.99 \pm 0.42$ & $-12.16 \pm 0.28$ & $1.79 \pm 1.25$ & $3.59 \pm 2.41$ & $1.05$ \\
    \hline
    Spirals (\texttt{L1\_M9}) & $-0.66 \pm 0.57$ & $1.09 \pm 0.39$ & $2.23 \pm 1.71$ & $-7.04 \pm 3.29$ & $1.03$ \\
    Ellipticals (\texttt{L1\_M9}) & $-8.89 \pm 0.48$ & $-13.53 \pm 0.33$ & $1.61 \pm 1.44$ & $3.78 \pm 2.75$ & $1.00$ \\
    ELG (\texttt{L1\_M9}) & $-0.65 \pm 0.45$ & $-0.25 \pm 0.30$ & $0.30 \pm 1.34$ & $1.23 \pm 2.57$ & $1.14$ \\
    LRG (\texttt{L1\_M9}) & $-8.37 \pm 0.43$ & $-12.22 \pm 0.29$ & $1.08 \pm 1.29$ & $3.42 \pm 2.48$ & $0.96$ \\
    \hline\hline
  \end{tabular}
  \caption{Linear and quadratic tidal-Galileon shape-bias coefficients for the FLAMINGO galaxy samples at high (\texttt{L1\_M8}) and normal ($\texttt{L1\_M9}$) resolution, for iterative, unreduced stellar inertia tensors (\texttt{sshape\_iter}). The reported uncertainties are marginalized over the 31 cubic and broad-band parameters. \flamingo analyses have ${\rm dof} = 568$}
  \label{tab:flamingo_gal_bias}
\end{table}
\section{Conclusions}
\label{sec:conclusions}
In this work, we have constructed the Lagrangian spin-2 bias expansion at the level of the full real-space 3D shape fluctuation field, including all operators to cubic order.  This is in contrast with past works that focused on summary statistics calculated within this model (or the accompanying Eulerian formulation of the EFT of IA). We stress-tested the field-level EFT's ability to describe the fluctuation field of dark matter halo and galaxy shapes in numerical simulations, finding a higher cross-correlation coefficient between the field-level EFT and halo shapes than any of the popular IA models used in the literature (NLA or TATT). Additionally, we find that for the volume of a single \quijote realization the distribution of reduced $\chi^2$ from fitting the field level model has a median value close to one, indicating the model is describing the structure of the field-level observables well for the set of scales considered. \par 
We also found the field-level formulation of shape statistics presents some additional subtleties not present in the scalar case -- namely, the transfer functions measured ( $\mathcal{C}_{m;i}(k)$ ) are helicity-dependent, and sensitive to different combinations of bias coefficients even in the low-$k$ limit. This makes a fully analytic / perturbative description of the transfer functions challenging, and motivates a lattice-based construction of the underlying correlations that set the relation between the coefficients of the model and the transfer functions. The accuracy of our lattice-based construction was validated using a set of Gaussian field correlators for which the analytic expectation can be calculated. \par 
We then showed that the same procedure used to stress-test the field-level EFT can be re-cast as a formal generalized least-squares regression problem. This led to an estimator for the shape bias coefficients of this expansion, as well as its accompanying uncertainties, which cancels cosmic variance in the measurement. Notably, compared to past field-level EFT publications, the GLS-based estimator for coefficients allows us to characterize the uncertainty and covariance on measured parameters with a simple analytic expression. This analytic expression is compared to the ensemble covariance matrix measured from 50 boxes and found to be in good agreement for both numerical value and the off-diagonal cross-correlation coefficients. \par 
Applying this estimator to dark matter halo shapes in the \quijote simulation leads to precise measurements of all ten coefficients through cubic order, in a way that is consistent with coefficients inferred from fitting to the one-loop shape auto-spectrum and shape-matter cross-spectrum. We find that the higher order halo alignment coefficients generically grow with halo mass, and trends with the linear coefficient $c_s$ are broadly independent of simulation suite or halo shape definition used. The coefficients associated with tidal torquing and velocity shear tensors $(c_{\mathcal{G}^s}, c_t)$ grow non-monotonically with halo mass, changing sign at $|c_s| \gtrsim 0.1$. This pattern is repeated in the cubic coefficients, where operators which contribute with a flat low-$k$ contribution to the error power spectrum are constrained, with monotonic growth in their amplitude, whereas contributions which enter at one loop ($c_{\mathcal{G}^{(3a)}}, c_{\mathcal{G}^{(3b)}}, c_{L_3}$ undergo a similar sign-change at a specific halo mass bin. \par 
As a final application we turned to the case of analyzing samples of galaxies in the \flamingo hydrodynamic simulation. Two sets of sample selections were made, one focusing on spectroscopic samples and another focusing on a morphological classification. While more massive / red galaxies were found to be similar across mass resolutions, the higher-resolution \texttt{L1\_m8} box is needed to resolve the properties of blue galaxies which are relevant for studies of ELG alignment and cosmic shear contamination. Our estimator recovers alignment amplitudes for LRGs consistent with those found from higher-volume \quijote boxes at similar signal-to-noise ratio, despite the $20\times$ lower volume used, as well as precise measurements of higher order coefficients. For our bluer samples of galaxies, we find no evidence for ELG alignment, consistent with past empirical studies at the current uncertainties, but for the `spiral' galaxy sample we consider we detect at high significance evidence for linear alignment, density-weighted alignment, and tidal torquing contributions. This detection of linear-and-nonlinear alignment in spiral galaxies motivates further characterization of their signatures. \par 
There are several directions to which we could extend our work. Perhaps the most pressing direction would be to characterize the dependence of alignment coefficients on hydrodynamical prescription, but we find that the necessity of using \texttt{L1\_m8} hinders the study of alignments for the various \flamingo realizations. Of direct relevance to characterizing the impact of intrinsic alignments in weak lensing surveys is to extend our studies to multiple redshifts, as the time-dependence of the alignment signal is essentially unknown. Another pressing avenue is using paired measurements from halos and galaxies to understand whether a `halo occupation distribution-like' construction can be made for shape galaxies~\cite{Fortuna_2020, Van_Alfen_2024, Pandya_2026, gatti2026halobasedintrinsicalignmentmodelsimulationbased}, from which high-fidelity halo alignment coefficient measurements could be used to build priors for galaxy alignments (as has been successfully done for scalar galaxy bias~\cite{ivanov2025simulationbasedpriorssimulationsanalytic}). Finally, we also note that one of the unique applications of galaxy shape correlations is the ability to probe anisotropic forms of primordial non-Gaussianity. The new operators induced in the shape bias expansion from these forms of PNG carry their own EFT coefficients, but field-level techniques can be used to measure these coefficients~\cite{sullivan2025localprimordialnongaussianbias, sharma2025equilateralnongaussianbiasfield}. It is a straightforward extension of the formalism developed out in this work to develop an estimator for these PNG shape bias coefficients, conditioned on simulations which have anisotropic PNG~\cite{Akitsu_2021}. It is our intention to return to many of these points in upcoming work. 
\acknowledgments
We thank Beatriz Tucci, Noah Sailer, Martin White, Francisco Villaescusa-Navarro, Elisabeth Krause and Matias Zaldarriaga for helpful conversations. We thank Kazuyuki Akitsu for insightful conversations and code comparison. NK acknowledges support from a potpourri of funding sources including Edward and Kiyomi Baird, the Bershadsky Fund, and the Fund for Memberships in Natural Sciences of the Institute for Advanced Study. NK gratefully acknowledges the Pacific Postdoctoral Program at the Dark Universe Science Center, University of Washington, where part of this work was carried out. The Pacific Postdoctoral Program is supported by a grant from the Simons Foundation (SFI-MPS-T-Institutes-00012000, ML). 
This work was initiated at the 
Aspen Center for Physics,
which is supported by National Science Foundation grant PHY-2210452 and a grant from the Simons Foundation (1161654, Troyer).
MMI is supported by the Office of High Energy Physics of U.S. Department of Energy under grant Contract Number  DE-SC0012567. 
JMS acknowledges support provided by The Brinson Foundation through a Brinson Prize Fellowship grant.
Calculations and figures in this work have been made using \texttt{nbodykit} \citep{Hand_2018} and the SciPy Stack \citep{2020NumPy-Array,2020SciPy-NMeth,4160265}. This research has made use of NASA's Astrophysics Data System and the arXiv preprint server. Large-language models provided by OpenAI under an educational license -- of the GPT-5.6 and GPT-6 family -- as well as a suite of Anthropic's Claude models were used to aid in some of the analytic and numerical aspects of this work. These uses were independently cross-checked by the authors, who take full responsibility. No LLMs were used for the purposes of writing text.

\appendix
\crefalias{section}{appendix}
\section{First-principles transfer function and orthogonalization:}
\label{app:tfortho}
The starting point for the orthogonalization procedure in the scalar case is the construction of the operator correlation matrix (as discussed in \cite{Schmittfull_2019})
\begin{equation}
    \mathbb{C}_{ij} (k) = \frac{\langle \mathcal{O}_i (\bk ) \mathcal{O}_j ^* (\bk)  \rangle }{\sqrt{\langle \mathcal{O}_i (\bk) \mathcal{O}_i^*(\bk) \rangle \langle \mathcal{O}_j (\bk) \mathcal{O}_j^* (\bk)\rangle}}.
\end{equation}
Let's consider, generically, the extension of this to the helicity modes $m$
\begin{equation}
    \mathbb{C}^{mm'}_{ij} (k) = \frac{\langle \mathcal{O}_i^m (\bk ) \mathcal{O}_j ^{m'*} (\bk)  \rangle }{\sqrt{\langle \mathcal{O}_i^m (\bk) \mathcal{O}_i^{m*}(\bk) \rangle \langle \mathcal{O}_j^{m'} (\bk) \mathcal{O}_j^{m'*} (\bk)\rangle}}.
\end{equation}
Since operators of different helicities in this basis are orthogonal, this matrix is diagonal in $\delta^K_{mm'}$. Thus the orthogonalization will be constructed for every helicity mode independently 
\begin{equation}
     \mathbb{C}^{m}_{ij} (k) = \frac{\langle \mathcal{O}_i (\bk ) \mathcal{O}_j^* (\bk)  \rangle }{\sqrt{\langle \mathcal{O}_i(\bk) \mathcal{O}_i^*(\bk) \rangle \langle \mathcal{O}_j (\bk) \mathcal{O}_j^{*} (\bk)\rangle}},
\end{equation}
where we've suppressed the helicity dependence of the operators since there's no additional complexity. The orthogonalization procedure then follows in exact analogy to the scalar case. Due to the field-level degeneracies discussed in \Cref{sec:galileons}, for some of the lowest $k$-bins the Cholesky decomposition may fail. Thus, we implement a ridge regularization of the correlation matrix $\mathbb{C}$, by applying the Cholesky decomposition to $\mathbb{C}' = \mathbb{C} + \lambda \mathbb{I}$, and in practice find that $\lambda=10^{-6}$ is sufficient for all $k$-bins which are affected. We have verified numerically that the resulting orthogonalized helicity fields are not only orthogonal in $m$ but also in operators $i,j$ to within numerical precision. 
\section{Validation of numerical Lagrangian basis and helicity decomposition}
\label{app:lagbasis}
In this appendix we validate the numerical implementation of the shape bias basis, as well as the implementation of the helicity decomposition. Our numerical implementation on the lattice is based on the usual Lagrangian (non-galileon) basis, which we then rotate to the galileon representation before measuring field-level transfer functions. \par 
In the absence of any displacements, the shape-shape power spectra of the Lagrangian bias basis assumes simple forms. The only non-zero spectra are the linear $(s_{ab}, s_{cd})$ spectrum, the cross-spectra of quadratic operators and two $P_{13}$ terms between $s_{ab}$ and cubic operators involving $t_{ab}$ or $L_{3{ab}}$. Furthermore, any spectrum involving either $s_{ab}$ or the `velocity shear' $t_{ab}$ will only possess an $m=0$ helicity spectrum. \par 
Given a realization of the linear density field $\delta_{\rm IC}$, we compute the full 10-operator basis. To validate this implementation, we check against known analytic one-loop results -- quadratic operator spectra, as well as the `31' one-loop contribution -- worked out in \cite{chenkokron2024}. For each helicity $m$, we compute all of the cross-spectra which are even in powers of the density field. Analytic calculations of these Lagrangian correlators are carried out using the FFTLog algorithm, with the same scheme as implemented in \texttt{spinosaurus}. Strictly for the purpose of validation, we apply a spherical sharp-k filter for all  $|\bk| \leq  k_{\rm UV} = 1 \ihmpc$ in the linear initial conditions. To match this cut in the analytic prediction, we use as an input linear power spectrum $P(k) \to {\rm erfc} \left ( \frac{k - k_{\rm UV}}{\Delta k} \right ) P(k)/2$, with $\Delta k = 0.05 \ihmpc$, which smoothly implements this cutoff. \par
\begin{figure}
    \centering
    \includegraphics[width=\linewidth]{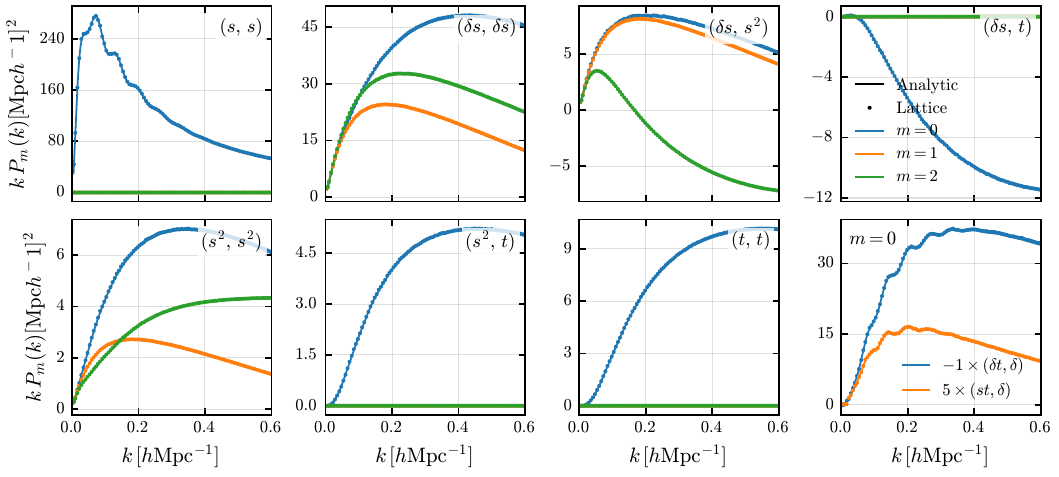}
    \caption{Comparison between analytic and lattice realizations of Lagrangian tensor correlators. The color indicates the helicity of the measurement, solid lines are the analytic prediction and the points are the measurements from ensemble of \quijote initial conditions. The final panel shows two cubic correlators cross-correlated with the linear density field, so there is only an $m=0$ component. }
    \label{fig:lagcheck}
\end{figure}

 
The comparison between the lattice correlators and analytic predictions is shown in Fig.~\ref{fig:lagcheck}. We observe very close agreement for all helicities between Lagrangian correlators in the analytic and lattice realizations. Other checks such as orthogonality of fields for different $m$ have also been verified numerically. We have also checked that for the remaining `Zeldovich' cubic operators, after proper renormalization, their cross-spectrum with $\delta$ is consistent with zero. Therefore, both the field-level Lagrangian operator basis and the spherical tensor basis are properly implemented.  \par 
For the de-facto field-level calculations in this paper we do not implement this spherical sharp-k cutoff, opting instead to use the natural grid cell of the \quijote initial conditions as a smoothing scale. To control for aliasing for cubic operators, we follow the prescription used in the \texttt{GridSPT}~\cite{Taruya_2018} code. A sharp-k cutoff at $k_{\rm UV} = 1 \ihmpc$ in the linear field is applied before the real-space product with the quadratic field is computed. 
\section{Reconstructing \flamingo inertia tensors}
\label{app:tensorrecon}
The publicly available inertia tensors for \flamingo are incomplete. The public documentation states ``When computing the inertia tensors, a filter was applied based on subhalo bound mass ($2\times10^{11} M_\odot$), instead of the normal particle number threshold. This means the tensors are not available for low mass subhalos.''\footnote{\url{https://dataweb.cosma.dur.ac.uk:8443/flamingo/support/support_issues.html\#missing-inertia-tensors}}. For the M9 resolution, this results in 98.9\% of subhalos with more than 20 star particles having no recorded inertia tensor (specifically, \texttt{StellarInertiaTensorNoniterative}). Selecting on subhalos with more than 300 star particles still results in 67.1\% of subhalos not having a recorded inertia tensor. This poses a challenge in studying the stellar shapes of lower-mass halos, such as those that host ELGs or spiral galaxies. \par 
To this end, we have re-implemented the halo shape calculation functionality of SOAP, the code used to compute shapes of halos in \flamingo , in order to recalculate missing shapes. Specifically, we aim to recompute the halo shapes associated with the \texttt{BoundSubhalo} attribute in SOAP catalogs, as opposed to shapes measured from various spherical aperture definitions. This is done by selecting halos with more than 20 star particles and recording their \texttt{HaloCatalogueIndex}. Given \flamingo particle membership files, for dark matter, baryon and star particles, we sub-select required particle IDs by selecting for particles whose \texttt{GroupNr\_bound} are found in the set of halo catalog indices. From these matched particle catalogs we record the required variables to compute both $\kappa_{\rm corot}$ and the resulting shapes. These include coordinates, masses (stellar and gas), velocities, and star formation rates. \par
\begin{figure}
    \centering
    \includegraphics[width=\linewidth]{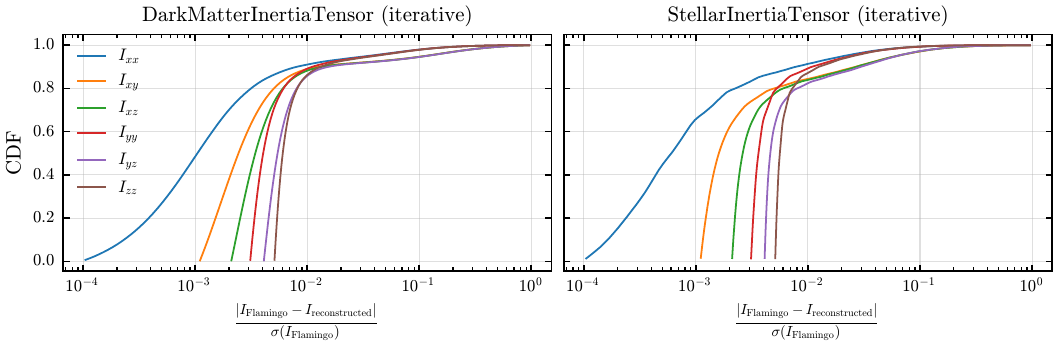}
    \caption{Cumulative distribution functions for components of the difference between public \flamingo and our reconstructed shape tensors, for iterative dark matter and stellar halo shapes.}
    \label{fig:tensorcdf}
\end{figure}

 
From these sub-selected particle catalogs, we proceed to compute inertia tensors through SOAP's built-in inertia tensor function\footnote{\url{https://github.com/SWIFTSIM/SOAP/blob/master/SOAP/property_calculation/inertia_tensors.py}}. In Fig.~\ref{fig:tensorcdf} we show the cumulative distribution functions of the absolute difference between the SOAP-stored inertia tensors and our independent reconstruction (for halos that have a non-zero tensor), normalized by the dispersion of each component in the original catalogs. We only show iterative, non-reduced tensors but the results are consistent across all definitions we have adopted. Our reconstruction of the original tensors are highly accurate, with $\sim 90\%$ of halos being reconstructed to better than $.1\sigma$. \par 
Part of the (very small) disagreements between halo shapes can be attributed to the compression algorithm applied to public SOAP catalogs from \flamingo. Compressing our reconstructed tensors with the same \texttt{FMantissa9} filter resolves these discrepancies, but we opt to use the uncompressed reconstructed shape tensors in this work.
\section{Analytic transfer function and parameter covariances}
\label{app:tferror}
\begin{figure}
    \centering
    \includegraphics[width=1\linewidth]{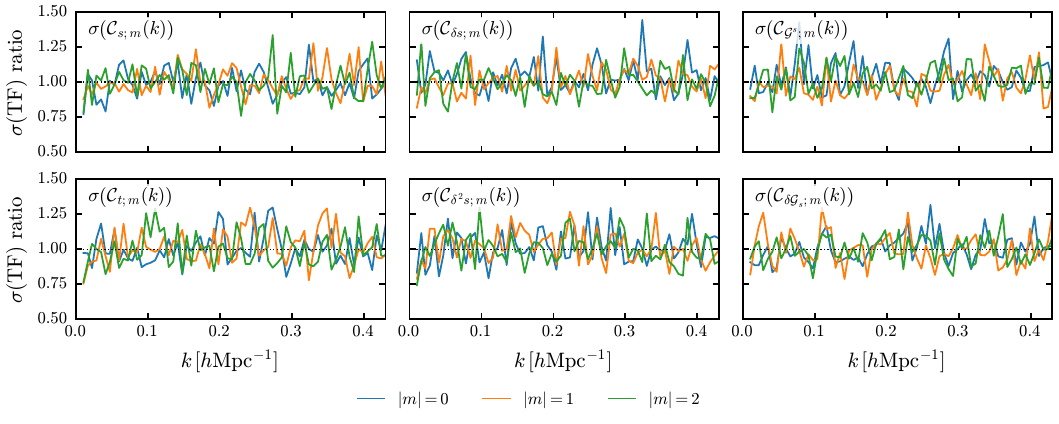}
    \caption{Ratio between the analytic (\Cref{eqn:tferr}) and empirical estimates for standard deviations of transfer functions. Each panel denotes a separate transfer function, while the colors denote the three different helicities. The empirical estimates are calculated using $N=50$ simulations.}
    \label{fig:tferror}
\end{figure}
In the limit of a Gaussian-distributed stochasticity field $\epsilon_m$ with variance given by the shape noise $\sigma_\gamma^2/\bar{n}$, the error of the transfer function $\mathcal{C}_{i,m}(k)$ is given by \Cref{eqn:tferr}. To see this, we follow the derivation of uncertainties in parameter covariances from Ref.~\cite{Kokron:2021faa}. Start from the estimator for the transfer function (we have suppressed the helicity index here, since all expressions are diagonal in $m$)
\begin{equation}
    \hat{\mathcal{C}}_{i}(k) = \frac{\sum_{\bk} \gamma (\bk) \mathcal{O}_i^{\perp;*} (\bk)}{\sum_\bk |\mathcal{O}_i^\perp|^2(\bk)},
\end{equation}
where the $\bk$ sum is over wavenumbers that fall in that bin. Substituting the field-level expression for $\gamma(\bk)$ gives 
\begin{equation}
    \hat{\mathcal{C}}_{i}(k) = \frac{\sum_{\bk,j} \left [\mathcal{C}_{j}(k) \mathcal{O}_j^\perp (\bk) + \epsilon(\bk) \right] \mathcal{O}_i^{\perp,*}(\bk)}{\sum_\bk |\mathcal{O}_i^\perp|^2(\bk)}.
\end{equation}
Under the assumption that the transfer function does not vary drastically within the bin width (and since the transfer functions are generically smooth, this is reasonable), and that numerical violations of orthogonality are small (which they are to machine precision) the first term is simply $\mathcal{C}_{i}(k)$, i.e. the estimator is unbiased. The variance of the transfer function is thus 
\begin{equation}
    \sigma^2 (\hat{\mathcal{C}}_{i})(k) = \frac{\sum_{\bk, \bk'}\mathcal{O}_i^\perp (\bk)\mathcal{O}_i^\perp (\bk') \langle \epsilon(\bk) \epsilon(\bk') \rangle}{ \left (\sum_\bk |\mathcal{O}_i^\perp|^2(\bk) \right)^2}.
\end{equation}
Assuming a white error power spectrum, and using the relation $\sum_\bk |\mathcal{O}_i^\perp|^2(\bk) = N_{\rm modes}(k) \langle |O_i^\perp |^2 \rangle (k)$ then gives \Cref{eqn:tferr}. The factor $a_m$ comes from averaging over the $\pm m$ helicity realizations. In Fig.~\ref{fig:tferror} we compare \Cref{eqn:tferr} to empirical estimates of the scatter in transfer functions. The empirical measurements use transfer functions from $N_{\rm sims}=50$ \quijote realizations, for halos in $M_h \in [10^{12.5}, 10^{13.0}]\,M_\odot h^{-1}$. The auto-spectra of orthonormal operators is calculated by averaging across the same number of simulations, although we note they can in principle be calculated analytically. For all helicities and operators we find a good agreement between the analytic expression for scatter in the transfer functions and the empirical estimate. There are deviations on the order of $\sim 20\%$, but no systematic trends are visible.   \par
\begin{figure}
    \centering
    \includegraphics[width=1\linewidth]{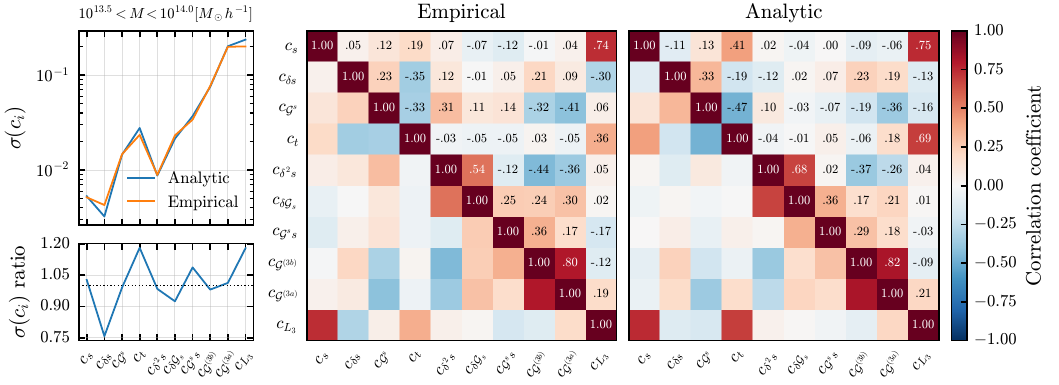}
    \caption{Comparison between empirical measurements of field-level IA parameters and the analytic expression given by generalized least-squares, for one of the \quijote + rockstar halo samples.}
    \label{fig:paramcov}
\end{figure}

 
Formulating the bias parameter estimation problem in the language of generalized least-squares also allows us to estimate the \emph{covariances} of the IA shape biases analytically. For a generalized least squares problem, the parameter Fisher matrix is given 
\begin{equation}
    \boldsymbol{F} = \boldsymbol{\mathcal{M}}^T \boldsymbol{C}^{-1} \boldsymbol{\mathcal{M}},
\end{equation} where $\boldsymbol{\mathcal{M}}$ is the full design matrix of the least-squares problem built from the mode-mixing matrix \Cref{eqn:designmatrix} and $\boldsymbol{C}$ is the covariance matrix of the transfer functions (empirical or analytic). In \Cref{fig:paramcov} we show a comparison between the analytic, GLS-based parameter covariance and the empirical covariance from studying $N_{\rm sims}=50$ \quijote-based realizations, for the mass bin of $M_h \in [10^{13.5}, 10^{14.0}]\,M_\odot h^{-1}$. We see that the \emph{amplitude} of parameter error bars are in good agreement for this mass bin -- the analytic estimator is within 20-25\% of the empirical result for all IA parameters. In the middle and right panels of \Cref{fig:paramcov} we show the correlation matrices between IA parameters for both estimates. While there are numerical differences, the qualitative agreement between both parameter covariances is strong. We can also see that the cubic parameters are relatively well-measured, although certain sets of cubic parameters are block-degenerate ($\{c_{\delta^2 s},c_{\delta \mathcal{G}_s}\}$), and the dynamical cubic galileons, which can be understood from the arguments in \S~\ref{sec:cubicparams} about the block sub-spaces which are well-constrained in the fit. \par 
The results of this appendix justify the use of analytic expressions for the errors of the transfer function, as well as the analytic expression for the covariance of our field-level IA parameters, which will be used extensively when applied to Flamingo due to the lack of independent realizations. 
\section{Halo transfer functions}
\label{app:transfers}
\begin{figure}[ht!]
    \centering
    \includegraphics[width=\linewidth]{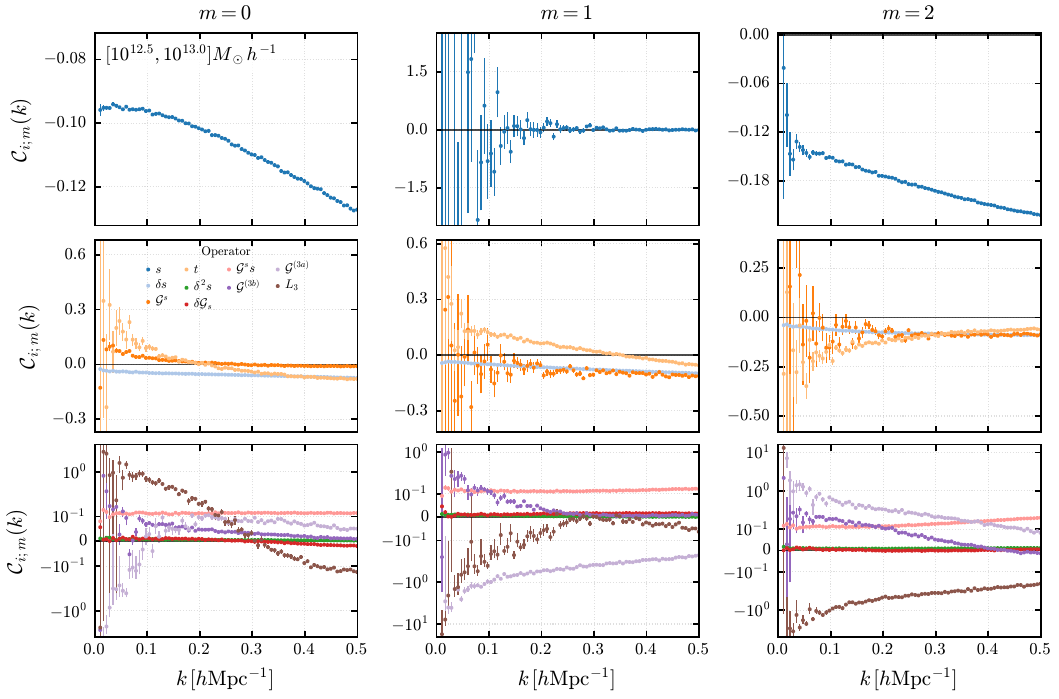}
    \caption{The same set of averaged transfer functions as shown in \Cref{fig:transfer_mbin3}, but for the $M\in [10^{12.5}, 10^{13.0}]M_\odot h^{-1}$ mass bin. }
    \label{fig:transfer_mbin1}
\end{figure}

\begin{figure}[ht!]
    \centering
    \includegraphics[width=\linewidth]{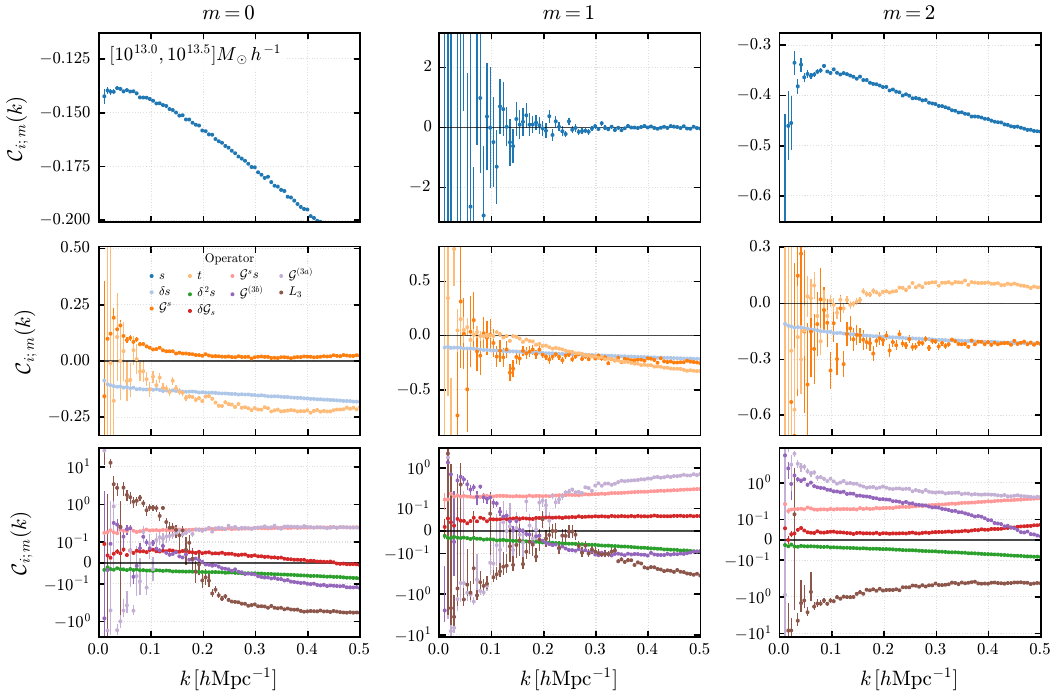}
    \caption{The same set of averaged transfer functions as shown in \Cref{fig:transfer_mbin3}, but for the $M\in [10^{13.0}, 10^{13.5}]M_\odot h^{-1}$ mass bin. }
    \label{fig:transfer_mbin2}
\end{figure}

\begin{figure}[ht!]
    \centering
    \includegraphics[width=\linewidth]{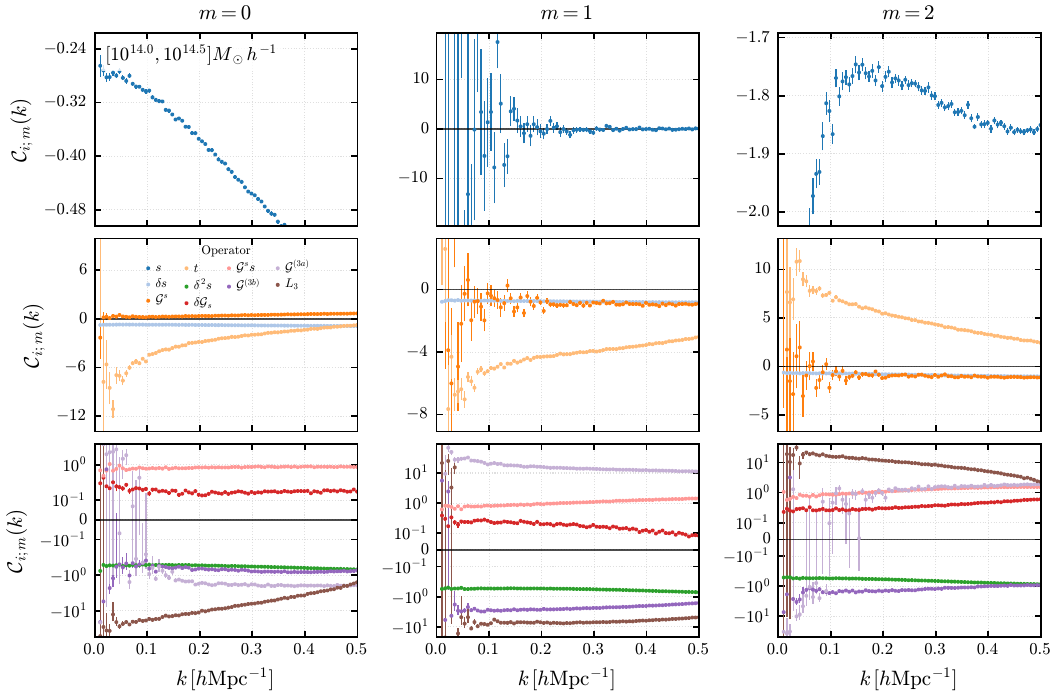}
    \caption{The same set of averaged transfer functions as shown in \Cref{fig:transfer_mbin3}, but for the $M\in [10^{14.0}, 10^{14.5}]M_\odot h^{-1}$ mass bin. }
    \label{fig:transfer_mbin4}
\end{figure}

\begin{figure}[ht!]
    \centering
    \includegraphics[width=\linewidth]{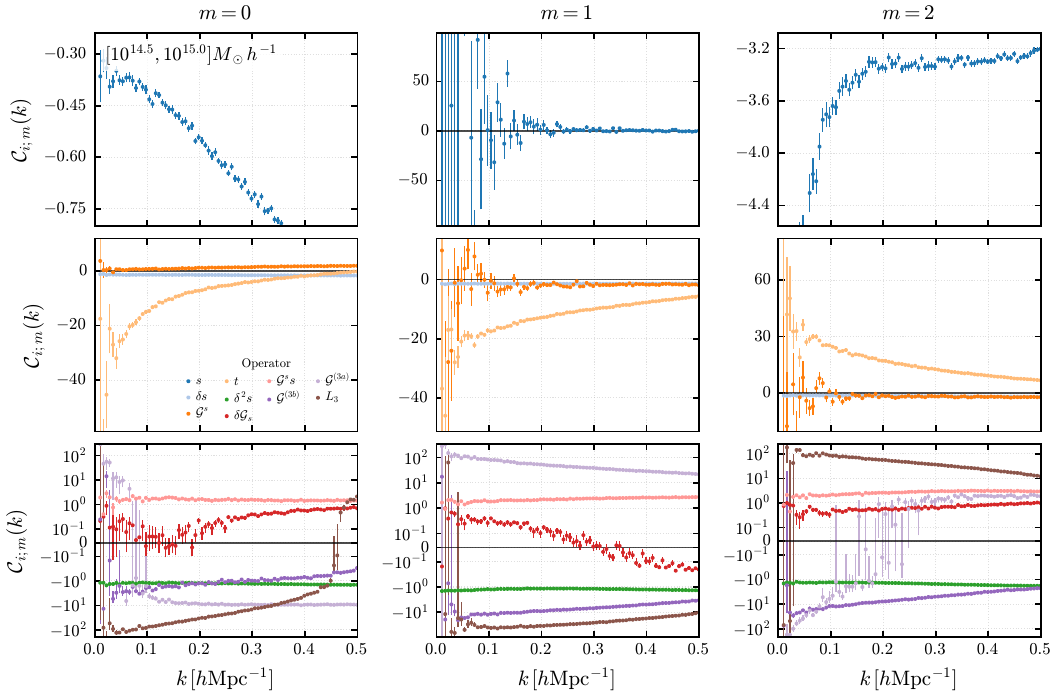}
    \caption{The same set of averaged transfer functions as shown in \Cref{fig:transfer_mbin3}, but for the $M\in [10^{14.5}, 10^{15.0}]M_\odot h^{-1}$ mass bin. }
    \label{fig:transfer_mbin5}
\end{figure}

In this appendix we show the remaining, 10-operator transfer functions for all halo mass bins. Figures~\ref{fig:transfer_mbin1}, \ref{fig:transfer_mbin2}, \ref{fig:transfer_mbin4} and Fig.~\ref{fig:transfer_mbin5} show transfer functions for the $[10^{12.5}, 10^{13.0}], [10^{13.0},10^{13.5}], [10^{14.0}, 10^{14.5}]$ and $[10^{14.5}, 10^{15.0}]$ mass bins, respectively. The format is the same as \Cref{fig:transfer_mbin3}. 
\section{Converting $c_s$ to $A_{1}$, and beyond}
\label{app:conversion}
The intrinsic alignment literature conventionally reports as the linear amplitude a coefficient `$A_{1}$' which is proportional to $c_s$ but with certain dimensionful factors, as the intrinsic ellipticity of galaxies is often normalized by a density $\bar{C}_1 = 5 \times 10^{-14} M_\odot^{-1} h^{-2} {\rm Mpc}^3$. This is the normalization used by past works studying alignments in hydrodynamic simulations, including \flamingo~\cite{herle2026intrinsicalignmentsflamingosimulations}, as well as the weak lensing community. The linear alignment amplitude $A_1$ is defined as (the factor of 2 coming from 3D shape v.s. E-mode scaling for the contribution~\cite{derose2026steelingweaklensingsource})
\begin{equation}
    c_s = - 2 A_1 \bar{C}_1 \rho_{\rm crit} \Omega_m D(z)^{-1} . 
\end{equation}
Since $\rho_{\rm crit} = 2.78 \times 10^{11} h^2 M_\odot {\rm Mpc}^{-3}$, we can take the product $\bar{C}_1 \rho_{\rm crit} \approx 0.014$. Assuming $\Omega_m = 0.306$ and $A_1 = 5$ we have 
\begin{equation}
    c_s \approx - \frac{4.3 \times 10^{-2}}{D(z)} \left ( \frac{\Omega_m}{0.306} \right )\left ( \frac{A_1}{5} \right ).
\end{equation}
As mentioned in the main text, we take $A_1 \approx 5 \pm 0.2$ for the \flamingo LRG sample at z=0 \cite{herle2026intrinsicalignmentsflamingosimulations}, as approximate uncertainties from fitting to $P_{gg}, P_{gI}$ with a $(2.8 {\rm Gpc})^3$ box. In \cite{herle2026assemblybiasredshiftevolution} they also provide a fitting function for the redshift evolution of linear alignments as 
\begin{equation}
    A_1(M,z) = \alpha_M (z) \left ( \frac{M}{10^{13.5} M_\odot } \right)^\beta_M(z),
\end{equation}
with $\alpha_M (z) = -2.18 z^2 + 12.69 z + 6.64,\,\beta_M(z) = 0.26 z^2 - 0.5z + 0.88$. Matching their value of $A_1 (M^*, z=0) = 5$ gives $M^* = 2.3 \times 10^{13}$. Fixing this mass and scaling the result in redshift gives $A_1 (M^*, z=0.5) = 9.9$. For this cosmology, $D(z=0.5) = 0.77$, and so we infer $100 c_s \approx -11.05 \pm 0.44$ from the 2-point, large-volume \flamingo analysis. In the main text we compare this constraint to our measurements. \par 
To convert the measured $c_{\mathcal{G}_s}, c_{\delta s}$ coefficients to the usual TATT amplitudes some additional care is required. The goal is to re-express the shifted operator field-level expansion in terms of quadratic Eulerian fields. Shifting a quadratic field introduces terms at cubic order, so we only need to consider the linear shift. To quadratic order, the shifted linear tidal field is written as
\begin{equation}
\label{eqn:linshift}
    \tilde{s}_{ab} = s_{ab} + \delta s_{ab} - \bPsi^{(1)} \cdot \nabla s_{ab} . 
\end{equation}
Conversion to Eulerian amplitudes follows through writing $\bPsi^{(1)} \cdot \nabla s_{ab}$ in terms of the EPT second-order tidal field 
\begin{equation*}
    s^{(2)}_{ab} = \mathcal{D}_{ab} \delta^{(2)} = s^2_{ab} + \frac{2}{3} \delta s_{ab} - \frac{5}{2} t_{ab} - \bPsi^{(1)} \cdot \nabla s_{ab},
\end{equation*}
which was originally derived in \cite{Schmitz_2018}. Collecting terms and equating the shifted-operator expansion to the Eulerian expansion at second order gives 
\begin{equation*}
    \tilde{s}_{ab} = s_{ab}^{(E)}  - \mathcal{G}^s_{ab} + \frac52 t_{ab}.
\end{equation*}
We can read off that the Lagrangian coefficients of the shape bias expansion are re-mapped to 
\begin{align*}
    &c_s^E = c_s^L, \quad  c_{\delta s}^E = c_{\delta s}^L\,, \\
    &c_{\mathcal{G}^s}^E = c_{\mathcal{G}^s}^L - c_s^L, \quad c_{t}^E = c_{t}^L +\frac52  c_s^L,
\end{align*}
but note the remapping is different for the non-Galileon basis. The usual Eulerian TATT amplitudes $A_{1\delta}, A_2$, are defined as~\cite{Blazek_2019}
\begin{align}
&C_{\delta s} = - A_{1\delta} \bar{C}_1\rho_{\rm crit} \Omega_m D(z)^{-1} \\
&C_{2} = A_2 \left ( \frac{5 \bar{C}_1 \rho_{\rm crit}}{\Omega_{m, {\rm fid}}} \right ) \Omega_m^2 D(z)^{-2}
\end{align}
These Eulerian `$C$' coefficients are defined with respect to the projected fields, so there is a similar $C_i = c_i/2$ conversion. Accounting for the basis rotation, we convert EFT shape biases to the TATT amplitudes and find
\begin{align}
    A_{1\delta} &\approx - \frac{c_{\delta s}^{(E)}D(z)}{8.5 \times 10^{-3}} \left ( \frac{0.306}{\Omega_m} \right ) \\ 
    &\approx - D(z) \left (\frac{c_{\delta s}^{(\mathcal{G})} - c_{\mathcal{G}^s}/3 + c_s/3}{8.5 \times10^{-3} } \right)  \left ( \frac{0.306}{\Omega_m} \right ),
\end{align}
where $c^{(E)}_{\delta s}$ is the Eulerian 3D $\delta s$ coefficient in the non-Galileon basis. For $A_2$ we find 
\begin{align}
    A_2 \approx D(z)^2 \left ( \frac{c_{\mathcal{G}^s} - c_s}{4.25 \times10^{-2}}\right ) \left ( \frac{0.306}{\Omega_m} \right ).
\end{align}

\bibliography{main}

@PREAMBLE{
 "\providecommand{\noopsort}[1]{}" 
 # "\providecommand{\singleletter}[1]{#1}%" 
}

@article{Ivanov:2021kcd,
    author = "Ivanov, Mikhail M. and Philcox, Oliver H. E. and Nishimichi, Takahiro and Simonovi{\'c}, Marko and Takada, Masahiro and Zaldarriaga, Matias",
    title = "{Precision analysis of the redshift-space galaxy bispectrum}",
    eprint = "2110.10161",
    archivePrefix = "arXiv",
    primaryClass = "astro-ph.CO",
    reportNumber = "YITP-21-120, CERN-TH-2021-155",
    doi = "10.1103/PhysRevD.105.063512",
    journal = "Phys. Rev. D",
    volume = "105",
    number = "6",
    pages = "063512",
    year = "2022"
}

@article{Bakx:2025pop,
    author = "Bakx, Thomas and Ivanov, Mikhail M. and Philcox, Oliver H. E. and Vlah, Zvonimir",
    title = "{One-loop galaxy bispectrum: Consistent theory, efficient analysis with COBRA, and implications for cosmological parameters}",
    eprint = "2507.22110",
    archivePrefix = "arXiv",
    primaryClass = "astro-ph.CO",
    reportNumber = "MIT-CTP/5891, RBI-ThPhys-2025-23",
    doi = "10.1103/zhcf-7cgh",
    journal = "Phys. Rev. D",
    volume = "114",
    number = "6",
    pages = "063532",
    year = "2026"
}

@article{Ivanov:2024hgq,
    author = "Ivanov, Mikhail M. and Cuesta-Lazaro, Carolina and Mishra-Sharma, Siddharth and Obuljen, Andrej and Toomey, Michael W.",
    title = "{Full-shape analysis with simulation-based priors: Constraints on single field inflation from BOSS}",
    eprint = "2402.13310",
    archivePrefix = "arXiv",
    primaryClass = "astro-ph.CO",
    reportNumber = "MIT-CTP/5684",
    doi = "10.1103/PhysRevD.110.063538",
    journal = "Phys. Rev. D",
    volume = "110",
    number = "6",
    pages = "063538",
    year = "2024"
}

@article{Ivanov:2024xgb,
    author = "Ivanov, Mikhail M. and Obuljen, Andrej and Cuesta-Lazaro, Carolina and Toomey, Michael W.",
    title = "{Full-shape analysis with simulation-based priors: Cosmological parameters and the structure growth anomaly}",
    eprint = "2409.10609",
    archivePrefix = "arXiv",
    primaryClass = "astro-ph.CO",
    reportNumber = "MIT-CTP/5762",
    doi = "10.1103/PhysRevD.111.063548",
    journal = "Phys. Rev. D",
    volume = "111",
    number = "6",
    pages = "063548",
    year = "2025"
}

@article{Taruya_2021,
   title={Lagrangian approach to super-sample effects on biased tracers at field level: galaxy density fields and intrinsic alignments},
   volume={2021},
   ISSN={1475-7516},
   url={http://dx.doi.org/10.1088/1475-7516/2021/11/061},
   DOI={10.1088/1475-7516/2021/11/061},
   number={11},
   journal={Journal of Cosmology and Astroparticle Physics},
   publisher={IOP Publishing},
   author={Taruya, Atsushi and Akitsu, Kazuyuki},
   year={2021},
   month=nov, pages={061} }

@article{kurita2025parityviolationgalaxyshapes,
    title = {Parity Violation in Galaxy Shapes: Primordial Non-Gaussianity},
    author = {Toshiki Kurita and Drew Jamieson and Eiichiro Komatsu and Fabian Schmidt},
    eprint = {2509.08787},
    archivePrefix = {arXiv},
    primaryClass = {astro-ph.CO},
    journal = "Phys. Rev. D",
    volume = "113",
    number = "6",
    pages = "063557",
    year = "2026",
    doi = "10.1103/fxh6-hpmk"
}

@article{Tokuue_2026,
   title={MPI-Rockstar: a Hybrid MPI and OpenMP Parallel Implementation of the Rockstar Halo finder},
   volume={11},
   ISSN={2475-9066},
   url={http://dx.doi.org/10.21105/joss.08295},
   DOI={10.21105/joss.08295},
   number={120},
   journal={Journal of Open Source Software},
   publisher={The Open Journal},
   author={Tokuue, Tomoyuki and Ishiyama, Tomoaki and Osato, Ken and Tanaka, Satoshi and Behroozi, Peter},
   year={2026},
   month=Apr, pages={8295} }

@article{chenkokron2024,
    title = {A Lagrangian theory for galaxy shape statistics},
    author = {Shi-Fan Chen and Nickolas Kokron},
    eprint = {2309.16761},
    archivePrefix = {arXiv},
    primaryClass = {astro-ph.CO},
    journal = "JCAP",
    volume = "01",
    pages = "027",
    year = "2024",
    doi = "10.1088/1475-7516/2024/01/027"
}

@article{Cabass_2020,
   title={The likelihood for LSS: stochasticity of bias coefficients at all orders},
   volume={2020},
   ISSN={1475-7516},
   url={http://dx.doi.org/10.1088/1475-7516/2020/07/051},
   DOI={10.1088/1475-7516/2020/07/051},
   number={07},
   journal={Journal of Cosmology and Astroparticle Physics},
   publisher={IOP Publishing},
   author={Cabass, Giovanni and Schmidt, Fabian},
   year={2020},
   month={Jul},
   pages={051–051}
}

@article{Schmidt_2019,
   title={A rigorous EFT-based forward model for large-scale structure},
   volume={2019},
   ISSN={1475-7516},
   url={http://dx.doi.org/10.1088/1475-7516/2019/01/042},
   DOI={10.1088/1475-7516/2019/01/042},
   number={01},
   journal={Journal of Cosmology and Astroparticle Physics},
   publisher={IOP Publishing},
   author={Schmidt, Fabian and Elsner, Franz and Jasche, Jens and Nguyen, Nhat Minh and Lavaux, Guilhem},
   year={2019},
   month={Jan},
   pages={042–042}
}

@article{Kokron_2021,
   title={The cosmology dependence of galaxy clustering and lensing from a hybrid N-body–perturbation theory model},
   volume={505},
   ISSN={1365-2966},
   url={http://dx.doi.org/10.1093/mnras/stab1358},
   DOI={10.1093/mnras/stab1358},
   number={1},
   journal={Monthly Notices of the Royal Astronomical Society},
   publisher={Oxford University Press (OUP)},
   author={Kokron, Nickolas and DeRose, Joseph and Chen, Shi-Fan and White, Martin and Wechsler, Risa H},
   year={2021},
   month={May},
   pages={1422–1440}
}

@article{eft_like_2020,
   title={Unbiased cosmology inference from biased tracers using the EFT likelihood},
   volume={2020},
   ISSN={1475-7516},
   url={http://dx.doi.org/10.1088/1475-7516/2020/11/008},
   DOI={10.1088/1475-7516/2020/11/008},
   number={11},
   journal={Journal of Cosmology and Astroparticle Physics},
   publisher={IOP Publishing},
   author={Schmidt, Fabian and Cabass, Giovanni and Jasche, Jens and Lavaux, Guilhem},
   year={2020},
   month={Nov},
   pages={008–008}
}

@article{schmidt2021,
   title={An n-th order Lagrangian forward model for large-scale structure},
   volume={2021},
   ISSN={1475-7516},
   url={http://dx.doi.org/10.1088/1475-7516/2021/04/033},
   DOI={10.1088/1475-7516/2021/04/033},
   number={04},
   journal={Journal of Cosmology and Astroparticle Physics},
   publisher={IOP Publishing},
   author={Schmidt, Fabian},
   year={2021},
   month={Apr},
   pages={033} }

@article{Carlson_2012,
   title={Convolution Lagrangian perturbation theory for biased tracers},
   volume={429},
   ISSN={1365-2966},
   url={http://dx.doi.org/10.1093/mnras/sts457},
   DOI={10.1093/mnras/sts457},
   number={2},
   journal={Monthly Notices of the Royal Astronomical Society},
   publisher={Oxford University Press (OUP)},
   author={Carlson, Jordan and Reid, Beth and White, Martin},
   year={2012},
   month={Dec},
   pages={1674–1685}
}

@article{Yuan_2022,
   title={Illustrating galaxy–halo connection in the DESI era with <scp>illustrisTNG</scp>},
   volume={512},
   ISSN={1365-2966},
   url={http://dx.doi.org/10.1093/mnras/stac830},
   DOI={10.1093/mnras/stac830},
   number={4},
   journal={Monthly Notices of the Royal Astronomical Society},
   publisher={Oxford University Press (OUP)},
   author={Yuan, Sihan and Hadzhiyska, Boryana and Bose, Sownak and Eisenstein, Daniel J},
   year={2022},
   month=Mar, pages={5793–5811} }

@article{bakx2026twolooprenormalizationrunninggalaxy,
    title = {Two-loop renormalization and running of galaxy bias},
    author = {Thomas Bakx and Mathias Garny and Henrique Rubira and Zvonimir Vlah},
    eprint = {2507.13905},
    archivePrefix = {arXiv},
    primaryClass = {astro-ph.CO},
    journal = "JCAP",
    volume = "03",
    pages = "056",
    year = "2026",
    doi = "10.1088/1475-7516/2026/03/056"
}

@article{gomes2026tidalalignmenttidaltorquing,
      title={Tidal alignment and tidal torquing modeling for the cosmic shear three-point correlation function and mass aperture skewness}, 
      author={Rafael C. H. Gomes and Kyle Miller and Sunao Sugiyama and Jonathan Blazek and Thomas Bakx and Bhuvnesh Jain},
      year={2026},
      eprint={2601.09133},
      archivePrefix={arXiv},
      primaryClass={astro-ph.CO},
      url={https://arxiv.org/abs/2601.09133}, 
}

@article{Delgado_2023,
   title={The MillenniumTNG project: intrinsic alignments of galaxies and haloes},
   volume={523},
   ISSN={1365-2966},
   url={http://dx.doi.org/10.1093/mnras/stad1781},
   DOI={10.1093/mnras/stad1781},
   number={4},
   journal={Monthly Notices of the Royal Astronomical Society},
   publisher={Oxford University Press (OUP)},
   author={Delgado, Ana Maria and Hadzhiyska, Boryana and Bose, Sownak and Springel, Volker and Hernquist, Lars and Barrera, Monica and Pakmor, Rüdiger and Ferlito, Fulvio and Kannan, Rahul and Hernández-Aguayo, César and White, Simon D M and Frenk, Carlos},
   year={2023},
   month=June, pages={5899–5914} }

@article{Vlah_2020,
   title={An EFT description of galaxy intrinsic alignments},
   volume={2020},
   ISSN={1475-7516},
   url={http://dx.doi.org/10.1088/1475-7516/2020/01/025},
   DOI={10.1088/1475-7516/2020/01/025},
   number={01},
   journal={Journal of Cosmology and Astroparticle Physics},
   publisher={IOP Publishing},
   author={Vlah, Zvonimir and Chisari, Nora Elisa and Schmidt, Fabian},
   year={2020},
   month=Jan, pages={025–025} }

@article{van_Heukelum_2026,
   title={Intrinsic alignment of disks and ellipticals across hydrodynamical simulations},
   volume={9},
   ISSN={2565-6120},
   url={http://dx.doi.org/10.33232/001c.158824},
   DOI={10.33232/001c.158824},
   journal={The Open Journal of Astrophysics},
   publisher={Maynooth University},
   author={van Heukelum, M. L. and Chisari, N. E.},
   year={2026},
   month=Mar }

@article{herle2026intrinsicalignmentsflamingosimulations,
    title = {Intrinsic alignments in the FLAMINGO simulations with two-point statistics},
    author = {A. Herle and N. E. Chisari and H. Hoekstra and D. Navarro-Gironés and M. Schaller and J. Schaye},
    eprint = {2601.15851},
    archivePrefix = {arXiv},
    primaryClass = {astro-ph.CO},
    journal = "Astron. Astrophys.",
    volume = "711",
    pages = "A235",
    year = "2026",
    doi = "10.1051/0004-6361/202659078"
}

@article{vedder2026threepointintrinsicalignmentsgalaxies,
    title = {Three-point intrinsic alignments of galaxies and haloes in the FLAMINGO simulations},
    author = {Casper Vedder and Thomas Bakx and Nora Elisa Chisari and Henk Hoekstra and Matthieu Schaller},
    eprint = {2601.17914},
    archivePrefix = {arXiv},
    primaryClass = {astro-ph.CO},
    journal = "Mon. Not. Roy. Astron. Soc.",
    volume = "550",
    number = "2",
    pages = "stag1008",
    year = "2026",
    doi = "10.1093/mnras/stag1008"
}

@article{herle2026assemblybiasredshiftevolution,
      title={Assembly bias and the redshift evolution of intrinsic alignments for LRGs}, 
      author={A. Herle and N. E. Chisari and H. Hoekstra and D. Neumann},
      year={2026},
      eprint={2607.00785},
      archivePrefix={arXiv},
      primaryClass={astro-ph.CO},
      url={https://arxiv.org/abs/2607.00785}, 
}

@article{Allgood:2005eu,
    author = "Allgood, Brandon and Flores, Ricardo A. and Primack, Joel R. and Kravtsov, Andrey V. and Wechsler, Risa H. and Faltenbacher, Andreas and Bullock, James S.",
    title = "{The shape of dark matter halos: dependence on mass, redshift, radius, and formation}",
    eprint = "astro-ph/0508497",
    archivePrefix = "arXiv",
    doi = "10.1111/j.1365-2966.2006.10094.x",
    journal = "Mon. Not. Roy. Astron. Soc.",
    volume = "367",
    pages = "1781--1796",
    year = "2006"
}

@article{Akitsu_2023,
   title={Quadratic shape biases in three-dimensional halo intrinsic alignments},
   volume={2023},
   ISSN={1475-7516},
   url={http://dx.doi.org/10.1088/1475-7516/2023/08/068},
   DOI={10.1088/1475-7516/2023/08/068},
   number={08},
   journal={Journal of Cosmology and Astroparticle Physics},
   publisher={IOP Publishing},
   author={Akitsu, Kazuyuki and Li, Yin and Okumura, Teppei},
   year={2023},
   month=Aug, pages={068} }

@article{Baldauf_2013,
   title={Halo stochasticity from exclusion and nonlinear clustering},
   volume={88},
   ISSN={1550-2368},
   url={http://dx.doi.org/10.1103/PhysRevD.88.083507},
   DOI={10.1103/physrevd.88.083507},
   number={8},
   journal={Physical Review D},
   publisher={American Physical Society (APS)},
   author={Baldauf, Tobias and Seljak, Uroš and Smith, Robert E. and Hamaus, Nico and Desjacques, Vincent},
   year={2013},
   month={Oct}
}

@article{derose2026steelingweaklensingsource,
      title={Steeling Weak Lensing Source Galaxy Samples against Systematics using Wide Field Spectroscopy}, 
      author={Joseph DeRose and Noah Weaverdyck and Martin White and Shi-Fan Chen and David Schlegel and Anže Slosar},
      year={2026},
      eprint={2603.10113},
      archivePrefix={arXiv},
      primaryClass={astro-ph.CO},
      url={https://arxiv.org/abs/2603.10113}, 
}

@article{Hirata_2007,
   title={Intrinsic galaxy alignments from the 2SLAQ and SDSS surveys: luminosity and redshift scalings and implications for weak lensing surveys: Intrinsic alignments},
   volume={381},
   ISSN={1365-2966},
   url={http://dx.doi.org/10.1111/j.1365-2966.2007.12312.x},
   DOI={10.1111/j.1365-2966.2007.12312.x},
   number={3},
   journal={Monthly Notices of the Royal Astronomical Society},
   publisher={Oxford University Press (OUP)},
   author={Hirata, Christopher M. and Mandelbaum, Rachel and Ishak, Mustapha and Seljak, Uroš and Nichol, Robert and Pimbblet, Kevin A. and Ross, Nicholas P. and Wake, David},
   year={2007},
   month=Sept, pages={1197–1218} }

@article{siegel2025intrinsicalignmentdemographicsnextgeneration,
    title = {Intrinsic alignment demographics for next-generation lensing: Revealing galaxy property trends with DESI Y1 direct measurements},
    author = {J. Siegel and J. McCullough and A. Amon and C. Lamman and N. Jeffrey and B. Joachimi and H. Hoekstra and S. Heydenreich and A. J. Ross and J. Aguilar and S. Ahlen and D. Bianchi and C. Blake and D. Brooks and F. J. Castander and T. Claybaugh and A. de la Macorra and J. DeRose and P. Doel and N. Emas and S. Ferraro and A. Font-Ribera and J. E. Forero-Romero and E. Gaztañaga and S. Gontcho A Gontcho and G. Gutierrez and K. Honscheid and M. Ishak and S. Joudaki and R. Kehoe and D. Kirkby and T. Kisner and A. Krolewski and O. Lahav and A. Lambert and M. Landriau and L. Le Guillou and M. E. Levi and M. Manera and A. Meisner and R. Miquel and J. Moustakas and S. Nadathur and J. A. Newman and G. Niz and N. Palanque-Delabrouille and W. J. Percival and A. Porredon and F. Prada and I. Pérez-Ràfols and G. Rossi and E. Sanchez and C. Saulder and D. Schlegel and M. Schubnell and A. Semenaite and J. Silber and D. Sprayberry and Z. Sun and G. Tarlé and B. A. Weaver and R. Zhou and H. Zou},
    eprint = {2507.11530},
    archivePrefix = {arXiv},
    primaryClass = {astro-ph.CO},
    journal = "Mon. Not. Roy. Astron. Soc.",
    volume = "552",
    number = "2",
    pages = "stag1691",
    year = "2026",
    doi = "10.1093/mnras/stag1691"
}

@article{Qin_2022,
   title={Effective bias expansion for 21-cm cosmology in redshift space},
   volume={106},
   ISSN={2470-0029},
   url={http://dx.doi.org/10.1103/PhysRevD.106.123506},
   DOI={10.1103/physrevd.106.123506},
   number={12},
   journal={Physical Review D},
   publisher={American Physical Society (APS)},
   author={Qin, Wenzer and Schutz, Katelin and Smith, Aaron and Garaldi, Enrico and Kannan, Rahul and Slatyer, Tracy R. and Vogelsberger, Mark},
   year={2022},
   month=Dec }

@article{Hand_2018,
   title={nbodykit: An Open-source, Massively Parallel Toolkit for Large-scale Structure},
   volume={156},
   ISSN={1538-3881},
   url={http://dx.doi.org/10.3847/1538-3881/aadae0},
   DOI={10.3847/1538-3881/aadae0},
   number={4},
   journal={The Astronomical Journal},
   publisher={American Astronomical Society},
   author={Hand, Nick and Feng, Yu and Beutler, Florian and Li, Yin and Modi, Chirag and Seljak, Uroš and Slepian, Zachary},
   year={2018},
   month={Sep},
   pages={160}
}

@ARTICLE{2020NumPy-Array,
  author  = {Harris, Charles R. and Millman, K. Jarrod and
            van der Walt, Stéfan J and Gommers, Ralf and
            Virtanen, Pauli and Cournapeau, David and
            Wieser, Eric and Taylor, Julian and Berg, Sebastian and
            Smith, Nathaniel J. and Kern, Robert and Picus, Matti and
            Hoyer, Stephan and van Kerkwijk, Marten H. and
            Brett, Matthew and Haldane, Allan and
            Fernández del Río, Jaime and Wiebe, Mark and
            Peterson, Pearu and Gérard-Marchant, Pierre and
            Sheppard, Kevin and Reddy, Tyler and Weckesser, Warren and
            Abbasi, Hameer and Gohlke, Christoph and
            Oliphant, Travis E.},
  title   = {Array programming with {NumPy}},
  journal = {Nature},
  year    = {2020},
  volume  = {585},
  pages   = {357–362},
  doi     = {10.1038/s41586-020-2649-2}
}

@ARTICLE{4160265,
  author={J. D. {Hunter}},
  journal={Computing in Science   Engineering}, 
  title={Matplotlib: A 2D Graphics Environment}, 
  year={2007},
  volume={9},
  number={3},
  pages={90-95},
  doi={10.1109/MCSE.2007.55}}

@ARTICLE{2020SciPy-NMeth,
  author  = {Virtanen, Pauli and Gommers, Ralf and Oliphant, Travis E. and
            Haberland, Matt and Reddy, Tyler and Cournapeau, David and
            Burovski, Evgeni and Peterson, Pearu and Weckesser, Warren and
            Bright, Jonathan and {van der Walt}, St{\'e}fan J. and
            Brett, Matthew and Wilson, Joshua and Millman, K. Jarrod and
            Mayorov, Nikolay and Nelson, Andrew R. J. and Jones, Eric and
            Kern, Robert and Larson, Eric and Carey, C J and
            Polat, {\.I}lhan and Feng, Yu and Moore, Eric W. and
            {VanderPlas}, Jake and Laxalde, Denis and Perktold, Josef and
            Cimrman, Robert and Henriksen, Ian and Quintero, E. A. and
            Harris, Charles R. and Archibald, Anne M. and
            Ribeiro, Ant{\^o}nio H. and Pedregosa, Fabian and
            {van Mulbregt}, Paul and {SciPy 1.0 Contributors}},
  title   = {{{SciPy} 1.0: Fundamental Algorithms for Scientific
            Computing in Python}},
  journal = {Nature Methods},
  year    = {2020},
  volume  = {17},
  pages   = {261--272},
  adsurl  = {https://rdcu.be/b08Wh},
  doi     = {10.1038/s41592-019-0686-2},
}

@article{Taruya_2018,
	doi = {10.1103/physrevd.98.103532},
  
	url = {https://doi.org/10.1103%2Fphysrevd.98.103532},
  
	year = 2018,
	month = {nov},
  
	publisher = {American Physical Society ({APS})},
  
	volume = {98},
  
	number = {10},
  
	author = {Atsushi Taruya and Takahiro Nishimichi and Donghui Jeong},
  
	title = {Grid-based calculation for perturbation theory of large-scale structure},
  
	journal = {Physical Review D}
}

@article{Mandelbaum:2018ouv,
    author = "Mandelbaum, Rachel and others",
    collaboration = "LSST Dark Energy Science",
    title = "{The LSST Dark Energy Science Collaboration (DESC) Science Requirements Document}",
    eprint = "1809.01669",
    archivePrefix = "arXiv",
    primaryClass = "astro-ph.CO",
    reportNumber = "FERMILAB-PUB-18-465-A",
    month = "9",
    year = "2018"
}

@article{Abidi:2018eyd,
    author = "Abidi, Muntazir Mehdi and Baldauf, Tobias",
    title = "{Cubic Halo Bias in Eulerian and Lagrangian Space}",
    eprint = "1802.07622",
    archivePrefix = "arXiv",
    primaryClass = "astro-ph.CO",
    doi = "10.1088/1475-7516/2018/07/029",
    journal = "JCAP",
    volume = "07",
    pages = "029",
    year = "2018"
}

@article{schmittfull2020modeling,
    title = {Modeling Galaxies in Redshift Space at the Field Level},
    author = {Marcel Schmittfull and Marko Simonović and Mikhail M. Ivanov and Oliver H. E. Philcox and Matias Zaldarriaga},
    eprint = {2012.03334},
    archivePrefix = {arXiv},
    primaryClass = {astro-ph.CO},
    journal = "JCAP",
    volume = "05",
    pages = "059",
    year = "2021",
    doi = "10.1088/1475-7516/2021/05/059"
}

@article{Van_Alfen_2024,
   title={An Empirical Model For Intrinsic Alignments: Insights From Cosmological Simulations},
   volume={7},
   ISSN={2565-6120},
   url={http://dx.doi.org/10.33232/001c.118783},
   DOI={10.33232/001c.118783},
   journal={The Open Journal of Astrophysics},
   publisher={Maynooth University},
   author={Van Alfen, Nicholas and Campbell, Duncan and Blazek, Jonathan and Leonard, C. Danielle and Lanusse, Francois and Hearin, Andrew and Mandelbaum, Rachel and },
   year={2024},
   month=June }

@article{Helly_2026,
   title={The FLAMINGO simulations data release},
   volume={57},
   ISSN={2213-1337},
   url={http://dx.doi.org/10.1016/j.ascom.2026.101159},
   DOI={10.1016/j.ascom.2026.101159},
   journal={Astronomy and Computing},
   publisher={Elsevier BV},
   author={Helly, John C. and McGibbon, Robert J. and Schaye, Joop and Schaller, Matthieu and McDonald, William and Braspenning, Joey and Broxterman, Jeger C. and Costello, Emily E. and Elbers, Willem and Forouhar Moreno, Victor J. and Frenk, Carlos S. and Jenkins, Adrian and Kugel, Roi and McCarthy, Ian G. and Salcido, Jaime and van Daalen, Marcel P. and Vandenbroucke, Bert and Yang, Tianyi},
   year={2026},
   month=Oct, pages={101159} }

@article{McGibbon_2025,
   title={SOAP: A Python Package for Calculating the Properties of Galaxies and Halos Formed in Cosmological Simulations},
   volume={10},
   ISSN={2475-9066},
   url={http://dx.doi.org/10.21105/joss.08252},
   DOI={10.21105/joss.08252},
   number={111},
   journal={Journal of Open Source Software},
   publisher={The Open Journal},
   author={McGibbon, Robert and Helly, John C. and Schaye, Joop and Schaller, Matthieu and Vandenbroucke, Bert},
   year={2025},
   month=July, pages={8252} }

@article{Hahn_2020,
   title={Higher order initial conditions for mixed baryon–CDM simulations},
   volume={503},
   ISSN={1365-2966},
   url={http://dx.doi.org/10.1093/mnras/staa3773},
   DOI={10.1093/mnras/staa3773},
   number={1},
   journal={Monthly Notices of the Royal Astronomical Society},
   publisher={Oxford University Press (OUP)},
   author={Hahn, Oliver and Rampf, Cornelius and Uhlemann, Cora},
   year={2020},
   month=Dec, pages={426–445} }

@article{jenkins2013panphasiauserguide,
      title={Panphasia: a user guide}, 
      author={Adrian Jenkins and Stephen Booth},
      year={2013},
      eprint={1306.5771},
      archivePrefix={arXiv},
      primaryClass={astro-ph.IM},
      url={https://arxiv.org/abs/1306.5771}, 
}

@article{DESI:2025yuk,
    author = "Valcin, D. and others",
    collaboration = "DESI",
    title = "{Combined tracer analysis for DESI 2024 BAO}",
    eprint = "2508.05467",
    archivePrefix = "arXiv",
    primaryClass = "astro-ph.CO",
    reportNumber = "FERMILAB-PUB-25-0587-PPD",
    doi = "10.1088/1475-7516/2026/04/058",
    journal = "JCAP",
    volume = "04",
    pages = "058",
    year = "2026"
}

@ARTICLE{1984ApJ...286...38W,
       author = {{White}, S.~D.~M.},
        title = "{Angular momentum growth in protogalaxies}",
      journal = {\apj},
         year = 1984,
        month = nov,
       volume = {286},
        pages = {38-41},
          doi = {10.1086/162573},
       adsurl = {https://ui.adsabs.harvard.edu/abs/1984ApJ...286...38W}
}

@article{Chisari_2015,
   title={Intrinsic alignments of galaxies in the Horizon-AGN cosmological hydrodynamical simulation},
   volume={454},
   ISSN={1365-2966},
   url={http://dx.doi.org/10.1093/mnras/stv2154},
   DOI={10.1093/mnras/stv2154},
   number={3},
   journal={Monthly Notices of the Royal Astronomical Society},
   publisher={Oxford University Press (OUP)},
   author={Chisari, N. and Codis, S. and Laigle, C. and Dubois, Y. and Pichon, C. and Devriendt, J. and Slyz, A. and Miller, L. and Gavazzi, R. and Benabed, K.},
   year={2015},
   month=Oct, pages={2736–2753} }

@ARTICLE{2007NJPh....9..444B,
       author = {{Bridle}, Sarah and {King}, Lindsay},
        title = "{Dark energy constraints from cosmic shear power spectra: impact of intrinsic alignments on photometric redshift requirements}",
      journal = {New Journal of Physics},
         year = 2007,
        month = dec,
       volume = {9},
       number = {12},
        pages = {444},
          doi = {10.1088/1367-2630/9/12/444},
archivePrefix = {arXiv},
       eprint = {0705.0166},
 primaryClass = {astro-ph},
       adsurl = {https://ui.adsabs.harvard.edu/abs/2007NJPh....9..444B}
}

@article{zjupa2020intrinsicalignmentsillustristngimplications,
    title = {Intrinsic Alignments in IllustrisTNG and their implications for weak lensing: Tidal shearing and tidal torquing mechanisms put to the test},
    author = {Jolanta Zjupa and Björn Malte Schäfer and Oliver Hahn},
    eprint = {2010.07951},
    archivePrefix = {arXiv},
    primaryClass = {astro-ph.GA},
    journal = "Mon. Not. Roy. Astron. Soc.",
    volume = "514",
    number = "2",
    pages = "2049--2072",
    year = "2022",
    doi = "10.1093/mnras/stac042"
}

@article{Schaye_2023,
   title={The FLAMINGO project: cosmological hydrodynamical simulations for large-scale structure and galaxy cluster surveys},
   volume={526},
   ISSN={1365-2966},
   url={http://dx.doi.org/10.1093/mnras/stad2419},
   DOI={10.1093/mnras/stad2419},
   number={4},
   journal={Monthly Notices of the Royal Astronomical Society},
   publisher={Oxford University Press (OUP)},
   author={Schaye, Joop and Kugel, Roi and Schaller, Matthieu and Helly, John C and Braspenning, Joey and Elbers, Willem and McCarthy, Ian G and van Daalen, Marcel P and Vandenbroucke, Bert and Frenk, Carlos S and Kwan, Juliana and Salcido, Jaime and Bahé, Yannick M and Borrow, Josh and Chaikin, Evgenii and Hahn, Oliver and Huško, Filip and Jenkins, Adrian and Lacey, Cedric G and Nobels, Folkert S J},
   year={2023},
   month=Aug, pages={4978–5020} }

@article{chudaykin2026reanalyzingdesidr15,
    title = {Reanalyzing DESI DR1: 5. Cosmological Constraints with Simulation-Based Priors},
    author = {Anton Chudaykin and Mikhail M. Ivanov and Oliver H. E. Philcox},
    eprint = {2602.18554},
    archivePrefix = {arXiv},
    primaryClass = {astro-ph.CO},
    journal = "Phys. Rev. D",
    volume = "114",
    number = "6",
    pages = "063527",
    year = "2026",
    doi = "10.1103/lg1n-k1rl"
}

@article{Samuroff_2021,
   title={Advances in constraining intrinsic alignment models with hydrodynamic simulations},
   volume={508},
   ISSN={1365-2966},
   url={http://dx.doi.org/10.1093/mnras/stab2520},
   DOI={10.1093/mnras/stab2520},
   number={1},
   journal={Monthly Notices of the Royal Astronomical Society},
   publisher={Oxford University Press (OUP)},
   author={Samuroff, S and Mandelbaum, R and Blazek, J},
   year={2021},
   month=Sept, pages={637–664} }

@article{Maion_2025,
   title={Probabilistic estimators of Lagrangian shape biases: Universal relations and physical insights},
   volume={699},
   ISSN={1432-0746},
   url={http://dx.doi.org/10.1051/0004-6361/202453212},
   DOI={10.1051/0004-6361/202453212},
   journal={Astronomy \& Astrophysics},
   publisher={EDP Sciences},
   author={Maion, F. and Stücker, J. and Angulo, R. E.},
   year={2025},
   month=July, pages={A271} }

@article{Fortuna_2020,
   title={The halo model as a versatile tool to predict intrinsic alignments},
   volume={501},
   ISSN={1365-2966},
   url={http://dx.doi.org/10.1093/mnras/staa3802},
   DOI={10.1093/mnras/staa3802},
   number={2},
   journal={Monthly Notices of the Royal Astronomical Society},
   publisher={Oxford University Press (OUP)},
   author={Fortuna, Maria Cristina and Hoekstra, Henk and Joachimi, Benjamin and Johnston, Harry and Chisari, Nora Elisa and Georgiou, Christos and Mahony, Constance},
   year={2020},
   month=Dec, pages={2983–3002} }

@article{sullivan2025localprimordialnongaussianbias,
    title = {Local Primordial Non-Gaussian Bias at the Field Level},
    author = {James M. Sullivan and Shi-Fan Chen},
    eprint = {2410.18039},
    archivePrefix = {arXiv},
    primaryClass = {astro-ph.CO},
    journal = "JCAP",
    volume = "03",
    pages = "016",
    year = "2025",
    doi = "10.1088/1475-7516/2025/03/016"
}

@article{sharma2025equilateralnongaussianbiasfield,
    title = {Equilateral non-Gaussian Bias at the Field Level},
    author = {Divij Sharma and James M. Sullivan and Kazuyuki Akitsu and Mikhail M. Ivanov},
    eprint = {2511.14677},
    archivePrefix = {arXiv},
    primaryClass = {astro-ph.CO},
    journal = "Phys. Rev. D",
    volume = "114",
    number = "4",
    pages = "043510",
    year = "2026",
    doi = "10.1103/zvgb-dbby"
}

@article{Pandya_2026,
   title={Differentiable Stochastic Halo Occupation Distribution with Galaxy Intrinsic Alignments},
   volume={9},
   ISSN={2565-6120},
   url={http://dx.doi.org/10.33232/001c.159889},
   DOI={10.33232/001c.159889},
   journal={The Open Journal of Astrophysics},
   publisher={Maynooth University},
   author={Pandya, Sneh and Blazek, Jonathan},
   year={2026},
   month=Mar }

@article{gatti2026halobasedintrinsicalignmentmodelsimulationbased,
      title={A halo-based intrinsic-alignment model for simulation-based inference}, 
      author={M. Gatti},
      year={2026},
      eprint={2609.37254},
      archivePrefix={arXiv},
      primaryClass={astro-ph.CO},
      url={https://arxiv.org/abs/2609.37254}, 
}

@article{sarkar2026simulationbasedpriorshibias,
      title={Simulation-Based Priors for HI Bias from Halo Occupation Physics}, 
      author={Debanjan Sarkar and Simon Foreman},
      year={2026},
      eprint={2607.14206},
      archivePrefix={arXiv},
      primaryClass={astro-ph.CO},
      url={https://arxiv.org/abs/2607.14206}, 
}

@article{zhang2025hodinformedprioreftbasedfullshape,
    title = {HOD-informed prior for EFT-based full-shape analyses of LSS},
    author = {Hanyu Zhang and Marco Bonici and Guido D'Amico and Simone Paradiso and Will J. Percival},
    eprint = {2409.12937},
    archivePrefix = {arXiv},
    primaryClass = {astro-ph.CO},
    journal = "JCAP",
    volume = "04",
    pages = "041",
    year = "2025",
    doi = "10.1088/1475-7516/2025/04/041"
}

@article{ivanov2025fullshapeanalysissimulationbasedpriors,
    title = {Full-shape analysis with simulation-based priors: cosmological parameters and the structure growth anomaly},
    author = {Mikhail M. Ivanov and Andrej Obuljen and Carolina Cuesta-Lazaro and Michael W. Toomey},
    eprint = {2409.10609},
    archivePrefix = {arXiv},
    primaryClass = {astro-ph.CO},
    journal = "Phys. Rev. D",
    volume = "111",
    number = "6",
    pages = "063548",
    year = "2025",
    doi = "10.1103/PhysRevD.111.063548"
}

@article{ivanov2025simulationbasedpriorssimulationsanalytic,
      title={Simulation-Based Priors without Simulations: an Analytic Perspective on EFT Parameters of Galaxies}, 
      author={Mikhail M. Ivanov},
      year={2025},
      eprint={2503.07270},
      archivePrefix={arXiv},
      primaryClass={astro-ph.CO},
      url={https://arxiv.org/abs/2503.07270}, 
}

@article{barreira2021galaxy,
    title = {Galaxy bias from forward models: linear and second-order bias of IllustrisTNG galaxies},
    author = {Alexandre Barreira and Titouan Lazeyras and Fabian Schmidt},
    eprint = {2105.02876},
    archivePrefix = {arXiv},
    primaryClass = {astro-ph.CO},
    journal = "JCAP",
    volume = "08",
    pages = "029",
    year = "2021",
    doi = "10.1088/1475-7516/2021/08/029"
}

@article{Schmittfull_2019,
   title={Modeling biased tracers at the field level},
   volume={100},
   ISSN={2470-0029},
   url={http://dx.doi.org/10.1103/PhysRevD.100.043514},
   DOI={10.1103/physrevd.100.043514},
   number={4},
   journal={Physical Review D},
   publisher={American Physical Society (APS)},
   author={Schmittfull, Marcel and Simonović, Marko and Assassi, Valentin and Zaldarriaga, Matias},
   year={2019},
   month={Aug}
}

@article{Michaux:2020yis,
    author = {Michaux, Micha\"el and Hahn, Oliver and Rampf, Cornelius and Angulo, Raul E.},
    title = "{Accurate initial conditions for cosmological N-body simulations: Minimizing truncation and discreteness errors}",
    eprint = "2008.09588",
    archivePrefix = "arXiv",
    primaryClass = "astro-ph.CO",
    doi = "10.1093/mnras/staa3149",
    journal = "Mon. Not. Roy. Astron. Soc.",
    volume = "500",
    number = "1",
    pages = "663--683",
    year = "2020"
}

@article{Catelan_2001,
   title={Intrinsic and extrinsic galaxy alignment},
   volume={320},
   ISSN={1365-2966},
   url={http://dx.doi.org/10.1046/j.1365-8711.2001.04105.x},
   DOI={10.1046/j.1365-8711.2001.04105.x},
   number={1},
   journal={Monthly Notices of the Royal Astronomical Society},
   publisher={Oxford University Press (OUP)},
   author={Catelan, P. and Kamionkowski, M. and Blandford, R. D.},
   year={2001},
   month=Jan, pages={L7–L13} }

@article{sanchezcid2026darkenergysurveyyear,
    title = {Dark Energy Survey Year 6 Results: Weak Lensing and Galaxy Clustering Cosmological Analysis Framework},
    author = {D. Sanchez-Cid and A. Ferté and J. Blazek and S. Samuroff and A. Amon and F. Andrade-Oliveira and J. M. Coloma-Nadal and J. Muir and A. Porredon and J. Prat and N. Weaverdyck and M. Yamamoto and D. Anbajagane and M. R. Becker and P. Carrilho and C. Chang and M. Crocce and G. Giannini and W. d'Assignies and J. DeRose and S. Dodelson and E. Krause and E. Legnani and J. Mena-Fernández and N. MacCrann and A. Pourtsidou and C. Preston and P. Rogozenski and M. Rodriguez-Monroy and R. Rosenfeld and E. Sanchez and I. Sevilla-Noarbe and M. Soares-Santos and C. To and M. A. Troxel and M. Tsedrik and B. Yin and J. Zuntz and T. M. C. Abbott and M. Aguena and S. Allam and O. Alves and S. Avila and D. Bacon and K. Bechtol and E. Bertin and S. Bocquet and D. Brooks and H. Camacho and R. Camilleri and A. Campos and A. Carnero Rosell and J. Carretero and F. J. Castander and R. Cawthon and A. Choi and L. N. da Costa and M. E. da Silva Pereira and T. M. Davis and J. De Vicente and S. Desai and C. Doux and A. Drlica-Wagner and T. Eifler and J. Elvin-Poole and S. Everett and A. E. Evrard and B. Flaugher and P. Fosalba and J. Frieman and J. García-Bellido and M. Gatti and E. Gaztanaga and P. Giles and K. Glazebrook and D. Gruen and G. Gutierrez and I. Harrison and K. Herner and S. R. Hinton and D. L. Hollowood and K. Honscheid and D. Huterer and B. Jain and D. J. James and N. Jeffrey and T. Kacprzak and K. Kuehn and O. Lahav and S. Lee and J. L. Marshall and F. Menanteau and R. Miquel and J. J. Mohr and J. Myles and R. C. Nichol and R. L. C. Ogando and A. Palmese and M. Paterno and W. J. Percival and A. A. Plazas Malagón and M. Raveri and A. Roodman and C. Sánchez and T. Schutt and E. Sheldon and N. Sherman and T. Shin and M. Smith and E. Suchyta and M. E. C. Swanson and M. Tabbutt and G. Tarle and D. Thomas and D. L. Tucker and V. Vikram and A. R. Walker and B. Yanny},
    eprint = {2601.14859},
    archivePrefix = {arXiv},
    primaryClass = {astro-ph.CO},
    journal = "Phys. Rev. D",
    volume = "114",
    number = "6",
    pages = "063506",
    year = "2026",
    doi = "10.1103/8n76-f6ln"
}

@article{spezzati2025equivalencefieldlevelinferenceconventional,
    title = {Equivalence of the field-level inference and conventional analyses on large scales},
    author = {Francesco Spezzati and Marco Marinucci and Marko Simonović},
    eprint = {2507.05378},
    archivePrefix = {arXiv},
    primaryClass = {astro-ph.CO},
    journal = "Phys. Rev. D",
    volume = "113",
    number = "6",
    pages = "063546",
    year = "2026",
    doi = "10.1103/yhg1-c63b"
}

@article{Nguyen_2024,
   title={How Much Information Can Be Extracted from Galaxy Clustering at the Field Level?},
   volume={133},
   ISSN={1079-7114},
   url={http://dx.doi.org/10.1103/PhysRevLett.133.221006},
   DOI={10.1103/physrevlett.133.221006},
   number={22},
   journal={Physical Review Letters},
   publisher={American Physical Society (APS)},
   author={Nguyen, Nhat-Minh and Schmidt, Fabian and Tucci, Beatriz and Reinecke, Martin and Kostić, Andrija},
   year={2024},
   month=Nov }

@article{akitsu2025cosmologyinferenceperturbativeforward,
      title={Cosmology inference with perturbative forward modeling at the field level: a comparison with joint power spectrum and bispectrum analyses}, 
      author={Kazuyuki Akitsu and Marko Simonović and Shi-Fan Chen and Giovanni Cabass and Matias Zaldarriaga},
      year={2025},
      eprint={2509.09673},
      archivePrefix={arXiv},
      primaryClass={astro-ph.CO},
      url={https://arxiv.org/abs/2509.09673}, 
}

@article{sullivan2025highredshiftmillenniumastridgalaxies,
    title = {High-redshift Millennium and Astrid galaxies in effective field theory at the field level},
    author = {James M. Sullivan and Carolina Cuesta-Lazaro and Mikhail M. Ivanov and Yueying Ni and Sownak Bose and Boryana Hadzhiyska and César Hernández-Aguayo and Lars Hernquist and Rahul Kannan},
    eprint = {2505.03626},
    archivePrefix = {arXiv},
    primaryClass = {astro-ph.CO},
    journal = "Phys. Rev. D",
    volume = "112",
    number = "8",
    pages = "083521",
    year = "2025",
    doi = "10.1103/1hps-3pf4"
}

@article{Shiferaw_2025,
   title={How Do Uncertainties in Galaxy Formation Physics Impact Field-level Galaxy Bias?},
   volume={989},
   ISSN={1538-4357},
   url={http://dx.doi.org/10.3847/1538-4357/ade66e},
   DOI={10.3847/1538-4357/ade66e},
   number={2},
   journal={The Astrophysical Journal},
   publisher={American Astronomical Society},
   author={Shiferaw, Mahlet and Kokron, Nickolas and Wechsler, Risa H.},
   year={2025},
   month=Aug, pages={218} }

@article{ivanov2025fundamentalphysicslymanalphaforest,
    title = {Fundamental physics with the Lyman-alpha forest: constraints on the growth of structure and neutrino masses from SDSS with effective field theory},
    author = {Mikhail M. Ivanov and Michael W. Toomey and Naim Göksel Karaçaylı},
    eprint = {2405.13208},
    archivePrefix = {arXiv},
    primaryClass = {astro-ph.CO},
    journal = "Phys. Rev. Lett.",
    volume = "134",
    number = "9",
    pages = "091001",
    year = "2025",
    doi = "10.1103/PhysRevLett.134.091001"
}

@article{ivanov2024millenniumastridgalaxieseffective,
      title={The Millennium and Astrid galaxies in effective field theory: comparison with galaxy-halo connection models at the field level}, 
      author={Mikhail M. Ivanov and Carolina Cuesta-Lazaro and Andrej Obuljen and Michael W. Toomey and Yueying Ni and Sownak Bose and Boryana Hadzhiyska and César Hernández-Aguayo and Lars Hernquist and Rahul Kannan and Volker Springel},
      year={2024},
      eprint={2412.01888},
      archivePrefix={arXiv},
      primaryClass={astro-ph.CO},
      url={https://arxiv.org/abs/2412.01888}, 
}

@article{Obuljen_2023,
   title={Modeling HI at the field level},
   volume={108},
   ISSN={2470-0029},
   url={http://dx.doi.org/10.1103/PhysRevD.108.083528},
   DOI={10.1103/physrevd.108.083528},
   number={8},
   journal={Physical Review D},
   publisher={American Physical Society (APS)},
   author={Obuljen, Andrej and Simonović, Marko and Schneider, Aurel and Feldmann, Robert},
   year={2023},
   month=Oct }

@article{de_Belsunce_2026,
   title={Modeling the Cosmological Lyman-$\alpha$
                    Forest at the Field Level},
   volume={136},
   ISSN={1079-7114},
   url={http://dx.doi.org/10.1103/8g9h-68ms},
   DOI={10.1103/8g9h-68ms},
   number={10},
   journal={Physical Review Letters},
   publisher={American Physical Society (APS)},
   author={de Belsunce, Roger and Ivanov, Mikhail M. and Sullivan, James M. and Akitsu, Kazuyuki and Chen, Shi-Fan},
   year={2026},
   month=Mar }

@article{bakx2023effectivefieldtheoryintrinsic,
    title = {Effective Field Theory of Intrinsic Alignments at One Loop Order: a Comparison to Dark Matter Simulations},
    author = {Thomas Bakx and Toshiki Kurita and Nora Elisa Chisari and Zvonimir Vlah and Fabian Schmidt},
    eprint = {2303.15565},
    archivePrefix = {arXiv},
    primaryClass = {astro-ph.CO},
    journal = "JCAP",
    volume = "10",
    pages = "005",
    year = "2023",
    doi = "10.1088/1475-7516/2023/10/005"
}

@article{Maion_2024,
   title={HYMALAIA: a hybrid lagrangian model for intrinsic alignments},
   volume={531},
   ISSN={1365-2966},
   url={http://dx.doi.org/10.1093/mnras/stae1331},
   DOI={10.1093/mnras/stae1331},
   number={2},
   journal={Monthly Notices of the Royal Astronomical Society},
   publisher={Oxford University Press (OUP)},
   author={Maion, Francisco and Angulo, Raul E and Bakx, Thomas and Chisari, Nora Elisa and Kurita, Toshiki and Pellejero-Ibáñez, Marcos},
   year={2024},
   month=May, pages={2684–2700} }

@article{Eifler_2021,
   title={Cosmology with the <i>Roman Space Telescope</i>: synergies with the Rubin Observatory Legacy Survey of Space and Time},
   volume={507},
   ISSN={1365-2966},
   url={http://dx.doi.org/10.1093/mnras/stab533},
   DOI={10.1093/mnras/stab533},
   number={1},
   journal={Monthly Notices of the Royal Astronomical Society},
   publisher={Oxford University Press (OUP)},
   author={Eifler, Tim and Simet, Melanie and Krause, Elisabeth and Hirata, Christopher and Huang, Hung-Jin and Fang, Xiao and Miranda, Vivian and Mandelbaum, Rachel and Doux, Cyrille and Heinrich, Chen and Huff, Eric and Miyatake, Hironao and Hemmati, Shoubaneh and Xu, Jiachuan and Rogozenski, Paul and Capak, Peter and Choi, Ami and Doré, Olivier and Jain, Bhuvnesh and Jarvis, Mike and Kruk, Jeffrey and MacCrann, Niall and Masters, Dan and Rozo, Eduardo and Spergel, David N and Troxel, Michael and von der Linden, Anja and Wang, Yun and Weinberg, David H and Wenzl, Lukas and Wu, Hao-Yi},
   year={2021},
   month=Mar, pages={1514–1527} }

@article{Vlah_2021,
   title={Galaxy shape statistics in the effective field theory},
   volume={2021},
   ISSN={1475-7516},
   url={http://dx.doi.org/10.1088/1475-7516/2021/05/061},
   DOI={10.1088/1475-7516/2021/05/061},
   number={05},
   journal={Journal of Cosmology and Astroparticle Physics},
   publisher={IOP Publishing},
   author={Vlah, Zvonimir and Chisari, Nora Elisa and Schmidt, Fabian},
   year={2021},
   month=May, pages={061} }

@article{Bakx_2025,
   title={The Bispectrum of Intrinsic Alignments: Theory Modelling and Forecasts for Stage IV Galaxy Surveys},
   volume={8},
   ISSN={2565-6120},
   url={http://dx.doi.org/10.33232/001c.144388},
   DOI={10.33232/001c.144388},
   journal={The Open Journal of Astrophysics},
   publisher={Maynooth University},
   author={Bakx, Thomas and Eggemeier, Alexander and Kurita, Toshiki and Chisari, Nora Elisa and Vlah, Zvonimir},
   year={2025},
   month=Sept }

@article{Bakx_2026,
   title={The Bispectrum of Intrinsic Alignments: II. Precision Comparison Against Dark Matter Simulations},
   volume={9},
   ISSN={2565-6120},
   url={http://dx.doi.org/10.33232/001c.156361},
   DOI={10.33232/001c.156361},
   journal={The Open Journal of Astrophysics},
   publisher={Maynooth University},
   author={Bakx, Thomas and Kurita, Toshiki and Eggemeier, Alexander and Chisari, Nora Elisa and Vlah, Zvonimir},
   year={2026},
   month=Feb }

@article{euclid2025,
   title={Euclid: I. Overview of the
                    Euclid
                    mission},
   volume={697},
   ISSN={1432-0746},
   url={http://dx.doi.org/10.1051/0004-6361/202450810},
   DOI={10.1051/0004-6361/202450810},
   journal={Astronomy andamp; Astrophysics},
   publisher={EDP Sciences},
   author={Euclid Collaboration},
   year={2025},
   month=Apr, pages={A1} }

@article{committee2025romanobservationstimeallocation,
      title={Roman Observations Time Allocation Committee: Final Report and Recommendations}, 
      author={Roman Observations Time Allocation Committee and Core Community Survey Definition Committees},
      year={2025},
      eprint={2505.10574},
      archivePrefix={arXiv},
      primaryClass={astro-ph.IM},
      url={https://arxiv.org/abs/2505.10574}, 
}

@article{moreno2025assessingsubhalofinderscosmological,
    title = {Assessing subhalo finders in cosmological hydrodynamical simulations},
    author = {Victor J. Forouhar Moreno and John Helly and Robert McGibbon and Joop Schaye and Matthieu Schaller and Jiaxin Han and Roi Kugel and Yannick M. Bahe},
    eprint = {2502.06932},
    archivePrefix = {arXiv},
    primaryClass = {astro-ph.CO},
    journal = "Mon. Not. Roy. Astron. Soc.",
    volume = "543",
    number = "2",
    pages = "1339--1372",
    year = "2025",
    doi = "10.1093/mnras/staf1478"
}

@article{cao2026fisherforecastscosmologicalyields,
    title = {Fisher Forecasts for Cosmological Yields from $3\!\times\!2$pt Analysis of the Roman Space Telescope High Latitude Imaging Survey},
    author = {Kaili Cao and David H. Weinberg and Vivian Miranda and Nihar Dalal and Tim Eifler and Jiachuan Xu and Haley Bowden and Katarina Markovič},
    eprint = {2601.00438},
    archivePrefix = {arXiv},
    primaryClass = {astro-ph.CO},
    journal = "Astrophys. J.",
    volume = "1010",
    number = "1",
    pages = "21",
    year = "2026",
    doi = "10.3847/1538-4357/ae9bc0"
}

@article{Hirata:2004gc,
    author = "Hirata, Christopher M. and Seljak, Uros",
    title = "{Intrinsic alignment-lensing interference as a contaminant of cosmic shear}",
    eprint = "astro-ph/0406275",
    archivePrefix = "arXiv",
    doi = "10.1103/PhysRevD.82.049901",
    journal = "Phys. Rev. D",
    volume = "70",
    pages = "063526",
    year = "2004",
    note = "[Erratum: Phys.Rev.D 82, 049901 (2010)]"
}

@article{McCullough_2026,
   title={Dark Energy Survey Year 3: Blue shear},
   volume={113},
   ISSN={2470-0029},
   url={http://dx.doi.org/10.1103/wmp3-qc3p},
   DOI={10.1103/wmp3-qc3p},
   number={10},
   journal={Physical Review D},
   publisher={American Physical Society (APS)},
   author={McCullough, J. and Amon, A. and Legnani, E. and Gruen, D. and Roodman, A. and Friedrich, O. and MacCrann, N. and Becker, M. and Myles, J. and Dodelson, S. and Samuroff, S. and Blazek, J. and Prat, J. and Pieres, A. and Ferté, A. and Alarcon, A. and Drlica-Wagner, A. and Choi, A. and Navarro-Alsina, A. and Campos, A. and Plazas Malagón, A. A. and Porredon, A. and Ross, A. J. and Carnero Rosell, A. and Yin, B. and Flaugher, B. and Yanny, B. and Sánchez, C. and Chang, C. and Davis, C. and To, C. and Doux, C. and Brooks, D. and James, D. J. and Sanchez Cid, D. and Hollowood, D. L. and Huterer, D. and Rykoff, E. S. and Gaztanaga, E. and Huff, E. M. and Suchyta, E. and Sheldon, E. and Sanchez, E. and Tarsitano, F. and Andrade-Oliveira, F. and Castander, F. J. and Bernstein, G. M. and Gutierrez, G. and Giannini, G. and Tarle, G. and Diehl, H. T. and Huang, H. and Harrison, I. and Sevilla-Noarbe, I. and Tutusaus, I. and Ferrero, I. and Elvin-Poole, J. and Marshall, J. L. and Muir, J. and Weller, J. and Zuntz, J. and Carretero, J. and DeRose, J. and Frieman, J. and Cordero, J. and De Vicente, J. and García-Bellido, J. and Mena-Fernández, J. and Eckert, K. and Romer, A. K. and Bechtol, K. and Herner, K. and Honscheid, K. and Kuehn, K. and Secco, L. F. and da Costa, L. N. and Paterno, M. and Soares-Santos, M. and Gatti, M. and Raveri, M. and Yamamoto, M. and Smith, M. and Carrasco Kind, M. and Troxel, M. A. and Jarvis, M. and Swanson, M. E. C. and Weaverdyck, N. and Lahav, O. and Doel, P. and Wiseman, P. and Miquel, R. and Gruendl, R. A. and Cawthon, R. and Allam, S. and Hinton, S. R. and Bridle, S. L. and Bocquet, S. and Desai, S. and Pandey, S. and Everett, S. and Lee, S. and Shin, T. and Aguena, M. and Alves, O. and Buckley-Geer, E. and Burke, D. L. and Conselice, C. and Crocce, M. and Pereira, M. E. S. and Farahi, A. and Honscheid, K. and Jeffrey, N. and Lima, M. and Palmese, A. and Schubnell, M. and Vikram, V. and Vincenzi, M. and Zhang, Y. and },
   year={2026},
   month=May }

@article{Akitsu_2021,
   title={Imprint of anisotropic primordial non-Gaussianity on halo intrinsic alignments in simulations},
   volume={103},
   ISSN={2470-0029},
   url={http://dx.doi.org/10.1103/PhysRevD.103.083508},
   DOI={10.1103/physrevd.103.083508},
   number={8},
   journal={Physical Review D},
   publisher={American Physical Society (APS)},
   author={Akitsu, Kazuyuki and Kurita, Toshiki and Nishimichi, Takahiro and Takada, Masahiro and Tanaka, Satoshi},
   year={2021},
   month=Apr }

@article{Kurita_2023,
   title={Constraints on anisotropic primordial non-Gaussianity from intrinsic alignments of SDSS-III BOSS galaxies},
   volume={108},
   ISSN={2470-0029},
   url={http://dx.doi.org/10.1103/PhysRevD.108.083533},
   DOI={10.1103/physrevd.108.083533},
   number={8},
   journal={Physical Review D},
   publisher={American Physical Society (APS)},
   author={Kurita, Toshiki and Takada, Masahiro},
   year={2023},
   month=Oct }

@article{Xu_2023,
   title={Evidence for baryon acoustic oscillations from galaxy–ellipticity correlations},
   volume={7},
   ISSN={2397-3366},
   url={http://dx.doi.org/10.1038/s41550-023-02035-4},
   DOI={10.1038/s41550-023-02035-4},
   number={10},
   journal={Nature Astronomy},
   publisher={Springer Science and Business Media LLC},
   author={Xu, Kun and Jing, Y. P. and Zhao, Gong-Bo and Cuesta, Antonio J.},
   year={2023},
   month=July, pages={1259–1264} }

@article{Shim_2025,
   title={Squeezing Full-shape Dynamical Dark Energy Constraints with Galaxy Alignments},
   volume={985},
   ISSN={2041-8213},
   url={http://dx.doi.org/10.3847/2041-8213/add1dd},
   DOI={10.3847/2041-8213/add1dd},
   number={1},
   journal={The Astrophysical Journal Letters},
   publisher={American Astronomical Society},
   author={Shim, Junsup and Okumura, Teppei and Taruya, Atsushi},
   year={2025},
   month=May, pages={L14} }

@article{Blazek_2019,
   title={Beyond linear galaxy alignments},
   volume={100},
   ISSN={2470-0029},
   url={http://dx.doi.org/10.1103/PhysRevD.100.103506},
   DOI={10.1103/physrevd.100.103506},
   number={10},
   journal={Physical Review D},
   publisher={American Physical Society (APS)},
   author={Blazek, Jonathan A. and MacCrann, Niall and Troxel, M. A. and Fang, Xiao},
   year={2019},
   month=Nov }

@article{McDonald_2009,
   title={Clustering of dark matter tracers: generalizing bias for the coming era of precision LSS},
   volume={2009},
   ISSN={1475-7516},
   url={http://dx.doi.org/10.1088/1475-7516/2009/08/020},
   DOI={10.1088/1475-7516/2009/08/020},
   number={08},
   journal={Journal of Cosmology and Astroparticle Physics},
   publisher={IOP Publishing},
   author={McDonald, Patrick and Roy, Arabindo},
   year={2009},
   month={Aug},
   pages={020–020}
}

@article{Schmitz_2018,
   title={Time evolution of intrinsic alignments of galaxies},
   volume={2018},
   ISSN={1475-7516},
   url={http://dx.doi.org/10.1088/1475-7516/2018/07/030},
   DOI={10.1088/1475-7516/2018/07/030},
   number={07},
   journal={Journal of Cosmology and Astroparticle Physics},
   publisher={IOP Publishing},
   author={Schmitz, D.M. and Hirata, C.M. and Blazek, J. and Krause, E.},
   year={2018},
   month=July, pages={030–030} }

@article{laureijs2011euclid,
      title={Euclid Definition Study Report}, 
      author={R. Laureijs and others},
      year={2011},
      collaboration="Euclid",
      eprint={1110.3193},
      archivePrefix={arXiv},
      primaryClass={astro-ph.CO}
}

@ARTICLE{2013PASP..125..306F,
       author = {{Foreman-Mackey}, Daniel and {Hogg}, David W. and {Lang}, Dustin and
         {Goodman}, Jonathan},
        title = "{emcee: The MCMC Hammer}",
      journal = {\pasp},
         year = 2013,
        month = mar,
       volume = {125},
       number = {925},
        pages = {306},
          doi = {10.1086/670067},
archivePrefix = {arXiv},
       eprint = {1202.3665},
 primaryClass = {astro-ph.IM},
       adsurl = {https://ui.adsabs.harvard.edu/abs/2013PASP..125..306F}
}

@article{Villaescusa_Navarro_2020,
	doi = {10.3847/1538-4365/ab9d82},
  
	url = {https://doi.org/10.3847%2F1538-4365%2Fab9d82},
  
	year = 2020,
	month = {aug},
  
	publisher = {American Astronomical Society},
  
	volume = {250},
  
	number = {1},
  
	pages = {2},
  
	author = {Francisco Villaescusa-Navarro and ChangHoon Hahn and Elena Massara and Arka Banerjee and Ana Maria Delgado and Doogesh Kodi Ramanah and Tom Charnock and Elena Giusarma and Yin Li and Erwan Allys and Antoine Brochard and Cora Uhlemann and Chi-Ting Chiang and Siyu He and Alice Pisani and Andrej Obuljen and Yu Feng and Emanuele Castorina and Gabriella Contardo and Christina D. Kreisch and Andrina Nicola and Justin Alsing and Roman Scoccimarro and Licia Verde and Matteo Viel and Shirley Ho and Stephane Mallat and Benjamin Wandelt and David N. Spergel},
  
	title = {The Quijote Simulations},
  
	journal = {The Astrophysical Journal Supplement Series}
}

@article{Kokron:2021faa,
    author = "Kokron, Nickolas and DeRose, Joseph and Chen, Shi-Fan and White, Martin and Wechsler, Risa H.",
    title = "{Priors on red galaxy stochasticity from hybrid effective field theory}",
    eprint = "2112.00012",
    archivePrefix = "arXiv",
    primaryClass = "astro-ph.CO",
    journal = "Mon. Not. Roy. Astron. Soc.",
    volume = "514",
    number = "2",
    pages = "2198--2213",
    year = "2022",
    doi = "10.1093/mnras/stac1420"
}

@article{chen2024lensinglowanalysisdesitimesdes,
    title = {Not all lensing is low: An analysis of DESI$\times$DES using the Lagrangian Effective Theory of LSS},
    author = {S. Chen and J. DeRose and R. Zhou and M. White and S. Ferraro and C. Blake and J. U. Lange and R. H. Wechsler and J. Aguilar and S. Ahlen and D. Brooks and T. Claybaugh and K. Dawson and A. de la Macorra and P. Doel and A. Font-Ribera and E. Gaztañaga and S. Gontcho A Gontcho and G. Gutierrez and K. Honscheid and C. Howlett and R. Kehoe and D. Kirkby and T. Kisner and A. Kremin and M. Landriau and L. Le Guillou and M. Manera and A. Meisner and R. Miquel and J. A. Newman and G. Niz and N. Palanque-Delabrouille and W. J. Percival and F. Prada and G. Rossi and E. Sanchez and D. Schlegel and M. Schubnell and D. Sprayberry and G. Tarlé and B. A. Weaver},
    eprint = {2407.04795},
    archivePrefix = {arXiv},
    primaryClass = {astro-ph.CO},
    journal = "Phys. Rev. D",
    volume = "110",
    number = "10",
    pages = "103518",
    year = "2024",
    doi = "10.1103/PhysRevD.110.103518"
}
\bibliographystyle{jhep}

\end{document}